\documentclass{article} 
\usepackage{iclr2025_conference,times}

\usepackage{amsmath,amsfonts,bm}

\def\eqref#1{equation~\ref{#1}}

\def\1{\bm{1}}

\DeclareMathAlphabet{\mathsfit}{\encodingdefault}{\sfdefault}{m}{sl}
\SetMathAlphabet{\mathsfit}{bold}{\encodingdefault}{\sfdefault}{bx}{n}

\usepackage{hyperref}
\usepackage{url}
\usepackage{algorithm}
\usepackage{algorithmic}

\usepackage{graphicx}
\usepackage[utf8]{inputenc} 
\usepackage[T1]{fontenc}    
\usepackage{hyperref}       
\usepackage{url}            
\usepackage{booktabs}       
\usepackage{amsfonts}       
\usepackage{nicefrac}       
\usepackage{microtype}      
\usepackage{xcolor}         
\usepackage{multicol}
\usepackage{multirow}
\usepackage{amssymb}
\usepackage{wrapfig}

\title{Self-Supervised Diffusion MRI Denoising via Iterative and Stable Refinement}

\author{Chenxu Wu\textsuperscript{\rm 1,2}, Qingpeng Kong\textsuperscript{\rm 1,2}, Zihang Jiang\textsuperscript{\rm 1,2,}\thanks{Corresponding authors} \& S. Kevin Zhou\textsuperscript{\rm 1,2,3,4,}\footnotemark[1] \\
\textsuperscript{\rm 1}School of Biomedical Engineering, Division of Life Sciences and Medicine, USTC\\
\textsuperscript{\rm 2}MIRACLE Center, Suzhou Institute for Advance Research, USTC\\
\textsuperscript{\rm 3}State Key Laboratory of Precision and Intelligent Chemistry, USTC\\
\textsuperscript{\rm 4}Key Laboratory of Intelligent Information Processing of CAS, Institute of Computing Technology, CAS\\
\texttt{\{wuchenxu,qpkong27\}@mail.ustc.edu.cn,\{jzh0103,s.kevin.zhou\}@gmail.com}\\
}

\iclrfinalcopy 
\begin{document}

\maketitle

\begin{abstract}
\vspace{-5pt}
Magnetic Resonance Imaging (MRI), including diffusion MRI (dMRI), serves as a ``microscope'' for anatomical structures and routinely mitigates the influence of low signal-to-noise ratio scans by compromising temporal or spatial resolution. However, these compromises fail to meet clinical demands for both efficiency and precision. Consequently, denoising is a vital preprocessing step, particularly for dMRI, where clean data is unavailable. In this paper, we introduce Di-Fusion, a fully self-supervised denoising method that leverages the latter diffusion steps and an adaptive sampling process. Unlike previous approaches, our single-stage framework achieves efficient and stable training without extra noise model training and offers adaptive and controllable results in the sampling process. Our thorough experiments on real and simulated data demonstrate that Di-Fusion achieves state-of-the-art performance in microstructure modeling, tractography tracking, and other downstream tasks. Code is available at \textcolor{blue}{https://github.com/FouierL/Di-Fusion}.
\vspace{-5pt}
\end{abstract}

\section{Introduction}
\vspace{-5pt}
Characterizing real-world noise using data distributions is difficult~\citep{huang2021neighbor2neighbor}, particularly in non-invasive imaging modalities such as Magnetic Resonance Imaging (MRI), where the noise predominantly originates from {numerous factors including thermal fluctuations}~\citep{fadnavis2020patch2self}. MRI, including its subtype Diffusion-weighted Magnetic Resonance Imaging (dMRI)~\citep{basser1994mr}, serves as a vital tool for observing inferred structures~\citep{le2003looking,le2006artifacts,schilling2019limits} and necessitates a high Signal-to-Noise Ratio (SNR) for better clinical decision making. 
While it is possible to improve the SNR by increasing the acquisition time or reducing the image resolution, either way hinders the clinical application of MRI. Therefore, much research has focused on processing techniques like denoising for dMRI to improve its SNR and reduce acquisition time, which holds a great significance for clinical efficiency and accuracy.

The dMRI typically consists of 4D data ($X \in {{\mathbb{R}}^{w \times h \times d \times l}}$), including 3D spatial coordinates ($w$,$h$ and $d$) and 1D diffusion vectors ($l$), in which diffusion is measured along different gradient directions~\citep{westin2016q}. Different clinical applications require varying numbers of diffusion vectors and acquisition strategies, {leading to diverse noise sources and distributions, which complicates noise modeling and denoising implementation}. For supervised methods~\citep{gibbons2019simultaneous,kaye2020accelerating}, not only is it non-practical to obtain paired data with high SNR and low SNR, but the diversity of dMRI also leads to distributional shifts among different datasets, resulting in a fundamental drop in their performances~\citep{pmlr-v139-darestani21a}. Different from these approaches, our method offers a self-supervised solution for dMRI denoising through a single-stage construction and an efficient adaptive sampling process. Without the need for paired training data or clean data, our method is capable of removing the noise from dMRI with a denoising diffusion model. To mitigate the drift problem, a {\bf Fusion} process is proposed to align the forward process. Moreover, as real-world noise is difficult to characterize, a ``{\bf Di-}'' process is introduced to represent the noise distribution in a more effective manner. Consequently, our method {\bf Di-Fusion} is able to achieve better denoising results while preserving the desired anatomical structures. 

The main contributions of our work are three-fold: \textbf{(i)} We propose Di-Fusion, a stable and self-supervised dMRI denoising method leveraging the latter diffusion steps (Section \ref{Intuition of conditional training}). Di-Fusion integrates the statistical self-supervised denoising techniques~\citep{batson2019noise2self} into the diffusion models through the Fusion process and ``Di-'' process (Section \ref{Modifications on forward process}). \textbf{(ii)} Di-Fusion enables iterative refinement through an adaptive sampling process (Section \ref{Reverse process}). \textbf{(iii)} With thorough comparisons on real and simulated data, Di-Fusion demonstrates state-of-the-art denoising performance in microstructure modeling, tractography, and other downstream tasks (Section \ref{expriments}).

\section{Background and related works}
\label{gen_inst}
\vspace{-5pt}




\subsection{Statistical self-supervised image denoising} \label{Statistical self-supervised Image Denoising}
\vspace{-5pt}

Built upon the assumption that additive noise is pixel-wise independent, Noise2Noise~\citep{lehtinen2018noise2noise} learns the process of image restoration solely by observing corrupted measurements:
\begin{equation} \label{eq:noise2noise}
   \mathop {\arg \min }\limits_\theta  \left\{ {\mathbb{E}{{\left\| {{f_\theta }\left( {x'} \right) - x} \right\|}^2}} \right\} \approx \mathop {\arg \min }\limits_\theta  \left\{ {\mathbb{E}{{\left\| {{f_\theta }\left( {x'} \right) - y} \right\|}^2} + \mathbb{E}{{\left\| {x - y} \right\|}^2}} \right\},
\end{equation}
where $x$ and $x'$ are independent corrupted measurements of the clean ground truth $y$ and ${f_\theta }$ is a denoising function which is parameterized by $\theta$. Due to the assumption of independent noise, $\mathbb{E}{{\left\| {x - y} \right\|}^2}$ is usually a constant. Furthermore, Noise2Self~\citep{batson2019noise2self} proposes the $\mathcal{J}$-invariant theory, using only the same corrupted measurement to perform denoising. Following this theory, Noise2Void~\citep{krull2019noise2void}, Laine \textit{et al.}~\citep{laine2019high} and Noise2Same~\citep{xie2020noise2same} focus on how to construct unorganized collections of corrupted images by masked-based blind spot networks. Noisier2Noise~\citep{moran2020noisier2noise} and Noisy-As-Clean~\citep{xu2020noisy} add additional noise to the original noisy image to generate training image pairs. Nevertheless, these methods exhibit a significant drop in performance when confronted with real-world noisy images, particularly when the explicit noise model is unknown~\citep{huang2021neighbor2neighbor,mansour2023zero}.

\subsection{Diffusion models}
\label{Diffusion models}
\vspace{-5pt}

Denoising Diffusion Probabilistic Model (DDPM)~\citep{ho2020denoising,sohl2015deep} emerges as a powerful generative model, which is composed of a parameterized Markov chain with $T$ diffusion steps to fit a given data distribution. The forward process $q\left( {{x_t}|{x_{t - 1}}} \right)$ serves to perturb the data by gradually adding Gaussian noise based on a pre-defined noise schedule ${\beta _{1, \cdots ,T}}$ (Following ~\citep{ho2020denoising}, $\sigma _t^2 := {\beta _t}$, ${\alpha _t}: = 1 - {\beta _t}$ and ${\bar \alpha _t}: = \prod\nolimits_{s = 1}^t {{\alpha _s}}$ are sets of predetermined constants in this paper) until the data distribution approaches a standard Gaussian distribution:
\begin{equation} \label{eq:forward}
   q\left( {{x_{1:T}}{\rm{|}}{x_0}} \right): = \prod\limits_{t = 1}^T q \left( {{x_t}|{x_{t - 1}}} \right) , \quad q\left( {{x_t}|{x_{t - 1}}} \right) := {\cal N}\left( {{x_t};\sqrt {1 - {\beta _t}} {x_{t - 1}},{\beta _t}{\mathbf{I}}} \right).
\end{equation}
The reverse process starts from a Gaussian distribution $z \sim {\cal N}\left( {\mathbf{0},\mathbf{I}} \right)$ and uses a parameterized Gaussian transformation kernel ${{F}_\theta }$ to learn the step-by-step restoration of the original data distribution:
\begin{equation} \label{eq:reverse ddpm}
   {p_{F}}\left( {{x_{0:T}}} \right) := p\left( {{x_T}} \right)\mathop \prod \limits_{t = 1}^T {p_{F}}\left( {{x_{t - 1}}|{x_t}} \right) , \quad {p_{F}}\left( {{x_{t - 1}}|{x_t}} \right) := {\cal N}\left( {{x_{t-1}};{{F}_\theta }({x_t},t),\sigma _t^2{\mathbf{I}}} \right).
\end{equation}
Recently, there has been a large interest in exploring ways to enhance the extensibility and sampling efficiency of DDPM. For enhancing extensibility, ~\citep{song2019generative} uses gradient of the log density as a force to pull a random sample across the data space towards regions with a high data density characterized by $p\left( x \right)$~\citep{croitoru2023diffusion} by adopting Langevin dynamics algorithm~\citep{hyvarinen2005estimation}. ~\citep{song2020score} further extend the score function as solutions to reverse-time Stochastic Differential Equation (SDE) and extends DDPM to continuous states. Cold diffusion~\citep{bansal2024cold} investigates the necessity of Gaussian noise or any form of randomness for diffusion models to work effectively in practical scenarios. ~\citep{zhoudenoising} introduces a family of processes that interpolate between two paired distributions given as endpoints. For accelerating sampling speed, ~\citep{song2020denoising} and ~\citep{watson2021learning} generalize DDPM by introducing a class of non-Markovian diffusion processes that achieves the same sampling objective.

Previous works have demonstrated that diffusion models can be effectively applied to image restoration tasks~\citep{kawar2022denoising,xia2023diffir,ozdenizci2023restoring,chung2022improving,saharia2022palette,fei2023generative}. Conditioned on a low-resolution input image, ~\citep{saharia2022image} performs image super-resolution via repeated refinement. ~\citep{chung2022mr} , ~\citep{song2021solving}, ~\citep{song2024diffusionblend} and ~\citep{gao2023corediff} extend diffusion models to inverse problems in medical imaging. However, these models require clean data (e.g., normal-dose CT) to capture their prior data distribution, which makes direct application of these methods to dMRI data impractical because no clean data is available in dMRI itself. Our method does not require extra noise model training or clean ground truth $y$ and can be applied to the aforementioned scenarios.

\subsection{Related works}
\label{Related work about dMRI denoising}
\vspace{-5pt}

The initial denoising methods employed for dMRI are adaptations of techniques developed for natural images, like non-local means (NL-means~\citep{coupe2008optimized} and its variants~\citep{chen2016xq,coupe2012adaptive}). Under the assumption that small spatial structures exhibit relative consistency across varied dMRI measurements, Local Principal Component Analysis (LPCA)~\citep{manjon2013diffusion} and its Marchenko-Pastur extension (MPPCA)~\citep{veraart2016denoising} project dMRI to a local low-rank approximation. Training the Noise2Noise~\citep{lehtinen2018noise2noise} model directly using the same slices from different volumes can result in excessively smooth outcomes (Shown in the experiments of ~\citep{xiang2023ddm}). So, utilizing the entire volumes, Patch2Self~\citep{fadnavis2020patch2self} trains a full-rank locally linear denoiser to perform volume-wise denoising. {Patch2Self2~\citep{fadnavis2024patch2self2} further enhances the computational efficiency of Patch2Self}. Recently, Corruption2Self~\citep{tu2025scorebased} extends denoising score matching to accommodate noisy observations and provides a framework for denoising MRI. A state-of-the-art self-supervised method DDM2~\citep{xiang2023ddm} is proposed for denoising dMRI, which incorporates statistical image denoising into the diffusion model in a three-stage framework. However, the results obtained by DDM2 are prone to over-denoising as its performances in downstream tasks are not satisfactory (Section \ref{Effect on tractography}). 

\section{Methods}
\label{Methods}
\vspace{-5pt}

4D dMRI consists of independent noisy samples acquired at different gradient directions. Considering $x = {X_{*,*,i,j}}$ ($i$: slice index, $j$: volume index) as the target slice to denoise, $x' = {X_{*,*,i,j - 1}}$ and $x$ are independent corrupted measurements of the clean ground truth $y$. In this section, we demonstrate how to decompose the single-step mapping from $x'$ to $x$ into $T$ steps using a parameterized Markov chain (We denote ${\cal{F}_\theta }$ as the parameterized transformation kernel in our method). \textbf{We provide a complete definition of the entire Di-Fusion in Appendix \ref{Di-Fusion}}.

There are five questions to be answered in our method. \textbf{Q1:} Since $x'$ and $x$ are still different, how can we obtain the forward process to construct the multi-step mapping between two endpoints? \textbf{Q2:} How can we represent the noise distribution without extra noise model training? \textbf{Q3:} How can training be conducted with only noisy data? \textbf{Q4:} Why does Di-Fusion only leverage the latter diffusion steps? \textbf{Q5:} How does the reverse process enable iterative refinement?

\subsection{Modifications of forward process}
\label{Modifications on forward process}
\vspace{-5pt}

\begin{figure}
  \vspace{-2mm}
  \centering
  \includegraphics[scale = 0.7]{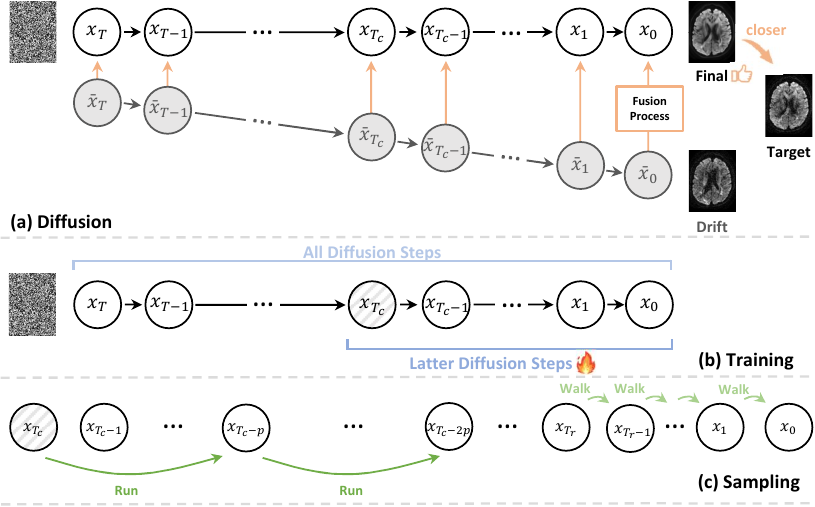}
  \caption{(a) Fusion process (Section \ref{Fusion process}) aligns $\left\{ {{{\bar x}_t}} \right\}_1^{{T}}$ to $\left\{ {{x_t}} \right\}_1^{{T}}$ and avoids drift (``Drift'' means drifted results, ``Final'' means the denoised version of ``Target''); (b) Training the latter diffusion steps (Section \ref{Intuition of conditional training}) imposes restrictions on the generation ability of diffusion models and decreases uncertainty; (c) \textit{Run-Walk} accelerated sampling (Section \ref{Run-Walk accelerated sampling}) accelerates the entire sampling process.}
  \label{key feature}
  \vspace{-2mm}
\end{figure}

\paragraph{Q1}
We use $\cal{F}_\theta$ to map from $x'$ to $x$, considering $x'$ as $x_T$ and $x$ as $x_0$, $\cal{F}_\theta$ should take $x_t$ and $t$ as input and output $x_{out}$ close to $x$:
\begin{equation} \label{eq:x'2x}
   x + {\epsilon_t} = {x_{out}}={{\cal F}_\theta }\left( {{x_t},t} \right),\quad{\left\| {x - {x_{out}}} \right\|^2} < \varepsilon, 
\end{equation}
where $\varepsilon$ represents a small positive value, $\epsilon_t$ is a perturbation term that depends on $t$, and $\epsilon_t$ decays as $t \to 0$. Performing the reverse process of DDPM, we find that $x_{t-1}$ should be a linear interpolation between ${x_{out}}$ and ${x_t}$ {plus a noise} instead of ${{\bar x}_{t-1}}$, the major difference is introduced by $x_{out}$ (See Appendix \ref{Proof: x barx diff} for detailed derivations):
\begin{equation} \label{eq:x'xbackward}
   {x_{t - 1}} = \underbrace {\frac{{\sqrt {{{\bar \alpha }_{t - 1}}} {\beta _t}}}{{1 - {{\bar \alpha }_t}}}({x + {\epsilon_t})}}_{{\rm{major}}{\kern 1pt} {\rm{difference}}} + \frac{{\sqrt {{\alpha _t}} \left( {1 - {{\bar \alpha }_{t - 1}}} \right)}}{{1 - {{\bar \alpha }_t}}}{x_t} + {\sigma _t}z \ne {{\bar x}_{t - 1}} = {\sqrt {{{\bar \alpha }_{t - 1}}} x' + \sqrt {1 - {{\bar \alpha }_{t - 1}}} z}, 
\end{equation}
where $z \sim {\cal N}\left( {\mathbf{0},\mathbf{I}} \right)$, $\left\{ {{{\bar x}_t}} \right\}_1^{{T}}$ are obtained by directly performing the forward process in DDPM and $\left\{ {{x_t}} \right\}_1^{{T}}$ are obtained from the reverse process of DDPM. Since the component $\epsilon_t \to 0$ as $t \to 0$, a larger proportion of $x_{t-1}$ aligns closer to $x$, {rather than merely being a noisy version of $x'$. If we still feed $x_{t-1}$ and $t-1$ into ${\cal F}_\theta$, it will cause output deviations, which accumulate in the sampling chain and ultimately} lead to the drift problem (Fig. \ref{key feature} (a)).

\paragraph{Fusion process (Q1)}\label{Fusion process}



Since ${\mathcal{F}_\theta }$ learns the mapping from $x'$ to $x$, $\left\{ {{x_t}} \right\}_1^{{T}}$ should be combinations of ${x}$ and ${x'}$, augmented with a sampled noise $z \sim {\cal N}\left( {\mathbf{0},\mathbf{I}} \right)$. These combinations can be approximated by utilizing the reverse process in DDPM to compute the linear interpolation between ${x'}$ to $x$:
\begin{equation} \label{eq:fusion}
   x_t^ *  = {\lambda _1^t} x + {\lambda _2^t} x',
\end{equation}
{\begin{equation} \label{eq:forwardour}
q\left( {{x_{t}}{\rm{|}}x_t^*} \right): = {\cal{N}}\left( {{x_t};\sqrt {{{\bar \alpha }_t}} x_t^*, ({1 - {{\bar \alpha }_t}}) {\mathbf{I}}} \right),
\end{equation}}

where we rewrite $\lambda _1^t=\frac{{\sqrt {{{\overline \alpha  }_{t - 1}}} {\beta _t}}}{{1 - {{\overline \alpha  }_t}}}$ and $\lambda _2^t=\frac{{\sqrt {{\alpha _t}} \left( {1 - {{\overline \alpha  }_{t - 1}}} \right)}}{{1 - {{\overline \alpha  }_t}}}$ for simplification. As $t$ decreases, ${x_t^ *}$ becomes closer to $x$ since $\lambda _1^t$ has a higher value. By substituting ${x_t^ *}$ for $x'$ in Eq. (\ref{eq:x'xbackward}), the Fusion process can be achieved, which obtain $x_t^ *$ with different $t$ as shown in Fig. \ref{linear interpolation}. {Intuitively, the Fusion process gradually introduces the target denoising slice $x$ to the model, guiding the model to optimize in a fixed direction, thereby mitigating the drift.} We thereby address \textbf{Q1} by defining the forward process $q\left( {{x_{t}}{\rm{|}}x_t^*} \right)$.

\paragraph{Q2}
Approximating noise as $z$ is definitely a feasible approach. However, the noise distribution in the real world often exhibits complex statistical properties, and thus cannot be easily captured mathematically (Section \ref{Statistical self-supervised Image Denoising}). Similar challenges also exist in dMRI.

\paragraph{``Di-'' process (Q2)}\label{"Di-" process}

To better characterize real-world noise, we represent the noise distribution involving the input noisy data. Since $x$ and $x'$ are independent corrupted measurements of the redundant part $y$ and have independent noise, directly calculating $x-x'$ leaves some linear combinations of noise ($x=y+n_1$, $x'=y+n_2$, $x-x'=n_1-n_2$, here $n_1$ and $n_2$ represent the noise in $x$ and $x'$, respectively), we perform a zero-mean operation on these linear combinations of noise to comply with the zero-mean constraint of $z$:
\begin{equation} \label{eq:Di-}
   {\xi _{x - x'}} = mess\left( {\left( {x - x'} \right) - {\mu _{x - x'}}} \right),\quad{\mu _{x - x'}} = \frac{{\sum\nolimits_{m = 1}^w {\sum\nolimits_{n = 1}^h {\left( {{x_{mn}} - {{x'}_{mn}}} \right)} } }}{{w \cdot h}},
\end{equation}
where $mess\left(  \cdot  \right)$ means spatial shuffling operation originated from DDM2~\citep{xiang2023ddm}, ${\mu _{x - x'}}$ is the mean of ${x - x'}$. ${\xi _{x - x'}}$ theoretically preserves the variance information of the noise (See Appendix \ref{Proof: variance information of nosie} for proof) and will serve as the noise distribution employed in both $q\left( {{x_{t}}{\rm{|}}x_t^*} \right)$ and ${p_\mathcal{F}}\left( {{x_{t - 1}}|{x_t}} \right)$. In this case, the forward process and reverse process no longer follow a Gaussian distribution, but they can be represented as Eq. (\ref{Di-Fusion forward}) and Eq. (\ref{eq:reverse}), respectively. In Fig. \ref{noise}, we demonstrate through experiments that ${\xi_{x - x'}}$ has different statistical properties from $z$. In Fig. \ref{Diprocess}, we show the impact of ${\xi_{x - x'}}$ and $z$ on the reverse process.

\subsection{Training process}
\label{training}
\vspace{-5pt}

\paragraph{$\mathcal{J}$-Invariance optimization (Q3)}\label{J-Invariance optimization}
When training ${\mathcal{F}_\theta }$, we first consider $x$ and ${x'}$ as ${\cal J} = \left\{ {x,x'} \right\}$. Assuming that the noise distributions of $x$ and ${x'}$ are mutually independent, the model with ${x'}$ as input and $x$ as the optimization target satisfies the property of input-output independence. According to the \textit{Proposition 1} declared in Noise2Self~\citep{batson2019noise2self}, the loss between ${{\mathcal{F}_\theta }\left( {x'} \right)}$ and ${x}$ will in expectation equal to the loss between ${{\mathcal{F}_\theta }\left( {x'} \right)}$ and clean ground truth ${y}$, plus a constant $\mathbb{E}{{\left\| {x - y} \right\|}^2}$(Eq. (\ref{eq:noise2noise})). Therefore, minimizing $\mathbb{E}{{\left\| {{\mathcal{F}_\theta }\left( {x'} \right) - x} \right\|}^2}$ is equivalent to minimizing $\mathbb{E}{{\left\| {{\mathcal{F}_\theta }\left( {x'} \right) - y} \right\|}^2}$ with respect to the clean ground truth ${y}$ and our simplified training objective is:
\begin{equation}
 {L_{{\rm{simple}}}}(\theta ): = {{\mathbb{E}}_{t,x_t^*,{\xi _{x - x'}}}}\left[ {{{\left\| {x - {\cal{F}_\theta }(\sqrt {{{\bar \alpha }_t}} x_t^* + \sqrt {1 - {{\bar \alpha }_t}} {\xi _{x - x'}},t)} \right\|}^2}} \right] 
 \label{eq:training_objective_simple}
\end{equation}

\paragraph{Intuition of training the latter diffusion steps (Q4)}\label{Intuition of conditional training}
In DDPM, it is shown that when conditioned on the same latent, the samples share high-level attributes (when conditioned on say ${x_{250}}$, the samples are close to each other)~\citep{ho2020denoising}. It is because of the thorough training in the former diffusion steps (${x_{{T}}} \to {x_{{T_c}}}$) that DDPM possesses diverse generative capabilities. Since we perform an image denoising task with such a strong prior (from one noisy volume to another noisy volume), training only the latter diffusion steps is possible to reduce the diverse generative capabilities of ${\mathcal{F}_\theta }$. More precisely, only the last $T_c$ steps in the Markov chain (${x_{{T_c}}} \to {x_{{0}}}$) are trained. In this way, a generative training task is simplified into a conditional generation task (${x_{{T_c}}} \to {x_{{0}}},{T_c} \le T$), with more $x_0$ information provided in $\left\{ {{x_t}} \right\}_1^{{T_c}}$ (Fig. \ref{key feature} (b)).

\begin{figure}
  \centering
  \vspace{-4mm}
  \includegraphics[scale = 0.48]{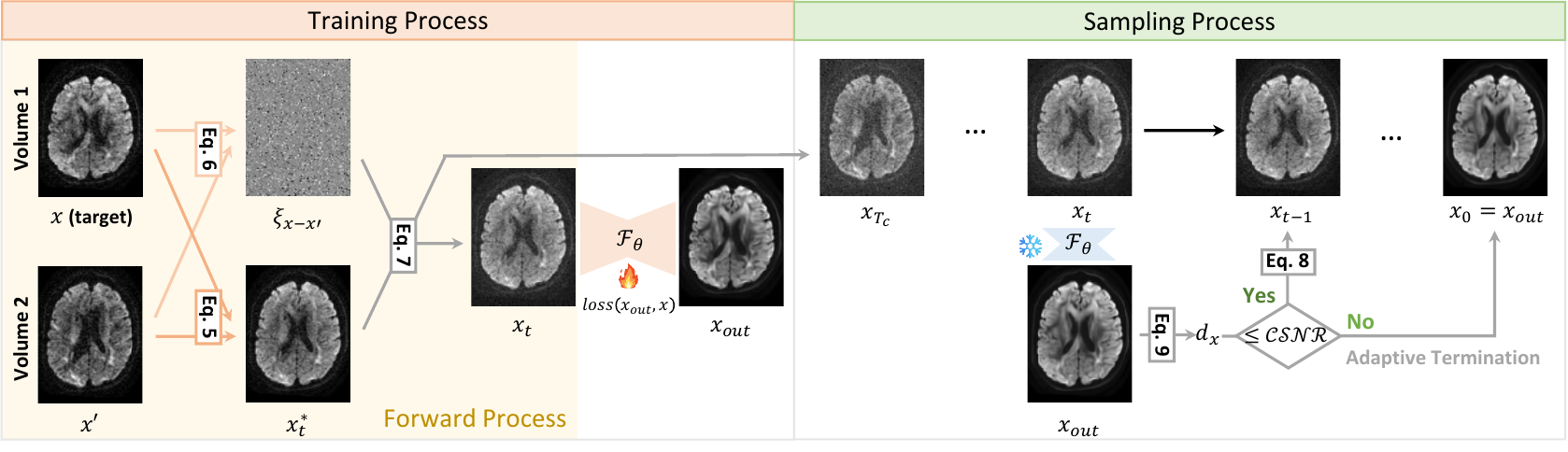}
  \vspace{-4mm}
  \caption{Overview of our single-stage Di-Fusion. The training process does not involve any extra model training apart from ${\cal F}_\theta$, and the sampling process offers adaptive and controllable results.}
  \label{framework}
  \vspace{-2mm}
\end{figure}

There are two main reasons for adopting this strategy. Firstly, training the latter diffusion steps weakens the generation capacity of the diffusion model, reducing its diversity. This, in turn, lowers the uncertainty in denoising results for our task. Secondly, with the same training time, obtaining a more stable ${\mathcal{F}_\theta }$ is possible. By training only the latter diffusion steps, each step receives more training iterations, resulting in improved stability for the model performance. Algorithm \ref{alg:train} outlines the training process, and Fig. \ref{framework} (left) provides an overview of the entire training process.

\begin{algorithm}
\caption{Training process}
\label{alg:train}
\begin{algorithmic}
\STATE Initialize ${\mathcal{F}_\theta}$ randomly; input 4D data: $X \in {\mathbb{R}^{w \times h \times d \times l}}$
\REPEAT 
\STATE $t \sim {\rm{Uniform}}\left( {\left\{ {1, \cdots ,T_c} \right\}} \right)$ \COMMENT{training the latter diffusion steps in Section \ref{Intuition of conditional training}}
\STATE $x = {X_{*,*,i,j}},x' = {X_{*,*,i,j-1}}$ \COMMENT{$i$: slice index, $j$: volume index}
\STATE ${\xi_{x - x'}} = mess\left( {\left( {x - x'} \right) - {\mu _{x - x'}}} \right)$ \COMMENT{Eq. (\ref{eq:Di-})}
\STATE ${x_t^ *} = \lambda _1^tx + \lambda _2^tx'$ \COMMENT{Eq. (\ref{eq:fusion})}
\STATE take gradient descent step on: ${\nabla _\theta }{\left\| {x - {{\mathcal{F}}_\theta }\left( {\sqrt {{{\bar \alpha }_t}} {x_t^ *} + \sqrt {1 - {{\bar \alpha }_t}} {\xi_{x - x'}},t} \right)} \right\|^2}$\COMMENT{Eq. (\ref{eq:training_objective_simple})}
\STATE resample $i$ and $j$ 
\UNTIL{converged}
\end{algorithmic}
\end{algorithm}
\vspace{-3mm}

\subsection{Sampling process}
\label{Reverse process}
\vspace{-5pt}

We make two specific modifications on ${p_\mathcal{F}}\left( {{x_{t - 1}}|{x_t}} \right)$ to achieve an adaptive sampling process and directly begin the sampling process at ${x_{T_c}}$. See Fig. \ref{framework} (right) for an overview of the entire sampling process and Algorithm \ref{alg:sample} for a detailed description of the complete sampling process.



\paragraph{\textit{Run-Walk} accelerated sampling}\label{Run-Walk accelerated sampling}
After substituting the standard normal distribution in Eq. (\ref{eq:reverse ddpm}) with ${\xi_{x - x'}}$, a typical reverse process ${p_{\mathcal{F}}}\left( {{x_{t - 1}}|{x_t}} \right)$ could be formulated as:
\begin{equation} \label{eq:reverse}
    {p_{\mathcal{F}}}\left( {{x_{t - 1}}|{x_t}} \right) \to {x_{t - 1}} = \lambda _1^t{{{\cal F}_\theta }\left( {{x_t},t} \right)} + \lambda _2^t{x_t} + \left( {{\sigma _t} \cdot \eta } \right){\xi_{x - x'}},
\end{equation}
where $\eta$ is a constant. DDIM~\citep{song2020denoising} notes a special case when \(\sigma_t = 0\) for all \(t\) \footnote{We do this by multiplying \({\sigma_t} \xi_{x - x'}\) with \(\eta\), where \(\eta = 0\) if no special instructions are provided.}; the forward process is deterministic given \(x_{t-1}\) and \(x_t^*\) except for \(t = 1\); in the sampling process, the coefficient before the noise \({\xi_{x - x'}}\) becomes zero, resulting in an implicit probabilistic model~\citep{mohamed2016learning}. However, we do not follow the uniform step strategy of DDIM in the sampling process; instead, we use \textit{Run-Walk} accelerated sampling. Consider a DDPM sampling process from $x_{T_c}$ to $x_0$, when $t$ is large ($t>{T_{r}}$,$1 \le {T_{r}} \le {T_c}$), the speed during each reverse process is slow; thus, acceleration can be applied (\textit{Run}). Conversely, when $t$ is small ($t<{T_{r}}$), the speed is fast, and deceleration is required (\textit{Walk}). In equation form, the difference between ${x_{t-1}}$ and ${x_{t}}$ can be represented as (See Appendix \ref{Proof: speed to x0} for additional derivations):
\begin{equation} \label{eq:step}
    \begin{array}{*{20}{l}}
{{x_{t - 1}} - {x_t}}& = &{\underbrace {\lambda _1^t\left( {x - {x_t}} \right)}_{{\rm{speed}}} + \underset{{\rm{perturbation}}}{\underbrace{{\lambda _1^t} {\epsilon_t}}}}.
    \end{array}
\end{equation}
When $t$ is large (e.g. $t>{T_{r}}$), $\lambda _1^t$ approaches zero and the speed ($\lambda _1^t\left( {{x} - {x_t}} \right)$) towards ${x_{0}}$ is relatively slow. This is when we perform accelerated sampling. When reaching the latter sampling process, $\lambda _1^t$ progressively increases and the speed towards ${x_{0}}$ is quite fast. This is when we stop accelerating. When $T_{r}=1$, \textit{Run-Walk} accelerated sampling degenerates into DDIM sampling. When $T_{r}=T_c$, \textit{Run-Walk} accelerated sampling degenerates into DDPM sampling.

Now let us consider the forward process as defined not on all $\left\{ {{x_t}} \right\}_1^{{T_c}}$, but on a subset $\{x_{\tau_1}, \ldots, x_{\tau_S}\}$, where $\tau$ is an increasing sub-sequence of $[1, \ldots, T_c]$ of length $S$. 
In particular, we define the sequential forward process over $x_{\tau_1}, \ldots, x_{\tau_S}$ (${x_{{\tau_k}}} = \sqrt {{{\bar \alpha }_{{\tau_k}}}} \left( {\lambda _1^{_{{\tau_k}}}x + \lambda _2^{_{{\tau_k}}}x'} \right) + \sqrt {1 - {{\bar \alpha }_{{\tau_k}}}} {\xi_{x - x'}}$, $1 \le k \le {S}$). The sampling process now samples according to $\text{reversed}(\tau)$ (In practice, $\tau  = \left\{ {1,2, \cdots ,{T_{r}} - 1,{T_{r}},{T_{r}} + p, \cdots ,{T_c} - p,{T_c}} \right\}$, where $p$ is an integer representing the acceleration factor). This can be more intuitively understood in Fig. \ref{key feature} (c).

\begin{algorithm}
\caption{Sampling process}
\label{alg:sample}
\begin{algorithmic}
\STATE Load pre-trained ${\mathcal{F}_\theta}$; input: $X \in {\mathbb{R}^{w \times h \times d \times l}}$, $i$, $j$ and $\cal CSNR$
\STATE $x = {X_{*,*,i,j}},x' = {X_{*,*,i,j-1}}$ \COMMENT{$i$: slice index, $j$: volume index}
\STATE ${\xi_{x - x'}} = mess\left( {\left( {x - x'} \right) - {\mu _{x - x'}}} \right)$ \COMMENT{Eq. (\ref{eq:Di-})}
\STATE ${x_{{T_c}}} = \sqrt {{{\bar \alpha }_{{T_c}}}} \left( {\lambda _1^{_{{T_c}}}x + \lambda _2^{_{{T_c}}}x'} \right) + \sqrt {1 - {{\bar \alpha }_{{T_c}}}} {\xi_{x - x'}}$ \COMMENT{Eq. (\ref{eq:forwardour})}
\STATE ${b_x = \frac{\sum_{m=1}^{w}\sum_{n=1}^{h}1}{2 \cdot \sum_{m=1}^{w}\sum_{n=1}^{h}{{\mathbb{I}}_{(x_{mn} > \rho_1)}}} + \frac{\sum_{m=1}^{w}\sum_{n=1}^{h}1}{2 \cdot \sum_{m=1}^{w}\sum_{n=1}^{h}{{\mathbb{I}}_{(x_{mn} > \rho_2)}}}}$ \COMMENT{Eq. (\ref{eq:brainvalue})}
\FOR{$\tau_k  = \rm{reversed}\left\{ {1,2, \cdots ,{T_{r}} - 1,{T_{r}},{T_{r}} + p, \cdots ,{T_c} - p,{T_c}} \right\}$}
\STATE ${\xi_{x - x'}} = mess\left( {{\xi_{x - x'}}} \right)$ \COMMENT{Shuffle ${\xi_{x - x'}}$ again}
\STATE ${x_{out}}={{\cal F}_\theta }\left( {{x_{\tau_k}},{\tau_k}} \right)$ \COMMENT{Eq. (\ref{eq:x'2x})}
\STATE ${d_x} = {\left\| {x - {x_{out}}} \right\|^2} \times {b_x}$ \COMMENT{Eq. (\ref{eq:reverseloss})}
\IF{${{d_x}>{\mathcal{CSNR}}}$} 
\STATE ${x_0}={x_{out}}$; \textbf{break}\COMMENT{In Section \ref{Towards iterative and adjustable refinement}}
\ELSE 
\STATE  ${x_{{\tau_{k-1}}}} = \lambda _1^{\tau_k}{x_{out}} + \lambda _2^{\tau_k}{x_{\tau_k}} + \left( {{\sigma _{\tau_k}} \cdot \eta } \right){\xi_{x - x'}}$ \COMMENT{Eq. (\ref{eq:reverse})}
\ENDIF 
\ENDFOR
\RETURN ${x_0}$
\end{algorithmic}
\end{algorithm}
\paragraph{Towards iterative and controllable refinement (Q5)}\label{Towards iterative and adjustable refinement}
During our experiments, we observe that the intermediate outputs, ${x_{out}}$, obtained during the sampling process demonstrate a substantial success in denoising. Therefore, we explore the feasibility of adaptive termination to stop sampling prematurely. More specifically, the degree of denoising in ${x_{out}}$ can be characterized by its distance from ${x}$. Nevertheless, directly computing this distance ${\left\| {x - {x_{out}}} \right\|^2}$ presents a problem. When the slice index $i$ is located at the edges, the resulting distance tends to be smaller due to the reduced amount of brain tissue in these edge slices. Hence, it would be preferable to calculate a coefficient ${b_x}$ that accounts for the ratio of brain tissue to the entire image. Here, we adopt a simple definition:
\begin{equation} \label{eq:brainvalue}
   {b_x = \frac{\sum_{m=1}^{w}\sum_{n=1}^{h}1}{2 \cdot \sum_{m=1}^{w}\sum_{n=1}^{h}{{\mathbb{I}}_{(x_{mn} > \rho_1)}}} + \frac{\sum_{m=1}^{w}\sum_{n=1}^{h}1}{2 \cdot \sum_{m=1}^{w}\sum_{n=1}^{h}{{\mathbb{I}}_{(x_{mn} > \rho_2)}}}},
\end{equation}
where ${\rho_1}$ and ${\rho_2}$ are constants depending on the data normalization methods employed\footnote{In our experiments, ${\rho_1}=-0.93$ and ${\rho_2}=-0.95$, changing their values has little impact on the results.} and $\mathbb{I}\left(  \cdot  \right)$ is an indicator function. $b_x$ can be used to correct ${d_x}$:
\begin{equation} \label{eq:reverseloss}
   {d_x} = {\left\| {x - {x_{out}}} \right\|^2} \times {b_x}.
\end{equation}
Since $d_x$ has been corrected, we can pre-define a universal value $\cal CSNR$ to perform the iterative and controllable refinement on each slice. During ${p_\mathcal{F}}\left( {{x_{t - 1}}|{x_t}} \right)$, we first get ${x_{out}}={{\cal F}_\theta }\left( {{x_t},t} \right)$ and compute ${d_x}$ (Eq. (\ref{eq:reverseloss})). Then if ${d_x}$ is greater than $\cal CSNR$, ${x_{0}} = {x_{out}}$ and the refinement iteration breaks. In contrast, the refinement iteration continues if ${d_x}$ is smaller than $\cal CSNR$. In extreme cases, when $\mathcal{CSNR} = 0$, the reverse process will immediately terminate and output ${x_{0}}$. When $\mathcal{CSNR} = 1$, the complete $\text{reversed}(\tau)$ will be executed until completion.

\section{Experiments}
\label{expriments}


\subsection{Datasets and competing methods}
\label{Datasets and competing methods}

\paragraph{Datasets}
To thoroughly evaluate Di-Fusion, we perform experiments on three publicly available brain dMRI datasets acquired using different, commonly-used acquisition schemes: \textit{(i)} High-Angular Resolution Diffusion Imaging (Stanford HARDI, $X \in {{\mathbb{R}}^{106 \times 81 \times 76 \times 150}}$~\citep{rokem2016stanford}); \textit{(ii)} Multi-Shell (Sherbrooke 3-Shell dataset, $X \in {{\mathbb{R}}^{128 \times 128 \times 64 \times 193}}$~\citep{garyfallidis2014dipy}); \textit{(iii)} Single-Shell (Parkinson’s Progression Markers Initiative (PPMI) dataset, $X \in {{\mathbb{R}}^{116 \times 116 \times 72 \times 64}}$~\citep{marek2011parkinson}). Simulated experiments are carried out on the fastMRI datasets~\citep{tibrewala2023fastmri,zbontar2018fastMRI}. We simulate noisy data with five different noise intensities.
\paragraph{Competing methods}
We compare Di-Fusion with five competing methods in the main paper (all experimental details are provided in Appendix \ref{Experiment and reproducibility details}): \textit{(i)} Adaptive Soft Coefficient Matching (ASCM), an improved extension of non-local means denoising~\citep{coupe2012adaptive}. \textit{(ii)} Deep Image Prior (DIP), a self-supervised denoising method~\citep{ulyanov2018deep}. \textit{(iii)} Noisier2Noise (Nr2N), a statistic-based denoising method~\citep{moran2020noisier2noise}. \textit{(iv)} Patch2Self (P2S), a multi-volume denoising method~\citep{fadnavis2020patch2self}. \textit{(v)} DDM2, state-of-the-art denoising method~\citep{xiang2023ddm}. {More comparisons with other denoising methods, including MPPCA \citep{veraart2016denoising}, Noise2Score \citep{kim2021noise2score}, Recorrupted2Recorrupted \citep{pang2021recorrupted}, and Patch2Self2 \citep{fadnavis2024patch2self2}, can be found in Appendix \ref{Supplementary experimental results}}.

\subsection{Impacts on downstream clinical tasks}
\label{Impacts on modeling}
\paragraph{Effect on tractography}
\label{Effect on tractography}
\begin{figure}
  \vspace{-4mm}
  \centering
  \includegraphics[scale = 0.345]{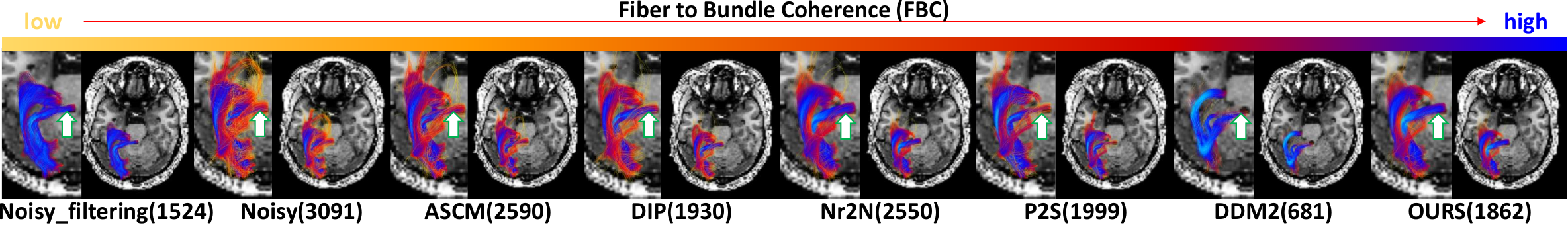}
  \vspace{-4mm}
  \caption{Density map of FBC projected on the streamlines of the OR bundles. The numbers in parentheses represent the number of streamlines. Di-Fusion generates the minimal number of streamlines while maintaining high FBCs (consider ``Noisy\_filtering'' as references for high FBCs).}
  \label{tractography}
  \vspace{-3mm}
\end{figure}
The noise in dMRI can impact tractography results, potentially causing the generation of spurious streamlines by the tracking algorithm~\citep{fadnavis2020patch2self,garyfallidis2014dipy,schilling2019limits}. We explore the effect of denoising on probabilistic tracking~\citep{girard2014towards} by employing the Fiber Bundle Coherency (FBC) metric~\citep{portegies2015improving} and reconstruct the optic radiation (OR) bundles (See Appendix \ref{Modeling implementation details} for details). Since low FBCs indicate which fibers are poorly aligned with their neighbors, we further clean the tractography results of noisy data (captioned by ``Noisy\_filtering'') using a stopping criterion~\citep{meesters2016cleaning}. In Fig. \ref{tractography}, we show the effect on the tractography of OR. Although DDM2 yields the fewest streamlines, noticeably, it misses the high FBCs indicated by the white arrow in Fig. \ref{tractography}. \textit{Di-Fusion generates the minimal number of streamlines while maintaining high FBCs}, which indicates that our method maximizes the denoising performance while preserving fiber bundle information.
\paragraph{Effect on microstructure model fitting}
\label{Effect on microstructure model fitting}
\begin{figure}
  \vspace{-4mm}
  \centering
  \includegraphics[scale = 0.50]{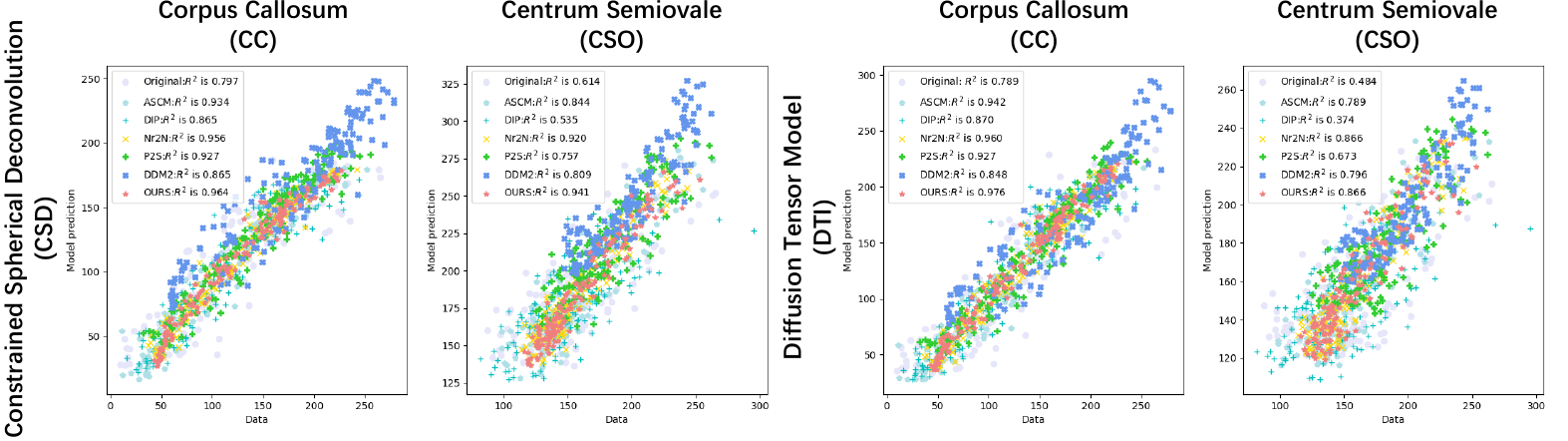}
  \vspace{-3mm}
  \caption{Scatter plots of the microstructure model predictions against input data. The top-left of each plot shows the quantitative $R^2$ metric computed from each model fit on the corresponding data. Our data points are more concentrated (higher $R^2$).}
  \label{microstructure model fitting}
  \vspace{-3mm}
\end{figure}
Denoising methods can be compared based on their accuracy in fitting the diffusion signal~\citep{ades2018evaluation}. We apply two commonly used microstructure fitting models, namely diffusion tensor model (DTI)~\citep{basser1994mr} and Constrained Spherical Deconvolution (CSD)~\citep{tournier2007robust}, on noisy and denoised data (Appendix \ref{Modeling implementation details} for details). We show the {quantitative} $R^2$ metric of microstructure predictions against the original data for Corpus Callosum (CC) and Centrum SemiOvale (CSO) in Table \ref{microstruture table} and the corresponding scatter plots are in Fig. \ref{microstructure model fitting}. \textit{As measured by $R^2$, Di-Fusion achieves the best results across all four different settings}. This means that Di-Fusion aids in the characterization of the microstructure.

\paragraph{Effect on diffusion signal estimates} 
We further examine how the denoising quality translates to creating {quantitative and clinically-relevant} DTI~\citep{basser1994mr} diffusion signal estimates (Details are in Appendix \ref{Modeling implementation details}). In Fig. \ref{DTI}, we show fractional anisotropy, axial diffusivity, mean diffusivity, and radial diffusivity comparisons. \textit{Our method effectively suppresses noise and reconstructs fiber tracts}.

\subsection{Quantitative and qualitative results on \textit{in-vivo} data}
\label{Quantitative and qualitative results}

\paragraph{Quantitative results on SNR/CNR metrics}\label{Quantitative results}
Considering the infeasibility of using metrics that need clean ground truth and their limited correlation with clinical utility~\citep{mason2019comparison}, computing metrics in downstream clinical regions of interest is more reasonable~\citep{adamson2021ssfd}. 
To quantify the denoising performance, we employ Signal-to-Noise Ratio (SNR) and Contrast-to-Noise Ratio (CNR) metrics (Details are in Appendix \ref{SNR and CNR implementation details}). The quantitative denoising results are reported as mean and standard deviation scores for the complete 4D volumes in Fig \ref{quantitative results}. \textit{Di-Fusion indicates better performance in terms of SNR/CNR metrics}.
\begin{wrapfigure}{r}{0.6\textwidth}
  \vspace{-2mm}
  \centering
  \includegraphics[scale = 0.5]{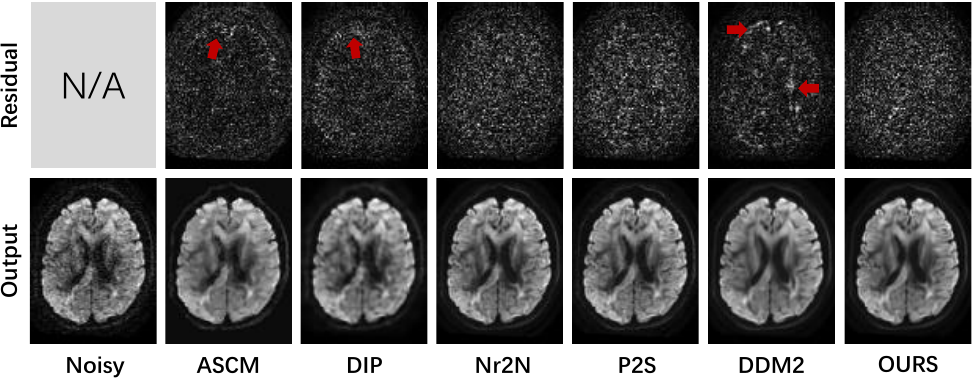}
  \vspace{-4mm}
  \caption{Qualitative results. ``OURS'' results are obtained when ${\mathcal{CSNR}}=0.040$. The area indicated by the red arrow does not appear in ``OURS'', indicating that Di-Fusion does not remove structural information during denoising.}
  \label{qualitative results}
    \vspace{-10mm}
\end{wrapfigure}
\paragraph{Qualitative results}\label{Qualitative results}
In Fig. \ref{qualitative results}, we visualize the denoising results and residuals on axial slices for Stanford HARDI (Fig. \ref{morequalitative_stanfordhardi}, \ref{morequalitative_s3sh} and Fig. \ref{morequalitative_ppmi} for more qualitative results). From the residuals of Fig. \ref{qualitative results}, the area indicated by the red arrow does not appear in ``OURS'', indicating that Di-Fusion does not remove any anatomical structure during denoising. All pictures are best viewed when zoomed in.


\subsection{Quantitative results on simulated data}
\label{Quantitative results on simulated data}

We show the PSNR and SSIM metrics \footnote{We have clean ground truth in simulation settings.} in Table \ref{simulatePSNRSSIM} (Implementation details are in Appendix \ref{Simulated data implementation details}, see Fig. \ref{simulated_experiment} for qualitative results). For Nr2N, DDM2, and Di-Fusion, three rounds of experiments are conducted to provide error bars. As P2S utilizes linear regressors, each round's results are consistently similar. Hence, P2S's error bars are not provided. When the noise intensity is high, our method performs the best and shows stable performance. Di-Fusion \textit{holds a tremendous potential for generalization and applicability for its stable performance and better performance under high noise intensity}. Moreover, these results provide evidence that \textit{Di-Fusion could be extended to self-supervised MRI denoising without relying on any clean data}.

\begin{table*}[h]\small
\vspace{-3mm}
\caption{Quantitative results on simulated data (Noisy means simulated data). PSNR (dB) and SSIM (\%) are reported. Numbers are presented as mean value with {\scriptsize{standard deviation}}. Di-Fusion exhibits more stable performance with a smaller standard deviation.} \label{simulatePSNRSSIM}
\begin{center}
\renewcommand{\arraystretch}{1.0}
\setlength\tabcolsep{2.5pt} 
\begin{tabular}{lcccccccccc}
\toprule
& \multicolumn{2}{c}{Simulation 1} & \multicolumn{2}{c}{Simulation 2} & \multicolumn{2}{c}{Simulation 3} & \multicolumn{2}{c}{Simulation 4} & \multicolumn{2}{c}{Simulation 5}\\
\cmidrule(l{2pt}r{2pt}){2-3}\cmidrule(l{2pt}r{2pt}){4-5}\cmidrule(l{2pt}r{2pt}){6-7}\cmidrule(l{2pt}r{2pt}){8-9}\cmidrule(l{2pt}r{2pt}){10-11}
Method & SSIM & PSNR & SSIM & PSNR & SSIM & PSNR & SSIM & PSNR & SSIM & PSNR \\

\midrule
Noisy & 11.38 & 13.72 & 20.65 & 16.09 & 37.41 & 20.38 & 52.02 & 23.76 & 64.62 & 26.63 \\
P2S & 24.53 & 11.20 & 43.94 & 17.63 & 65.34 & 24.91 & \underline{78.61} & \textbf{30.26} & 86.13 & \textbf{33.87} \\ 
Nr2N & 23.72\tiny{0.93} & 16.78\tiny{0.22} & \underline{50.83\tiny{2.27}} & \underline{22.13\tiny{0.94}} & \underline{71.77\tiny{6.94}} & \underline{26.80\tiny{1.69}} & 70.32\tiny{7.99} & 26.40\tiny{3.17} & \underline{87.31\tiny{4.70}} & \underline{32.82\tiny{0.72}} \\
DDM2 & \underline{26.02\tiny{2.63}} & \underline{17.92\tiny{1.06}} & 49.57\tiny{15.6} & 21.76\tiny{1.64} & 59.64\tiny{8.47} & 23.25\tiny{4.05} & 77.96\tiny{2.16} & \underline{29.14\tiny{1.01}} & 81.43\tiny{2.23} & 31.35\tiny{2.23} \\
OURS & \textbf{39.05\tiny{1.42}} & \textbf{19.91\tiny{0.32}} & \textbf{62.02\tiny{1.11}} & \textbf{23.60\tiny{0.05}} & \textbf{77.36\tiny{0.61}} & \textbf{26.96\tiny{0.37}} & \textbf{83.43\tiny{1.39}} & 28.69\tiny{0.67} & \textbf{89.52\tiny{0.18}} & 30.63\tiny{0.27} \\

\bottomrule
\end{tabular}
\end{center}
\vspace{-5mm}
\end{table*}

\subsection{Ablation studies}
\label{Ablation studies}

Without the \textbf{Fusion process}, the results in the early sampling phase do not deviate significantly (Fig. \ref{ablation_nosample}), but when we do not perform an adaptive termination, the results show areas that are absent in the noisy data (Fig. \ref{ablation}), indicating that they have indeed drifted; Without the \textbf{``Di-'' process}, the results lack some high-frequency information, and the overall gray value of the denoised images has also changed (Case 1 in Fig. \ref{ablation_nosample}), In Fig. \ref{noise}, we demonstrate through experiments that the noise computed by ``Di-'' process has different statistical properties from $z$; Without \textbf{training the latter diffusion steps}, the denoising results are noticeably smoother and have more hallucinations (Fig. \ref{ablation_nosample} and Fig. \ref{ablation}). The details of the above ablation studies can be found in Appendix \ref{On training in Di-Fusion}.

\begin{wrapfigure}{r}{0.58\textwidth}
  \vspace{-6mm}
  \centering
  \includegraphics[scale=0.24]{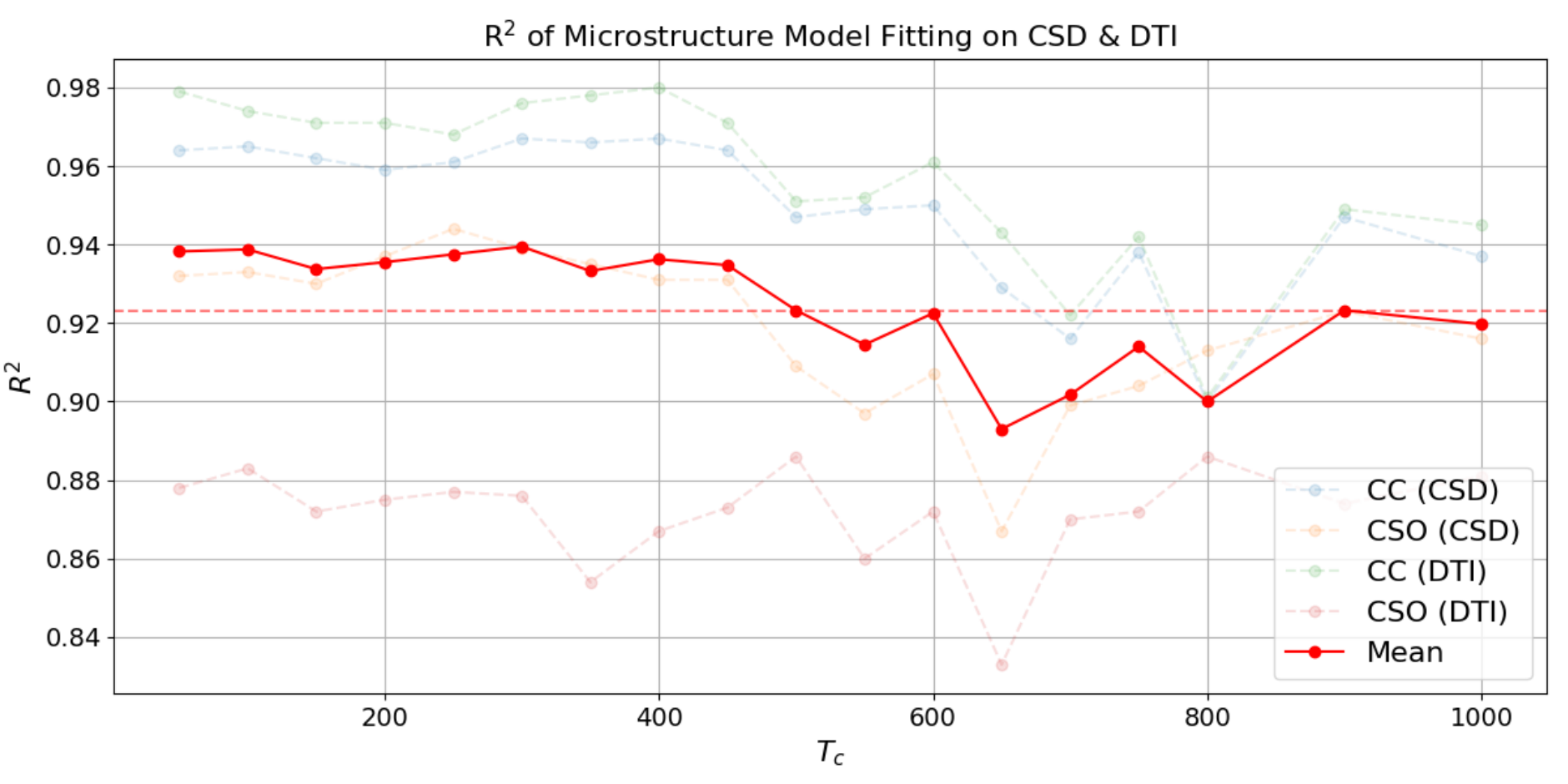}
  \vspace{-5mm}
  \caption{$R^2$ of microstructure model fitting on CSD \& DTI obtained when $T_c$ is different. When $T_c<500$, the performance is consistent.}
  \label{Tc}
  \vspace{-5mm}
\end{wrapfigure}

Furthermore, we balance the training epochs for different $T_c$\footnote{$1 \times 10^5$ epochs for $T_c=300$, $2 \times 10^5$ epochs for $T_c=600$, etc.} and show $R^2$ of microstructure model fitting results in Fig. \ref{Tc}. The results show that choosing any $T_c$ within a reasonable range ($T_c<500$) will not have a significant influence on the denoising results and the training difficulty of every step is relatively consistent.

During the reverse process, \textit{Run-Walk} accelerated sampling not only enables accelerated sampling, but also ensures that the sampling quality remains mainly unchanged (Fig. \ref{runwalk_sampling}). In Table \ref{Sampling Time per Slice}, the sampling time indicates that the adaptive termination and \textit{Run-Walk} accelerated sampling (Section \ref{Reverse process}) together greatly reduce the sampling time.  The details of the above ablation studies can be found in Appendix \ref{On sampling in Di-Fusion}.



\section{Discussions}
\label{Discussion}
\paragraph{On comparisons with related work}
\label{On comparisons with other methods}
In Appendix \ref{Comparisons with related works}, we discuss the differences between Di-Fusion and Patch2Self, as well as DDM2. In essence, Di-Fusion surpasses Patch2Self by \textit{significantly reducing the dependence on the number of input volumes}, thereby expanding its scope of application. In comparison to DDM2, Di-Fusion not only \textit{simplifies the method to a single-stage framework} but also implements \textit{iterative and controllable refinement through methods mentioned in} Section \ref{Methods}. Moreover, in Section \ref{expriments}, Appendix \ref{Supplementary experimental results} and \ref{More comparisons with competing methods}, Di-Fusion {achieves state-of-the-art performance in the conducted experiments}, showcasing its superior and stable results.
\paragraph{On limitations}
\label{On limitations}
\textit{(i)} Possible longer inference duration. The inference time of diffusion models is already relatively long, and there are concerns that the additional computation introduced by the adaptive termination in Section \ref{Towards iterative and adjustable refinement} may further increase the inference duration. In Appendix \ref{About sampling time}, we discuss the sampling burden associated with Di-Fusion. \textit{(ii)} Possible hallucinations. As a class of generative models, diffusion models inevitably raise concerns regarding generating fake anatomical details (hallucinations). However, we find that with the methods introduced in Section \ref{Methods}, particularly the training the latter diffusion steps mentioned in Section \ref{Intuition of conditional training}, the generative capacity of diffusion models can be restricted (Appendix \ref{On conditional training}), which helps reduce hallucinations. \textit{(iii)} Noise-artifact conflation and additive noise assumptions. Our method explicitly targets thermal noise modeled as additive Gaussian processes, consistent with common denoising frameworks \citep{chen2019noise,ramos2021snr,cordero2019complex}. However, this intentionally excludes spatially varying distortions (e.g., cardiac pulsation) often classified as "physiological noise" \citep{chang2005restore,chang2012informed,walker2011effects} but better characterized as artifacts. While our assumption enables tractable solutions, it may limit effectiveness for signal-dependent noise.

\section{Conclusion}
\label{Conclusion}

This paper proposes Di-Fusion, an end-to-end self-supervised MRI denoising method that achieves iterative and stable refinement without relying on extra noise model training or clean data. The Fusion process aligns the trajectory of the forward process and avoids drifted results. The ``Di-'' process characterizes real-world noise, enabling the model to capture statistical properties of the real-world noise. By training the latter diffusion steps, our model achieves enhanced stability and performance. During the inference stage, Di-Fusion offers controllable results through an adaptive sampling process. Comprehensive experiments on real and simulated data demonstrate that Di-Fusion achieves state-of-the-art performance in microstructure modeling, tractography tracking, and other downstream tasks.




\subsubsection*{Acknowledgments}
This research was supported by Natural Science Foundation of China under Grant 62271465, Suzhou Basic Research Program under Grant SYG202338, and Open Fund Project of Guangdong Academy of Medical Sciences, China (No. YKY-KF202206).

\bibliography{iclr2025_conference}

\begin{thebibliography}{84}
\providecommand{\natexlab}[1]{#1}
\providecommand{\url}[1]{\texttt{#1}}
\expandafter\ifx\csname urlstyle\endcsname\relax
  \providecommand{\doi}[1]{doi: #1}\else
  \providecommand{\doi}{doi: \begingroup \urlstyle{rm}\Url}\fi

\bibitem[Adamson et~al.(2021)Adamson, Gunel, Dominic, Desai, Spielman, Vasanawala, Pauly, and Chaudhari]{adamson2021ssfd}
Philip~M Adamson, Beliz Gunel, Jeffrey Dominic, Arjun~D Desai, Daniel Spielman, Shreyas Vasanawala, John~M Pauly, and Akshay Chaudhari.
\newblock Ssfd: Self-supervised feature distance as an mr image reconstruction quality metric.
\newblock In \emph{NeurIPS 2021 Workshop on Deep Learning and Inverse Problems}, 2021.

\bibitem[Ades-Aron et~al.(2018)Ades-Aron, Veraart, Kochunov, McGuire, Sherman, Kellner, Novikov, and Fieremans]{ades2018evaluation}
Benjamin Ades-Aron, Jelle Veraart, Peter Kochunov, Stephen McGuire, Paul Sherman, Elias Kellner, Dmitry~S Novikov, and Els Fieremans.
\newblock Evaluation of the accuracy and precision of the diffusion parameter estimation with gibbs and noise removal pipeline.
\newblock \emph{Neuroimage}, 183:\penalty0 532--543, 2018.

\bibitem[Bansal et~al.(2024)Bansal, Borgnia, Chu, Li, Kazemi, Huang, Goldblum, Geiping, and Goldstein]{bansal2024cold}
Arpit Bansal, Eitan Borgnia, Hong-Min Chu, Jie Li, Hamid Kazemi, Furong Huang, Micah Goldblum, Jonas Geiping, and Tom Goldstein.
\newblock Cold diffusion: Inverting arbitrary image transforms without noise.
\newblock \emph{Advances in Neural Information Processing Systems}, 36, 2024.

\bibitem[Basser et~al.(1994)Basser, Mattiello, and LeBihan]{basser1994mr}
Peter~J Basser, James Mattiello, and Denis LeBihan.
\newblock Mr diffusion tensor spectroscopy and imaging.
\newblock \emph{Biophysical journal}, 66\penalty0 (1):\penalty0 259--267, 1994.

\bibitem[Batson \& Royer(2019)Batson and Royer]{batson2019noise2self}
Joshua Batson and Loic Royer.
\newblock Noise2self: Blind denoising by self-supervision.
\newblock In \emph{International Conference on Machine Learning}, pp.\  524--533. PMLR, 2019.

\bibitem[Behrens et~al.(2014)Behrens, Sotiropoulos, and Jbabdi]{behrens2014mr}
Timothy~EJ Behrens, Stamatios~N Sotiropoulos, and Saad Jbabdi.
\newblock Mr diffusion tractography.
\newblock In \emph{Diffusion MRI}, pp.\  429--451. Elsevier, 2014.

\bibitem[Chang et~al.(2005)Chang, Jones, and Pierpaoli]{chang2005restore}
Lin-Ching Chang, Derek~K Jones, and Carlo Pierpaoli.
\newblock Restore: robust estimation of tensors by outlier rejection.
\newblock \emph{Magnetic Resonance in Medicine: An Official Journal of the International Society for Magnetic Resonance in Medicine}, 53\penalty0 (5):\penalty0 1088--1095, 2005.

\bibitem[Chang et~al.(2012)Chang, Walker, and Pierpaoli]{chang2012informed}
Lin-Ching Chang, Lindsay Walker, and Carlo Pierpaoli.
\newblock Informed restore: a method for robust estimation of diffusion tensor from low redundancy datasets in the presence of physiological noise artifacts.
\newblock \emph{Magnetic resonance in medicine}, 68\penalty0 (5):\penalty0 1654--1663, 2012.

\bibitem[Chaudhari et~al.(2020)Chaudhari, Stevens, Wood, Chakraborty, Gibbons, Fang, Desai, Lee, Gold, and Hargreaves]{chaudhari2020utility}
Akshay~S Chaudhari, Kathryn~J Stevens, Jeff~P Wood, Amit~K Chakraborty, Eric~K Gibbons, Zhongnan Fang, Arjun~D Desai, Jin~Hyung Lee, Garry~E Gold, and Brian~A Hargreaves.
\newblock Utility of deep learning super-resolution in the context of osteoarthritis mri biomarkers.
\newblock \emph{Journal of Magnetic Resonance Imaging}, 51\penalty0 (3):\penalty0 768--779, 2020.

\bibitem[Chen et~al.(2016)Chen, Wu, Shen, and Yap]{chen2016xq}
Geng Chen, Yafeng Wu, Dinggang Shen, and Pew-Thian Yap.
\newblock Xq-nlm: denoising diffusion mri data via x-q space non-local patch matching.
\newblock In \emph{Medical Image Computing and Computer-Assisted Intervention-MICCAI 2016: 19th International Conference, Athens, Greece, October 17-21, 2016, Proceedings, Part III 19}, pp.\  587--595. Springer, 2016.

\bibitem[Chen et~al.(2019)Chen, Wu, Shen, and Yap]{chen2019noise}
Geng Chen, Yafeng Wu, Dinggang Shen, and Pew-Thian Yap.
\newblock Noise reduction in diffusion mri using non-local self-similar information in joint x- q space.
\newblock \emph{Medical image analysis}, 53:\penalty0 79--94, 2019.

\bibitem[Chen et~al.(2020)Chen, Zhang, Zen, Weiss, Norouzi, and Chan]{chen2020wavegrad}
Nanxin Chen, Yu~Zhang, Heiga Zen, Ron~J Weiss, Mohammad Norouzi, and William Chan.
\newblock Wavegrad: Estimating gradients for waveform generation.
\newblock \emph{arXiv preprint arXiv:2009.00713}, 2020.

\bibitem[Chung et~al.(2022{\natexlab{a}})Chung, Lee, and Ye]{chung2022mr}
Hyungjin Chung, Eun~Sun Lee, and Jong~Chul Ye.
\newblock Mr image denoising and super-resolution using regularized reverse diffusion.
\newblock \emph{IEEE Transactions on Medical Imaging}, 42\penalty0 (4):\penalty0 922--934, 2022{\natexlab{a}}.

\bibitem[Chung et~al.(2022{\natexlab{b}})Chung, Sim, Ryu, and Ye]{chung2022improving}
Hyungjin Chung, Byeongsu Sim, Dohoon Ryu, and Jong~Chul Ye.
\newblock Improving diffusion models for inverse problems using manifold constraints.
\newblock \emph{Advances in Neural Information Processing Systems}, 35:\penalty0 25683--25696, 2022{\natexlab{b}}.

\bibitem[Cordero-Grande et~al.(2019)Cordero-Grande, Christiaens, Hutter, Price, and Hajnal]{cordero2019complex}
Lucilio Cordero-Grande, Daan Christiaens, Jana Hutter, Anthony~N Price, and Jo~V Hajnal.
\newblock Complex diffusion-weighted image estimation via matrix recovery under general noise models.
\newblock \emph{Neuroimage}, 200:\penalty0 391--404, 2019.

\bibitem[Coup{\'e} et~al.(2008)Coup{\'e}, Yger, Prima, Hellier, Kervrann, and Barillot]{coupe2008optimized}
Pierrick Coup{\'e}, Pierre Yger, Sylvain Prima, Pierre Hellier, Charles Kervrann, and Christian Barillot.
\newblock An optimized blockwise nonlocal means denoising filter for 3-d magnetic resonance images.
\newblock \emph{IEEE transactions on medical imaging}, 27\penalty0 (4):\penalty0 425--441, 2008.

\bibitem[Coup{\'e} et~al.(2012)Coup{\'e}, Manj{\'o}n, Robles, and Collins]{coupe2012adaptive}
Pierrick Coup{\'e}, Jos{\'e}~V Manj{\'o}n, Montserrat Robles, and D~Louis Collins.
\newblock Adaptive multiresolution non-local means filter for three-dimensional magnetic resonance image denoising.
\newblock \emph{IET image Processing}, 6\penalty0 (5):\penalty0 558--568, 2012.

\bibitem[Croitoru et~al.(2023)Croitoru, Hondru, Ionescu, and Shah]{croitoru2023diffusion}
Florinel-Alin Croitoru, Vlad Hondru, Radu~Tudor Ionescu, and Mubarak Shah.
\newblock Diffusion models in vision: A survey.
\newblock \emph{IEEE Transactions on Pattern Analysis and Machine Intelligence}, 2023.

\bibitem[Darestani et~al.(2021)Darestani, Chaudhari, and Heckel]{pmlr-v139-darestani21a}
Mohammad~Zalbagi Darestani, Akshay~S Chaudhari, and Reinhard Heckel.
\newblock Measuring robustness in deep learning based compressive sensing.
\newblock In Marina Meila and Tong Zhang (eds.), \emph{Proceedings of the 38th International Conference on Machine Learning}, volume 139 of \emph{Proceedings of Machine Learning Research}, pp.\  2433--2444. PMLR, 18--24 Jul 2021.
\newblock URL \url{https://proceedings.mlr.press/v139/darestani21a.html}.

\bibitem[Desai et~al.(2021{\natexlab{a}})Desai, Gunel, Ozturkler, Beg, Vasanawala, Hargreaves, R{\'e}, Pauly, and Chaudhari]{desai2021vortex}
Arjun~D Desai, Beliz Gunel, Batu Ozturkler, Harris Beg, Shreyas Vasanawala, Brian Hargreaves, Christopher R{\'e}, John~M Pauly, and Akshay Chaudhari.
\newblock Vortex: Physics-driven data augmentations for consistency training for robust accelerated mri reconstruction.
\newblock \emph{Medical Imaging with Deep Learning, 2022}, 2021{\natexlab{a}}.

\bibitem[Desai et~al.(2021{\natexlab{b}})Desai, Ozturkler, Sandino, Boutin, Willis, Vasanawala, Hargreaves, R{\'e}, Pauly, and Chaudhari]{desai2021noise2recon}
Arjun~D Desai, Batu~M Ozturkler, Christopher~M Sandino, Robert Boutin, Marc Willis, Shreyas Vasanawala, Brian~A Hargreaves, Christopher~M R{\'e}, John~M Pauly, and Akshay~S Chaudhari.
\newblock Noise2recon: Enabling joint mri reconstruction and denoising with semi-supervised and self-supervised learning.
\newblock \emph{arXiv preprint arXiv:2110.00075}, 2021{\natexlab{b}}.

\bibitem[Duits \& Franken(2011)Duits and Franken]{duits2011left}
Remco Duits and Erik Franken.
\newblock Left-invariant diffusions on the space of positions and orientations and their application to crossing-preserving smoothing of hardi images.
\newblock \emph{International Journal of Computer Vision}, 92:\penalty0 231--264, 2011.

\bibitem[Fadnavis et~al.(2020{\natexlab{a}})Fadnavis, Batson, and Garyfallidis]{fadnavis2020patch2self}
Shreyas Fadnavis, Joshua Batson, and Eleftherios Garyfallidis.
\newblock Patch2self: Denoising diffusion mri with self-supervised learning.
\newblock \emph{Advances in Neural Information Processing Systems}, 33:\penalty0 16293--16303, 2020{\natexlab{a}}.

\bibitem[Fadnavis et~al.(2020{\natexlab{b}})Fadnavis, Batson, and Garyfallidis]{fadnavispatch2self}
Shreyas Fadnavis, Joshua Batson, and Eleftherios Garyfallidis.
\newblock Patch2self: Denoising diffusion mri with self-supervised learning.
\newblock \emph{Advances in Neural Information Processing Systems}, 33:\penalty0 16293--16303, 2020{\natexlab{b}}.

\bibitem[Fadnavis et~al.(2024)Fadnavis, Chowdhury, Batson, Drineas, and Garyfallidis]{fadnavis2024patch2self2}
Shreyas Fadnavis, Agniva Chowdhury, Joshua Batson, Petros Drineas, and Eleftherios Garyfallidis.
\newblock Patch2self2: Self-supervised denoising on coresets via matrix sketching.
\newblock In \emph{Proceedings of the IEEE/CVF Conference on Computer Vision and Pattern Recognition}, pp.\  27641--27651, 2024.

\bibitem[Fei et~al.(2023)Fei, Lyu, Pan, Zhang, Yang, Luo, Zhang, and Dai]{fei2023generative}
Ben Fei, Zhaoyang Lyu, Liang Pan, Junzhe Zhang, Weidong Yang, Tianyue Luo, Bo~Zhang, and Bo~Dai.
\newblock Generative diffusion prior for unified image restoration and enhancement.
\newblock In \emph{Proceedings of the IEEE/CVF Conference on Computer Vision and Pattern Recognition}, pp.\  9935--9946, 2023.

\bibitem[Gao et~al.(2023)Gao, Li, Zhang, Zhang, and Shan]{gao2023corediff}
Qi~Gao, Zilong Li, Junping Zhang, Yi~Zhang, and Hongming Shan.
\newblock Corediff: Contextual error-modulated generalized diffusion model for low-dose ct denoising and generalization.
\newblock \emph{IEEE Transactions on Medical Imaging}, 2023.

\bibitem[Garyfallidis et~al.(2014)Garyfallidis, Brett, Amirbekian, Rokem, Van Der~Walt, Descoteaux, Nimmo-Smith, and Contributors]{garyfallidis2014dipy}
Eleftherios Garyfallidis, Matthew Brett, Bagrat Amirbekian, Ariel Rokem, Stefan Van Der~Walt, Maxime Descoteaux, Ian Nimmo-Smith, and Dipy Contributors.
\newblock Dipy, a library for the analysis of diffusion mri data.
\newblock \emph{Frontiers in neuroinformatics}, 8:\penalty0 8, 2014.

\bibitem[Gibbons et~al.(2019)Gibbons, Hodgson, Chaudhari, Richards, Majersik, Adluru, and DiBella]{gibbons2019simultaneous}
Eric~K Gibbons, Kyler~K Hodgson, Akshay~S Chaudhari, Lorie~G Richards, Jennifer~J Majersik, Ganesh Adluru, and Edward~VR DiBella.
\newblock Simultaneous noddi and gfa parameter map generation from subsampled q-space imaging using deep learning.
\newblock \emph{Magnetic resonance in medicine}, 81\penalty0 (4):\penalty0 2399--2411, 2019.

\bibitem[Girard et~al.(2014)Girard, Whittingstall, Deriche, and Descoteaux]{girard2014towards}
Gabriel Girard, Kevin Whittingstall, Rachid Deriche, and Maxime Descoteaux.
\newblock Towards quantitative connectivity analysis: reducing tractography biases.
\newblock \emph{Neuroimage}, 98:\penalty0 266--278, 2014.

\bibitem[Hastie et~al.(2009)Hastie, Tibshirani, Friedman, and Friedman]{hastie2009elements}
Trevor Hastie, Robert Tibshirani, Jerome~H Friedman, and Jerome~H Friedman.
\newblock \emph{The elements of statistical learning: data mining, inference, and prediction}, volume~2.
\newblock Springer, 2009.

\bibitem[Ho et~al.(2020)Ho, Jain, and Abbeel]{ho2020denoising}
Jonathan Ho, Ajay Jain, and Pieter Abbeel.
\newblock Denoising diffusion probabilistic models.
\newblock \emph{Advances in neural information processing systems}, 33:\penalty0 6840--6851, 2020.

\bibitem[Huang et~al.(2021)Huang, Li, Jia, Lu, and Liu]{huang2021neighbor2neighbor}
Tao Huang, Songjiang Li, Xu~Jia, Huchuan Lu, and Jianzhuang Liu.
\newblock Neighbor2neighbor: Self-supervised denoising from single noisy images.
\newblock In \emph{Proceedings of the IEEE/CVF conference on computer vision and pattern recognition}, pp.\  14781--14790, 2021.

\bibitem[Hyv{\"a}rinen \& Dayan(2005)Hyv{\"a}rinen and Dayan]{hyvarinen2005estimation}
Aapo Hyv{\"a}rinen and Peter Dayan.
\newblock Estimation of non-normalized statistical models by score matching.
\newblock \emph{Journal of Machine Learning Research}, 6\penalty0 (4), 2005.

\bibitem[Karayumak et~al.(2019)Karayumak, Bouix, Ning, James, Crow, Shenton, Kubicki, and Rathi]{karayumak2019retrospective}
Suheyla~Cetin Karayumak, Sylvain Bouix, Lipeng Ning, Anthony James, Tim Crow, Martha Shenton, Marek Kubicki, and Yogesh Rathi.
\newblock Retrospective harmonization of multi-site diffusion mri data acquired with different acquisition parameters.
\newblock \emph{Neuroimage}, 184:\penalty0 180--200, 2019.

\bibitem[Kawar et~al.(2022)Kawar, Elad, Ermon, and Song]{kawar2022denoising}
Bahjat Kawar, Michael Elad, Stefano Ermon, and Jiaming Song.
\newblock Denoising diffusion restoration models.
\newblock \emph{Advances in Neural Information Processing Systems}, 35:\penalty0 23593--23606, 2022.

\bibitem[Kaye et~al.(2020)Kaye, Aherne, Duzgol, H{\"a}ggstr{\"o}m, Kobler, Mazaheri, Fung, Zhang, Otazo, Vargas, et~al.]{kaye2020accelerating}
Elena~A Kaye, Emily~A Aherne, Cihan Duzgol, Ida H{\"a}ggstr{\"o}m, Erich Kobler, Yousef Mazaheri, Maggie~M Fung, Zhigang Zhang, Ricardo Otazo, Hebert~A Vargas, et~al.
\newblock Accelerating prostate diffusion-weighted mri using a guided denoising convolutional neural network: retrospective feasibility study.
\newblock \emph{Radiology: Artificial Intelligence}, 2\penalty0 (5):\penalty0 e200007, 2020.

\bibitem[Kim \& Ye(2021)Kim and Ye]{kim2021noise2score}
Kwanyoung Kim and Jong~Chul Ye.
\newblock Noise2score: tweedie’s approach to self-supervised image denoising without clean images.
\newblock \emph{Advances in Neural Information Processing Systems}, 34:\penalty0 864--874, 2021.

\bibitem[Krull et~al.(2019)Krull, Buchholz, and Jug]{krull2019noise2void}
Alexander Krull, Tim-Oliver Buchholz, and Florian Jug.
\newblock Noise2void-learning denoising from single noisy images.
\newblock In \emph{Proceedings of the IEEE/CVF conference on computer vision and pattern recognition}, pp.\  2129--2137, 2019.

\bibitem[Laine et~al.(2019)Laine, Karras, Lehtinen, and Aila]{laine2019high}
Samuli Laine, Tero Karras, Jaakko Lehtinen, and Timo Aila.
\newblock High-quality self-supervised deep image denoising.
\newblock \emph{Advances in Neural Information Processing Systems}, 32, 2019.

\bibitem[Le~Bihan(2003)]{le2003looking}
Denis Le~Bihan.
\newblock Looking into the functional architecture of the brain with diffusion mri.
\newblock \emph{Nature reviews neuroscience}, 4\penalty0 (6):\penalty0 469--480, 2003.

\bibitem[Le~Bihan et~al.(2006)Le~Bihan, Poupon, Amadon, and Lethimonnier]{le2006artifacts}
Denis Le~Bihan, Cyril Poupon, Alexis Amadon, and Franck Lethimonnier.
\newblock Artifacts and pitfalls in diffusion mri.
\newblock \emph{Journal of Magnetic Resonance Imaging: An Official Journal of the International Society for Magnetic Resonance in Medicine}, 24\penalty0 (3):\penalty0 478--488, 2006.

\bibitem[Lehtinen et~al.(2018)Lehtinen, Munkberg, Hasselgren, Laine, Karras, Aittala, and Aila]{lehtinen2018noise2noise}
Jaakko Lehtinen, Jacob Munkberg, Jon Hasselgren, Samuli Laine, Tero Karras, Miika Aittala, and Timo Aila.
\newblock Noise2noise: Learning image restoration without clean data.
\newblock \emph{arXiv preprint arXiv:1803.04189}, 2018.

\bibitem[Manj{\'o}n et~al.(2013)Manj{\'o}n, Coup{\'e}, Concha, Buades, Collins, and Robles]{manjon2013diffusion}
Jos{\'e}~V Manj{\'o}n, Pierrick Coup{\'e}, Luis Concha, Antonio Buades, D~Louis Collins, and Montserrat Robles.
\newblock Diffusion weighted image denoising using overcomplete local pca.
\newblock \emph{PloS one}, 8\penalty0 (9):\penalty0 e73021, 2013.

\bibitem[Mansour \& Heckel(2023)Mansour and Heckel]{mansour2023zero}
Youssef Mansour and Reinhard Heckel.
\newblock Zero-shot noise2noise: Efficient image denoising without any data.
\newblock In \emph{Proceedings of the IEEE/CVF Conference on Computer Vision and Pattern Recognition}, pp.\  14018--14027, 2023.

\bibitem[Marek et~al.(2011)Marek, Jennings, Lasch, Siderowf, Tanner, Simuni, Coffey, Kieburtz, Flagg, Chowdhury, et~al.]{marek2011parkinson}
Kenneth Marek, Danna Jennings, Shirley Lasch, Andrew Siderowf, Caroline Tanner, Tanya Simuni, Chris Coffey, Karl Kieburtz, Emily Flagg, Sohini Chowdhury, et~al.
\newblock The parkinson progression marker initiative (ppmi).
\newblock \emph{Progress in neurobiology}, 95\penalty0 (4):\penalty0 629--635, 2011.

\bibitem[Mason et~al.(2019)Mason, Rioux, Clarke, Costa, Schmidt, Keough, Huynh, and Beyea]{mason2019comparison}
Allister Mason, James Rioux, Sharon~E Clarke, Andreu Costa, Matthias Schmidt, Valerie Keough, Thien Huynh, and Steven Beyea.
\newblock Comparison of objective image quality metrics to expert radiologists’ scoring of diagnostic quality of mr images.
\newblock \emph{IEEE transactions on medical imaging}, 39\penalty0 (4):\penalty0 1064--1072, 2019.

\bibitem[Meesters et~al.(2016)Meesters, Sanguinetti, Garyfallidis, Portegies, Ossenblok, and Duits]{meesters2016cleaning}
SPL Meesters, GR~Sanguinetti, Eleftherios Garyfallidis, JM~Portegies, PPW Ossenblok, and Remco Duits.
\newblock Cleaning output of tractography via fiber to bundle coherence, a new open source implementation.
\newblock \emph{a new open source implementation}, 2016.

\bibitem[Mittal et~al.(2011)Mittal, Moorthy, and Bovik]{mittal2011blind}
Anish Mittal, Anush~K Moorthy, and Alan~C Bovik.
\newblock Blind/referenceless image spatial quality evaluator.
\newblock In \emph{2011 conference record of the forty fifth asilomar conference on signals, systems and computers (ASILOMAR)}, pp.\  723--727. IEEE, 2011.

\bibitem[Mohamed \& Lakshminarayanan(2016)Mohamed and Lakshminarayanan]{mohamed2016learning}
Shakir Mohamed and Balaji Lakshminarayanan.
\newblock Learning in implicit generative models.
\newblock \emph{arXiv preprint arXiv:1610.03483}, 2016.

\bibitem[Moran et~al.(2020)Moran, Schmidt, Zhong, and Coady]{moran2020noisier2noise}
Nick Moran, Dan Schmidt, Yu~Zhong, and Patrick Coady.
\newblock Noisier2noise: Learning to denoise from unpaired noisy data.
\newblock In \emph{Proceedings of the IEEE/CVF Conference on Computer Vision and Pattern Recognition}, pp.\  12064--12072, 2020.

\bibitem[Novikov et~al.(2018)Novikov, Kiselev, and Jespersen]{novikov2018modeling}
Dmitry~S Novikov, Valerij~G Kiselev, and Sune~N Jespersen.
\newblock On modeling.
\newblock \emph{Magnetic resonance in medicine}, 79\penalty0 (6):\penalty0 3172--3193, 2018.

\bibitem[Ostu(1979)]{ostu1979threshold}
Nobuyuki Ostu.
\newblock A threshold selection method from gray-level histograms.
\newblock \emph{IEEE Trans SMC}, 9:\penalty0 62, 1979.

\bibitem[{\"O}zdenizci \& Legenstein(2023){\"O}zdenizci and Legenstein]{ozdenizci2023restoring}
Ozan {\"O}zdenizci and Robert Legenstein.
\newblock Restoring vision in adverse weather conditions with patch-based denoising diffusion models.
\newblock \emph{IEEE Transactions on Pattern Analysis and Machine Intelligence}, 2023.

\bibitem[Pang et~al.(2021)Pang, Zheng, Quan, and Ji]{pang2021recorrupted}
Tongyao Pang, Huan Zheng, Yuhui Quan, and Hui Ji.
\newblock Recorrupted-to-recorrupted: Unsupervised deep learning for image denoising.
\newblock In \emph{Proceedings of the IEEE/CVF conference on computer vision and pattern recognition}, pp.\  2043--2052, 2021.

\bibitem[Paszke et~al.(2019)Paszke, Gross, Massa, Lerer, Bradbury, Chanan, Killeen, Lin, Gimelshein, Antiga, et~al.]{paszke2019pytorch}
Adam Paszke, Sam Gross, Francisco Massa, Adam Lerer, James Bradbury, Gregory Chanan, Trevor Killeen, Zeming Lin, Natalia Gimelshein, Luca Antiga, et~al.
\newblock Pytorch: An imperative style, high-performance deep learning library.
\newblock \emph{Advances in neural information processing systems}, 32, 2019.

\bibitem[Portegies et~al.(2015)Portegies, Fick, Sanguinetti, Meesters, Girard, and Duits]{portegies2015improving}
Jorg~M Portegies, Rutger Henri~Jacques Fick, Gonzalo~R Sanguinetti, Stephan~PL Meesters, Gabriel Girard, and Remco Duits.
\newblock Improving fiber alignment in hardi by combining contextual pde flow with constrained spherical deconvolution.
\newblock \emph{PloS one}, 10\penalty0 (10):\penalty0 e0138122, 2015.

\bibitem[Ramos-Llord{\'e}n et~al.(2021)Ramos-Llord{\'e}n, Vegas-S{\'a}nchez-Ferrero, Liao, Westin, Setsompop, and Rathi]{ramos2021snr}
Gabriel Ramos-Llord{\'e}n, Gonzalo Vegas-S{\'a}nchez-Ferrero, Congyu Liao, Carl-Fredrik Westin, Kawin Setsompop, and Yogesh Rathi.
\newblock Snr-enhanced diffusion mri with structure-preserving low-rank denoising in reproducing kernel hilbert spaces.
\newblock \emph{Magnetic resonance in medicine}, 86\penalty0 (3):\penalty0 1614--1632, 2021.

\bibitem[Rokem(2016)]{rokem2016stanford}
Ariel Rokem.
\newblock Stanford hardi surfaces, 2016.

\bibitem[Ronneberger et~al.(2015)Ronneberger, Fischer, and Brox]{ronneberger2015u}
Olaf Ronneberger, Philipp Fischer, and Thomas Brox.
\newblock U-net: Convolutional networks for biomedical image segmentation.
\newblock In \emph{Medical image computing and computer-assisted intervention--MICCAI 2015: 18th international conference, Munich, Germany, October 5-9, 2015, proceedings, part III 18}, pp.\  234--241. Springer, 2015.

\bibitem[Saharia et~al.(2022{\natexlab{a}})Saharia, Chan, Chang, Lee, Ho, Salimans, Fleet, and Norouzi]{saharia2022palette}
Chitwan Saharia, William Chan, Huiwen Chang, Chris Lee, Jonathan Ho, Tim Salimans, David Fleet, and Mohammad Norouzi.
\newblock Palette: Image-to-image diffusion models.
\newblock In \emph{ACM SIGGRAPH 2022 conference proceedings}, pp.\  1--10, 2022{\natexlab{a}}.

\bibitem[Saharia et~al.(2022{\natexlab{b}})Saharia, Ho, Chan, Salimans, Fleet, and Norouzi]{saharia2022image}
Chitwan Saharia, Jonathan Ho, William Chan, Tim Salimans, David~J Fleet, and Mohammad Norouzi.
\newblock Image super-resolution via iterative refinement.
\newblock \emph{IEEE transactions on pattern analysis and machine intelligence}, 45\penalty0 (4):\penalty0 4713--4726, 2022{\natexlab{b}}.

\bibitem[Schilling et~al.(2019)Schilling, Nath, Hansen, Parvathaneni, Blaber, Gao, Neher, Aydogan, Shi, Ocampo-Pineda, et~al.]{schilling2019limits}
Kurt~G Schilling, Vishwesh Nath, Colin Hansen, Prasanna Parvathaneni, Justin Blaber, Yurui Gao, Peter Neher, Dogu~Baran Aydogan, Yonggang Shi, Mario Ocampo-Pineda, et~al.
\newblock Limits to anatomical accuracy of diffusion tractography using modern approaches.
\newblock \emph{Neuroimage}, 185:\penalty0 1--11, 2019.

\bibitem[Sohl-Dickstein et~al.(2015)Sohl-Dickstein, Weiss, Maheswaranathan, and Ganguli]{sohl2015deep}
Jascha Sohl-Dickstein, Eric Weiss, Niru Maheswaranathan, and Surya Ganguli.
\newblock Deep unsupervised learning using nonequilibrium thermodynamics.
\newblock In \emph{International conference on machine learning}, pp.\  2256--2265. PMLR, 2015.

\bibitem[Song et~al.(2024)Song, Hu, Luo, Fessler, and Shen]{song2024diffusionblend}
Bowen Song, Jason Hu, Zhaoxu Luo, Jeffrey~A Fessler, and Liyue Shen.
\newblock Diffusionblend: Learning 3d image prior through position-aware diffusion score blending for 3d computed tomography reconstruction.
\newblock \emph{arXiv preprint arXiv:2406.10211}, 2024.

\bibitem[Song et~al.(2020{\natexlab{a}})Song, Meng, and Ermon]{song2020denoising}
Jiaming Song, Chenlin Meng, and Stefano Ermon.
\newblock Denoising diffusion implicit models.
\newblock \emph{arXiv preprint arXiv:2010.02502}, 2020{\natexlab{a}}.

\bibitem[Song \& Ermon(2019)Song and Ermon]{song2019generative}
Yang Song and Stefano Ermon.
\newblock Generative modeling by estimating gradients of the data distribution.
\newblock \emph{Advances in neural information processing systems}, 32, 2019.

\bibitem[Song et~al.(2020{\natexlab{b}})Song, Sohl-Dickstein, Kingma, Kumar, Ermon, and Poole]{song2020score}
Yang Song, Jascha Sohl-Dickstein, Diederik~P Kingma, Abhishek Kumar, Stefano Ermon, and Ben Poole.
\newblock Score-based generative modeling through stochastic differential equations.
\newblock \emph{arXiv preprint arXiv:2011.13456}, 2020{\natexlab{b}}.

\bibitem[Song et~al.(2021)Song, Shen, Xing, and Ermon]{song2021solving}
Yang Song, Liyue Shen, Lei Xing, and Stefano Ermon.
\newblock Solving inverse problems in medical imaging with score-based generative models.
\newblock \emph{arXiv preprint arXiv:2111.08005}, 2021.

\bibitem[Tibrewala et~al.(2023)Tibrewala, Dutt, Tong, Ginocchio, Keerthivasan, Baete, Chopra, Lui, Sodickson, Chandarana, and Johnson]{tibrewala2023fastmri}
Radhika Tibrewala, Tarun Dutt, Angela Tong, Luke Ginocchio, Mahesh~B Keerthivasan, Steven~H Baete, Sumit Chopra, Yvonne~W Lui, Daniel~K Sodickson, Hersh Chandarana, and Patricia~M Johnson.
\newblock {FastMRI Prostate}: A publicly available, biparametric {MRI} dataset to advance machine learning for prostate cancer imaging, 2023.

\bibitem[Tournier et~al.(2007)Tournier, Calamante, and Connelly]{tournier2007robust}
J-Donald Tournier, Fernando Calamante, and Alan Connelly.
\newblock Robust determination of the fibre orientation distribution in diffusion mri: non-negativity constrained super-resolved spherical deconvolution.
\newblock \emph{Neuroimage}, 35\penalty0 (4):\penalty0 1459--1472, 2007.

\bibitem[Tu et~al.(2025)Tu, Shi, and Lam]{tu2025scorebased}
Jiachen Tu, Yaokun Shi, and Fan Lam.
\newblock Score-based self-supervised mri denoising.
\newblock In \emph{International Conference on Learning Representations}, 2025.

\bibitem[Ulyanov et~al.(2018)Ulyanov, Vedaldi, and Lempitsky]{ulyanov2018deep}
Dmitry Ulyanov, Andrea Vedaldi, and Victor Lempitsky.
\newblock Deep image prior.
\newblock In \emph{Proceedings of the IEEE conference on computer vision and pattern recognition}, pp.\  9446--9454, 2018.

\bibitem[Veraart et~al.(2016)Veraart, Novikov, Christiaens, Ades-Aron, Sijbers, and Fieremans]{veraart2016denoising}
Jelle Veraart, Dmitry~S Novikov, Daan Christiaens, Benjamin Ades-Aron, Jan Sijbers, and Els Fieremans.
\newblock Denoising of diffusion mri using random matrix theory.
\newblock \emph{Neuroimage}, 142:\penalty0 394--406, 2016.

\bibitem[Walker et~al.(2011)Walker, Chang, Koay, Sharma, Cohen, Verma, and Pierpaoli]{walker2011effects}
Lindsay Walker, Lin-Ching Chang, Cheng~Guan Koay, Nik Sharma, Leonardo Cohen, Ragini Verma, and Carlo Pierpaoli.
\newblock Effects of physiological noise in population analysis of diffusion tensor mri data.
\newblock \emph{Neuroimage}, 54\penalty0 (2):\penalty0 1168--1177, 2011.

\bibitem[Watson et~al.(2021)Watson, Chan, Ho, and Norouzi]{watson2021learning}
Daniel Watson, William Chan, Jonathan Ho, and Mohammad Norouzi.
\newblock Learning fast samplers for diffusion models by differentiating through sample quality.
\newblock In \emph{International Conference on Learning Representations}, 2021.

\bibitem[Westin et~al.(2016)Westin, Knutsson, Pasternak, Szczepankiewicz, {\"O}zarslan, van Westen, Mattisson, Bogren, O'Donnell, Kubicki, et~al.]{westin2016q}
Carl-Fredrik Westin, Hans Knutsson, Ofer Pasternak, Filip Szczepankiewicz, Evren {\"O}zarslan, Danielle van Westen, Cecilia Mattisson, Mats Bogren, Lauren~J O'Donnell, Marek Kubicki, et~al.
\newblock Q-space trajectory imaging for multidimensional diffusion mri of the human brain.
\newblock \emph{Neuroimage}, 135:\penalty0 345--362, 2016.

\bibitem[Woodard \& Carley-Spencer(2006)Woodard and Carley-Spencer]{woodard2006no}
Jeffrey~P Woodard and Monica~P Carley-Spencer.
\newblock No-reference image quality metrics for structural mri.
\newblock \emph{Neuroinformatics}, 4:\penalty0 243--262, 2006.

\bibitem[Xia et~al.(2023)Xia, Zhang, Wang, Wang, Wu, Tian, Yang, and Van~Gool]{xia2023diffir}
Bin Xia, Yulun Zhang, Shiyin Wang, Yitong Wang, Xinglong Wu, Yapeng Tian, Wenming Yang, and Luc Van~Gool.
\newblock Diffir: Efficient diffusion model for image restoration.
\newblock In \emph{Proceedings of the IEEE/CVF International Conference on Computer Vision}, pp.\  13095--13105, 2023.

\bibitem[Xiang et~al.(2023)Xiang, Yurt, Syed, Setsompop, and Chaudhari]{xiang2023ddm}
Tiange Xiang, Mahmut Yurt, Ali~B Syed, Kawin Setsompop, and Akshay Chaudhari.
\newblock Ddm $^2$: Self-supervised diffusion mri denoising with generative diffusion models.
\newblock \emph{arXiv preprint arXiv:2302.03018}, 2023.

\bibitem[Xie et~al.(2020)Xie, Wang, and Ji]{xie2020noise2same}
Yaochen Xie, Zhengyang Wang, and Shuiwang Ji.
\newblock Noise2same: Optimizing a self-supervised bound for image denoising.
\newblock \emph{Advances in neural information processing systems}, 33:\penalty0 20320--20330, 2020.

\bibitem[Xu et~al.(2020)Xu, Huang, Cheng, Liu, Zhu, Xu, and Shao]{xu2020noisy}
Jun Xu, Yuan Huang, Ming-Ming Cheng, Li~Liu, Fan Zhu, Zhou Xu, and Ling Shao.
\newblock Noisy-as-clean: Learning self-supervised denoising from corrupted image.
\newblock \emph{IEEE Transactions on Image Processing}, 29:\penalty0 9316--9329, 2020.

\bibitem[Zbontar et~al.(2018)Zbontar, Knoll, Sriram, Murrell, Huang, Muckley, Defazio, Stern, Johnson, Bruno, Parente, Geras, Katsnelson, Chandarana, Zhang, Drozdzal, Romero, Rabbat, Vincent, Yakubova, Pinkerton, Wang, Owens, Zitnick, Recht, Sodickson, and Lui]{zbontar2018fastMRI}
Jure Zbontar, Florian Knoll, Anuroop Sriram, Tullie Murrell, Zhengnan Huang, Matthew~J. Muckley, Aaron Defazio, Ruben Stern, Patricia Johnson, Mary Bruno, Marc Parente, Krzysztof~J. Geras, Joe Katsnelson, Hersh Chandarana, Zizhao Zhang, Michal Drozdzal, Adriana Romero, Michael Rabbat, Pascal Vincent, Nafissa Yakubova, James Pinkerton, Duo Wang, Erich Owens, C.~Lawrence Zitnick, Michael~P. Recht, Daniel~K. Sodickson, and Yvonne~W. Lui.
\newblock {fastMRI}: An open dataset and benchmarks for accelerated {MRI}, 2018.

\bibitem[Zhou et~al.(2024)Zhou, Lou, Khanna, and Ermon]{zhoudenoising}
Linqi Zhou, Aaron Lou, Samar Khanna, and Stefano Ermon.
\newblock Denoising diffusion bridge models.
\newblock In \emph{The Twelfth International Conference on Learning Representations}, 2024.

\end{thebibliography}
\bibliographystyle{iclr2025_conference}

\appendix

\renewcommand{\thesection}{\Alph{section}}
\renewcommand{\thetable}{S\arabic{table}}
\renewcommand{\thefigure}{S\arabic{figure}}

\section{Comparisons with related works}
\label{Comparisons with related works}

\paragraph{Comparison with Patch2Self}
Patch2Self~\citep{fadnavis2020patch2self} requires a minimum of ten additional diffusion vector volumes to denoise a single diffusion vector volume. Instead, our work only needs one additional volume, which is clinically meaningful as common clinical dMRI often scans fewer than ten diffusion vector volumes~\citep{karayumak2019retrospective,xiang2023ddm}. Moreover, our model does not require repetitive training, whereas Patch2Self necessitates training multiple regressors to perform voxel-by-voxel denoising. However, despite achieving better results, Di-Fusion takes relatively longer time than Patch2Self, which is due to the training time of the diffusion models.

\paragraph{Comparison with DDM2}
Our work only requires a single stage for denoising, whereas DDM2 typically involves three stages. Furthermore, it's worth noting that the noise model in the first stage of DDM2 critically influences the ultimate denoising results, and finding an optimal solution that simultaneously maximizes evaluation metrics scores and minimizes training time can be challenging (See Fig. \ref{DDM2BAD} for details).

\section{Di-Fusion}
\label{Di-Fusion}

\subsection{Forward process}

Consider $x = {X_{*,*,i,j}}$ ($i$: slice index, $j$: volume index) as the target slice to denoise, $x' = {X_{*,*,i,j - 1}}$. ${\beta _{1, \cdots ,T}}$ is a pre-defined noise schedule, $\sigma _t^2 := {\beta _t}$, ${\alpha _t}: = 1 - {\beta _t}$ and ${\bar \alpha _t}: = \prod\nolimits_{s = 1}^t {{\alpha _s}}$. We rewrite $\lambda _1^t=\frac{{\sqrt {{{\overline \alpha  }_{t - 1}}} {\beta _t}}}{{1 - {{\overline \alpha  }_t}}}$ and $\lambda _2^t=\frac{{\sqrt {{\alpha _t}} \left( {1 - {{\overline \alpha  }_{t - 1}}} \right)}}{{1 - {{\overline \alpha  }_t}}}$ for simplification.

Perform the Fusion process:
\begin{equation}
   x_t^ *  = {\lambda _1^t} x + {\lambda _2^t} x'.
\end{equation}

Then we get a linear interpolation between $x$ and $x'$, we compute $x_t$ based on $q\left( {{x_{t}}{\rm{|}}x_t^*} \right)$:

\begin{equation}
    {x_t} = \sqrt {{{\bar \alpha }_t}} x_t^* + \sqrt {1 - {{\bar \alpha }_t}} z.
\end{equation}

The forward process can be defined if using $z \sim {\cal N}\left( {\mathbf{0},\mathbf{I}} \right)$ for perturbing data distribution:

\begin{equation}
q\left( {{x_{t}}{\rm{|}}x_t^*} \right): = {\cal{N}}\left( {{x_t};\sqrt {{{\bar \alpha }_t}} x_t^*, ({1 - {{\bar \alpha }_t}}) {\mathbf{I}}} \right).
\end{equation}

However, we use ``Di-'' process to compute a noise ${\xi _{x - x'}}$ to substitute for $z$:

\begin{equation}
   {\xi _{x - x'}} = mess\left( {\left( {x - x'} \right) - {\mu _{x - x'}}} \right),\quad{\mu _{x - x'}} = \frac{{\sum\nolimits_{m = 1}^w {\sum\nolimits_{n = 1}^h {\left( {{x_{mn}} - {{x'}_{mn}}} \right)} } }}{{w \cdot h}}.
\end{equation}

So the forward process can't be represented as ${\cal{N}}\left( {{x_t};\sqrt {{{\bar \alpha }_t}} x_t^*, ({1 - {{\bar \alpha }_t}}) {\mathbf{I}}} \right)$, but could be computed using the following formula:
{\begin{equation}\label{Di-Fusion forward}
    q\left( {{x_{t}}{\rm{|}}x_t^*} \right) \to {x_t} = \sqrt {{{\bar \alpha }_t}} x_t^* + \sqrt {1 - {{\bar \alpha }_t}} {\xi _{x - x'}}.
\end{equation}}

{We leverage a dynamic combination (the Fusion process) and continuously varying noise (the "Di-" process) to provide the model with more augmented training data, thereby enhancing its robustness. This idea is similar to those in Noise2Void~\citep{krull2019noise2void}, Noisier2Noise~\citep{moran2020noisier2noise}, and Noisy-as-Clean~\citep{xu2020noisy}, which also utilize data augmentation to construct training data.}

\subsection{Training process}

Our simplified training objective is:
\begin{equation}
 {L_{{\rm{simple}}}}(\theta ): = {{\mathbb{E}}_{t,x_t^*,{\xi _{x - x'}}}}\left[ {{{\left\| {x - {\cal{F}_\theta }(\sqrt {{{\bar \alpha }_t}} x_t^* + \sqrt {1 - {{\bar \alpha }_t}} {\xi _{x - x'}},t)} \right\|}^2}} \right].
\end{equation}

We perform training the latter diffusion steps by sample $t \sim {\rm{Uniform}}\left( {\left\{ {1, \cdots ,T_c} \right\}} \right)$.

\subsection{Reverse process}

The details of how to perform the reverse process in DDPM if a data predictor is used are in Appendix \ref{Proof: x barx diff}. If it is a data predictor $\cal{F}_\theta$ that directly predict $x_0$, the reverse process for DDPM becomes:
\begin{equation}
{x_{t-1}} = \lambda _1^{t}{\cal{F}_\theta }\left( {{x_{t}},{t}} \right) + \lambda _2^{t}{x_{t}} + \left( {{\sigma _{t}} \cdot \eta } \right){\xi_{x - x'}}.
\end{equation}

And ${p_\mathcal{F}}\left( {{x_{t - 1}}|{x_t}} \right)$ can be defined as:
\begin{equation}
    {p_{\mathcal{F}}}\left( {{x_{t - 1}}|{x_t}} \right) \to {x_{t - 1}} = \lambda _1^t{{{\cal F}_\theta }\left( {{x_t},t} \right)} + \lambda _2^t{x_t} + \left( {{\sigma _t} \cdot \eta } \right){\xi_{x - x'}}.
\end{equation}

Now let us consider the forward process as defined not on all $\left\{ {{x_t}} \right\}_1^{{T_c}}$, but on a subset $\{x_{\tau_1}, \ldots, x_{\tau_S}\}$, where $\tau$ is an increasing sub-sequence of $[1, \ldots, T_c]$ of length $S$. In particular, we define the sequential forward process over $x_{\tau_1}, \ldots, x_{\tau_S}$ (${x_{{\tau_k}}} = \sqrt {{{\bar \alpha }_{{\tau_k}}}} \left( {\lambda _1^{_{{\tau_k}}}x + \lambda _2^{_{{\tau_k}}}x'} \right) + \sqrt {1 - {{\bar \alpha }_{{\tau_k}}}} {\xi_{x - x'}}$, $1 \le k \le {S}$). 

The \textit{Run-Walk} accelerated sampling now sample according to $\text{reversed}(\tau)$ (In practice, $\tau  = \left\{ {1,2, \cdots ,{T_{r}} - 1,{T_{r}},{T_{r}} + p, \cdots ,{T_c} - p,{T_c}} \right\}$), then the reverse process become:

\begin{equation}
{p_{\mathcal{F}_\theta}}\left( {{x_{{\tau _{k - 1}}}}|{x_{{\tau _k}}}} \right)\to{x_{{\tau_{k-1}}}} = \lambda _1^{\tau_k}{\cal{F}_\theta }\left( {{x_{\tau_k}},{\tau_k}} \right) + \lambda _2^{\tau_k}{x_{\tau_k}} + \left( {{\sigma _{\tau_k}} \cdot \eta } \right){\xi_{x - x'}}.
\end{equation}

Before sampling, we define an universal value $\cal{CSNR}$ and compute $b_x$:

\begin{equation}
   {b_x = \frac{\sum_{m=1}^{w}\sum_{n=1}^{h}1}{2 \cdot \sum_{m=1}^{w}\sum_{n=1}^{h}{{\mathbb{I}}_{(x_{mn} > \beta_1)}}} + \frac{\sum_{m=1}^{w}\sum_{n=1}^{h}1}{2 \cdot \sum_{m=1}^{w}\sum_{n=1}^{h}{{\mathbb{I}}_{(x_{mn} > \beta_2)}}}},
\end{equation}

During every ${p_{{\mathcal{F}_\theta}}}\left( {{x_{{\tau _{k - 1}}}}|{x_{{\tau _k}}}} \right)$, we first get $x_{out}={\cal{F}_\theta }\left( {{x_{\tau_k}},{\tau_k}} \right)$, and ${d_x} = {\left\| {x - {x_{out}}} \right\|^2} \times {b_x}$.

Then if ${d_x}$ is greater than $\cal CSNR$, the output ${x_{0}} = {x_{out}}$ and the refinement iteration breaks. In contrast, the refinement iteration continues if ${d_x}$ is smaller than $\cal CSNR$.

\section{Additional derivations}

\subsection{The difference between the two trajectories}
\label{Proof: x barx diff}

The original sampling process in the \textit{Algorithm 2} of DDPM~\citep{ho2020denoising} is:
\begin{equation}
x_{t-1} = \frac{1}{\sqrt{\alpha_t}}\left( x_t - \frac{1-\alpha_t}{\sqrt{1-\bar\alpha_t}} \epsilon_\theta(x_t, t) \right) + \sigma_t z,
\end{equation}

where $\epsilon_\theta$ is a noise predictor. However, we use a data predictor $\cal{F_\theta}$ to directly predict $x_0$ in our paper. We will demonstrate how to perform the reverse process in DDPM if a data predictor is used.

Given a data point sampled from a real data distribution $x_0 \sim q(x)$, let us define a forward diffusion process in which we add small amount of Gaussian noise to the sample in $T$ steps, producing a sequence of noisy samples $x_1, \dots, x_T$. The step sizes are governed by a variance schedule \( \{\beta_t \in (0, 1)\}_{t=1}^T \):
\begin{equation}
q(x_t \vert x_{t-1}) = \mathcal{N}(x_t; \sqrt{1 - \beta_t} x_{t-1}, \beta_t\mathbf{I}) ,\quad
q(x_{1:T} \vert x_0) = \prod^T_{t=1} q(x_t \vert x_{t-1}).
\end{equation}
As the step \( t \) increases, the data sample \( x_0 \) gradually loses its distinguishable features. Ultimately, when \( T \to \infty \), \( x_T \) converges to an isotropic Gaussian distribution.

Let \( \alpha_t = 1 - \beta_t \) and \( \bar{\alpha}_t = \prod_{i=1}^t \alpha_i \). A nice property of the aforementioned process is that we can sample \( x_t \) at any arbitrary time step \( t \) in closed form using the reparameterization trick:

\begin{equation}
\begin{aligned}
{x}_t 
&= \sqrt{\alpha_t}{x}_{t-1} + \sqrt{1 - \alpha_t}\boldsymbol{z}_{t-1} & \text{ ;where } \boldsymbol{z}_{t-1}, \boldsymbol{z}_{t-2}, \dots \sim \mathcal{N}(\mathbf{0}, \mathbf{I}). \\
&= \sqrt{\alpha_t \alpha_{t-1}} {x}_{t-2} + \sqrt{1 - \alpha_t \alpha_{t-1}} \bar{\boldsymbol{z}}_{t-2} & \text{ ;where } \bar{\boldsymbol{z}}_{t-2} \text{ merges two Gaussians.} \\
&= \dots \\
&= \sqrt{\bar{\alpha}_t}{x}_0 + \sqrt{1 - \bar{\alpha}_t}\boldsymbol{z},
\end{aligned}
\end{equation}
where we merge two Gaussians with different variances, \( \mathcal{N}(\mathbf{0}, \sigma_1^2 \mathbf{I}) \) and \( \mathcal{N}(\mathbf{0}, \sigma_2^2 \mathbf{I}) \), resulting in a new distribution \( \mathcal{N}(\mathbf{0}, (\sigma_1^2 + \sigma_2^2) \mathbf{I}) \). Here, the merged standard deviation is given by \( \sqrt{(1 - \alpha_t) + \alpha_t (1 - \alpha_{t-1})} = \sqrt{1 - \alpha_t \alpha_{t-1}} \). We can then derive:

\begin{equation}
    q({x}_t \vert {x}_0) = \mathcal{N}({x}_t; \sqrt{\bar{\alpha}_t} {x}_0, (1 - \bar{\alpha}_t)\mathbf{I}).
\end{equation}

Consider a reverse process, it is noteworthy that the reverse conditional probability is tractable when conditioned on ${x}_0$:
\begin{equation}
q({x}_{t-1} \vert {x}_t, {x}_0) = \mathcal{N}({x}_{t-1}; {\tilde{\boldsymbol{\mu}}}({x}_t, {x}_0),{\sigma _t^2{\mathbf{I}}}).    
\end{equation}

Using Bayes' rule, we then have~\citep{ho2020denoising}:

\begin{equation}
\begin{aligned}
&q({x}_{t-1} \vert {x}_t, {x}_0) \\
&= q({x}_t \vert {x}_{t-1}, {x}_0) \frac{ q({x}_{t-1} \vert {x}_0) }{ q({x}_t \vert {x}_0) } \\
&\propto \exp \Big(-\frac{1}{2} \big(\frac{({x}_t - \sqrt{\alpha_t} {x}_{t-1})^2}{\beta_t} + \frac{({x}_{t-1} - \sqrt{\bar{\alpha}_{t-1}} {x}_0)^2}{1-\bar{\alpha}_{t-1}} - \frac{({x}_t - \sqrt{\bar{\alpha}_t} {x}_0)^2}{1-\bar{\alpha}_t} \big) \Big) \\
&= \exp \Big(-\frac{1}{2} \big(\frac{{x}_t^2 - 2\sqrt{\alpha_t} {x}_t {{x}_{t-1}} {+ \alpha_t} {{x}_{t-1}^2} }{\beta_t} + \frac{ {{x}_{t-1}^2} {- 2 \sqrt{\bar{\alpha}_{t-1}} {x}_0} {{x}_{t-1}} {+ \bar{\alpha}_{t-1} {x}_0^2}  }{1-\bar{\alpha}_{t-1}} - \frac{({x}_t - \sqrt{\bar{\alpha}_t} {x}_0)^2}{1-\bar{\alpha}_t} \big) \Big) \\
&= \exp\Big( -\frac{1}{2} \big( {(\frac{\alpha_t}{\beta_t} + \frac{1}{1 - \bar{\alpha}_{t-1}})} {x}_{t-1}^2 - {(\frac{2\sqrt{\alpha_t}}{\beta_t} {x}_t + \frac{2\sqrt{\bar{\alpha}_{t-1}}}{1 - \bar{\alpha}_{t-1}} {x}_0)} {x}_{t-1} { + C({x}_t, {x}_0) \big) \Big)},
\end{aligned}
\end{equation}

where \( C(x_t, x_0) \) is some function that does not involve \( x_{t-1} \), and the details are omitted. Following the standard Gaussian density function, the mean and variance can be parameterized as follows (recall that \( \alpha_t = 1 - \beta_t \) and \( \bar{\alpha}_t = \prod_{i=1}^T \alpha_i \)):

\begin{equation}
\begin{aligned}
\sigma _t^2 
&= 1/(\frac{\alpha_t}{\beta_t} + \frac{1}{1 - \bar{\alpha}_{t-1}}) 
= 1/(\frac{\alpha_t - \bar{\alpha}_t + \beta_t}{\beta_t(1 - \bar{\alpha}_{t-1})})
= {\frac{1 - \bar{\alpha}_{t-1}}{1 - \bar{\alpha}_t} \cdot \beta_t}.
\end{aligned}
\end{equation}
\begin{equation}
\begin{aligned}
\tilde{\boldsymbol{\mu}}_t ({x}_t, {x}_0)
&= (\frac{\sqrt{\alpha_t}}{\beta_t} {x}_t + \frac{\sqrt{\bar{\alpha}_{t-1} }}{1 - \bar{\alpha}_{t-1}} {x}_0)/(\frac{\alpha_t}{\beta_t} + \frac{1}{1 - \bar{\alpha}_{t-1}}) \\
&= (\frac{\sqrt{\alpha_t}}{\beta_t} {x}_t + \frac{\sqrt{\bar{\alpha}_{t-1} }}{1 - \bar{\alpha}_{t-1}} {x}_0) {\frac{1 - \bar{\alpha}_{t-1}}{1 - \bar{\alpha}_t} \cdot \beta_t} \\
&= \frac{\sqrt{\alpha_t}(1 - \bar{\alpha}_{t-1})}{1 - \bar{\alpha}_t} {x}_t + \frac{\sqrt{\bar{\alpha}_{t-1}}\beta_t}{1 - \bar{\alpha}_t} {x}_0.
\end{aligned}
\end{equation}

Thus, if it is a data predictor $\cal{F}_\theta$ that directly predict $x_0$, based on $q({x}_{t-1} \vert {x}_t, {x}_0)$, the reverse process for DDPM becomes:

\begin{equation}
\begin{aligned}
{x_{t - 1}} 
&= {\tilde {\boldsymbol{\mu }}_t}({x_t},{x_0}) + \sigma _t^2 z\\
&= \frac{{\sqrt {{\alpha _t}} \left( {1 - {{\bar \alpha }_{t - 1}}} \right)}}{{1 - {{\bar \alpha }_t}}}{x_t} + \frac{{\sqrt {{{\bar \alpha }_{t - 1}}} {\beta _t}}}{{1 - {{\bar \alpha }_t}}}{\cal{F}_\theta }\left( {{x_t},t} \right) + {\sigma _t^2}z.
\end{aligned}
\end{equation}
DDPM~\citep{ho2020denoising} found that both $\sigma_t^2 = \beta_t$ and $\sigma_t^2 = \frac{1-\bar\alpha_{t-1}}{1-\bar\alpha_t}\beta_t$ had similar results through experiments. We set $\sigma _t^2 = {\beta _{1, \cdots ,T}}$ and hold ${\beta _{1, \cdots ,T}}$ as hyperparameters. 

Now we know how to perform the reverse process if a data predictor is used. According to Eq. (\ref{eq:x'2x}), we know $x + {\epsilon_t} = {x_{out}}={\cal{F}_\theta }\left( {{x_t},t} \right)$, then we can get:

\begin{equation}
\begin{aligned}
{x_{t - 1}} 
&= \frac{{\sqrt {{\alpha _t}} \left( {1 - {{\bar \alpha }_{t - 1}}} \right)}}{{1 - {{\bar \alpha }_t}}}{x_t} + \frac{{\sqrt {{{\bar \alpha }_{t - 1}}} {\beta _t}}}{{1 - {{\bar \alpha }_t}}} \left( x + {\epsilon_t} \right)+ {\sigma _t^2}z,
\end{aligned}
\end{equation}

Now let us consider directly performing the forward process ($q({x}_t \vert {x}_0)$) on $x'$($x = {X_{*,*,i,j}},x' = {X_{*,*,i,j-1}}$) without the Fusion process (Eq. (\ref{eq:fusion})):

\begin{equation}
{{\bar x}_{t - 1}}=\sqrt {{{\bar \alpha }_{t - 1}}} x' + \sqrt {1 - {{\bar \alpha }_{t - 1}}} z,
\end{equation}

thus the trajectory $\left\{ {{{\bar x}_t}} \right\}_1^{{T}}$ obtained by directly performing the forward process in DDPM and the trajectory $\left\{ {{x_t}} \right\}_1^{{T}}$ obtained from the reverse process of DDPM are different, and the major difference is brought by $(x + {\epsilon_t})$:

\begin{equation}
   {x_{t - 1}} = \underbrace {\frac{{\sqrt {{{\bar \alpha }_{t - 1}}} {\beta _t}}}{{1 - {{\bar \alpha }_t}}}({x + {\epsilon_t})}}_{{\rm{major}}{\kern 1pt} {\rm{difference}}} + \frac{{\sqrt {{\alpha _t}} \left( {1 - {{\bar \alpha }_{t - 1}}} \right)}}{{1 - {{\bar \alpha }_t}}}{x_t} + {\sigma _t}z \ne {{\bar x}_{t - 1}} = \sqrt {{{\bar \alpha }_{t - 1}}} x' + \sqrt {1 - {{\bar \alpha }_{t - 1}}} z. 
\end{equation}
This is because component $\epsilon_t$ decays as $t \to 0$, then a larger proportion of components in $x_{t-1}$ becomes closer to $x$.

{If we directly feed $x _ {{t}-1}$ and ${t}-1$ into ${{\cal F} _ \theta }$, the output would deviate slightly further from $x$. This occurs because during training, ${{\cal F} _ \theta }$ is optimized only with the objective:
${\left\| {x - {\cal{F} _ \theta }(\sqrt {{{\bar \alpha }_{{t-1}}}} x' + \sqrt {1 - {{\bar \alpha } _ {{t-1}}}} {z},{{t-1}})} \right\|}^2$ (the training objective without the Fusion process). Importantly, $x _ {{t-1}}$ is one step closer to $x$. ($x _ {{t-1}}= {\frac{{\sqrt {{{\bar \alpha }_{t - 1}}} {\beta _t}}}{{1 - {{\bar \alpha }_t}}}({x + {\epsilon_t})}} + \frac{{\sqrt {{\alpha _t}} \left( {1 - {{\bar \alpha }_{t - 1}}} \right)}}{{1 - {{\bar \alpha }_t}}}{x_t} + {\sigma _t}z$), rather than simply being a noisy version of $x'$. This drift accumulates over the sampling chain, ultimately leading the result to drift toward another slice.}

\subsection{Variance information of noise in ``Di-'' process}
\label{Proof: variance information of nosie}

${\xi _{x - x'}}$ theoretically preserves the variance information of the noise:

\begin{equation}
\begin{array}{*{20}{l}}
{{\rm{Var}}\left( {x - x'} \right)}& = &{\rm{Var}}\left( {y + {n_1} - \left( {y + {n_2}} \right)} \right) \\
{}& = &{{\rm{Var}}\left( {{n_1} - {n_2}} \right)}\\
{}& = &{{\rm{Var}}\left( {{n_1}} \right) + {\rm{Var}}\left( {{n_2}} \right) - 2{\rm{Cov}}\left( {{n_1},{n_2}} \right)}\\
{}& = &{{\rm{Var}}\left( {{n_1}} \right) + {\rm{Var}}\left( {{n_2}} \right),}
\end{array}
\end{equation}

where ${\rm{Cov}}\left(  \cdot  \right)$ is the covariance, ${\rm{Var}}\left(  \cdot  \right)$ is the variance, ${\rm{Cov}}\left( {n_1},n_2 \right)=0$ since $n_1$ and $n_1$ are independent. Assuming that $n_1$ and $n_2$ follow the same distribution, the variance information of this distribution is retained.

In Fig. \ref{noise}, we show that the noise in ``Di-'' process has different statistical properties compared to Gaussian noise.

\subsection{Speed towards the target}
\label{Proof: speed to x0}

The difference between ${x_{t-1}}$ and ${x_{t}}$ can be formulated as:

\begin{equation} \label{ddddddd}
    \begin{array}{*{20}{l}}
{{x_{t - 1}} - {x_t}}& = &{\lambda _1^t{{\cal F}_\theta }\left( {{x_t},t} \right) + \lambda _2^t{x_t} + \left( {{\sigma _t} \cdot \eta } \right){\xi_{x - x'}} - {x_t}}\\
{}& = &{\lambda _1^t{x_{out}} + \left( {1 - \lambda _1^t} \right){x_t} + \left( {{\sigma _t} \cdot \eta } \right){\xi_{x - x'}} - {x_t}}.
\end{array}
\end{equation}

According to Eq. (\ref{eq:x'2x}), we know $x + {\epsilon_t} = {x_{out}}$, then we can substitute ${x_{out}}$ into the Eq. (\ref{ddddddd}) and get:
\begin{equation} 
    \begin{array}{*{20}{l}}
{{x_{t - 1}} - {x_t}}& = &{\lambda _1^t{x_{out}} + \left( {1 - \lambda _1^t} \right){x_t} + \left( {{\sigma _t} \cdot \eta } \right){\xi _{x - x'}} - {x_t}}\\
{}& = &{\lambda _1^t\left( {x + {_t}} \right) + \left( {1 - \lambda _1^t} \right){x_t} + \left( {{\sigma _t} \cdot \eta } \right){\xi _{x - x'}} - {x_t}}\\
{}& = &{\underbrace {\lambda _1^t\left( {x - {x_t}} \right)}_{{\rm{speed}}} + {\underbrace{\left( {{\sigma _t} \cdot \eta } \right){\xi _{x - x'}}}_{{\rm{noise}}}} + \underset{{\rm{perturbation}}}{\underbrace{{\lambda _1^t} {\epsilon_t}}}}.
\end{array}
\end{equation}

In DDIM~\citep{song2020denoising}, \(\eta=0\), so typically, the term ``noise'' disappears, leading to the following expression:

\begin{equation} 
    \begin{array}{*{20}{l}}
{{x_{t - 1}} - {x_t}}& = &{\underbrace {\lambda _1^t\left( {x - {x_t}} \right)}_{{\rm{speed}}} + \underset{{\rm{perturbation}}}{\underbrace{{\lambda _1^t} {\epsilon_t}}}}.
\end{array}
\end{equation}

We know that $\epsilon_t$ is a perturbation term that depends on $t$, as $t \to 0$, $\epsilon_t \to 0$, ${\lambda _1^t} \to 1$ at the same time. So the value of ``perturbation'' item does not change significantly when $t$ decreases; thus difference between ${x_{t-1}}$ and ${x_{t}}$ are mainly caused by the ``speed'' item.

\section{Experimental details}
\label{A-Experimental details}

\subsection{Experiment and reproducibility details}
\label{Experiment and reproducibility details}
\paragraph{Noise schedule}
\label{Noise schedule}
A typical noise schedule ~\citep{ho2020denoising,saharia2022image} follows a ``warm-up'' scheduling strategy. Inspired by DDM2, we implement a reverse ``warm-up'' strategy where $\beta_t$ remains at $5e^{-5}$ for the first 300 iterations and then linearly increases to $1e^2$ between $(300, 1000]$ iterations~\citep{xiang2023ddm}.

\paragraph{Training details}
\label{Training details}

Following DDPM~\citep{ho2020denoising}, we set $\sigma _t^2 = {\beta _{1, \cdots ,T}}$ and hold ${\beta _{1, \cdots ,T}}$ as hyperparameters. Since we are performing a deterministic sampling process, $\eta$ in Eq. (\ref{eq:reverse}) is set to 0 (we talk about how $\eta$ influences the final results in Appendix \ref{A-Di- process: Different noise distribution}). We implement denoising functions ${\mathcal{F}_\theta }$ via U-Net~\citep{ronneberger2015u} with modifications suggested in ~\citep{saharia2022image,song2020score}. Inspired by ~\citep{chen2020wavegrad,song2019generative}, we train ${\mathcal{F}_\theta }$ to condition on ${{\overline \alpha  }_{t}}$, $t \sim {\rm{Uniform}}\left( {\left\{ {1, \cdots ,T_c} \right\}} \right)$, $T_c=300$. Adam optimizer was used to optimize $\theta$ with a fixed learning rate of $1e^{-4}$ and a batch size of 32, and ${\mathcal{F}_\theta }$ was trained $1e^{5}$ steps from scratch. All experiments were performed on RTX GeForce 3090 GPUs in PyTorch~\citep{paszke2019pytorch}. {The training duration for one ${\mathcal{F}_\theta }$ is approximately five hours on a single RTX GeForce 3090 GPU with 5578MB of VRAM.}

\paragraph{Sampling details}
\label{Sampling details}

During sampling, $T_{r}=50$. ${\beta _1}=-0.93$ and ${\beta _2}=-0.95$ and changing their values has little impact on the results (We set these two factors as a simple way to extract the brain mask). \(\eta = 0\) and $p=10$ if no special instructions are provided. $\cal{CSNR}$ are provided in the figure caption.

\paragraph{Competing methods details}
\label{Evaluation details}

In the main paper, Di-Fusion is compared against four state-of-the-art self-supervised deep learning-based denoising methods (ASCM isn't deep learning-based). For fair comparisons DIP, Nr2N, and DDM2 adopt the architecture used in Di-Fusion. We follow the official repository\footnote{https://github.com/ShreyasFadnavis/patch2self} and use the parameters that should give the optimal denoising performance for P2S~\citep{fadnavis2020patch2self}. \textit{(i)} Deep Image Prior~\citep{ulyanov2018deep} trains a network on a random input to target a noisy image. Thus, network parameter optimization must be performed for each image. In our experiments, we use their official repository\footnote{https://github.com/DmitryUlyanov/deep-image-prior} to identify the optimal training iterations on a single image and then apply the same number of iterations for denoising the entire volume. \textit{(ii)} Noisier2Noise~\citep{moran2020noisier2noise} trains a network on a noisier input to target a noisy image. In our experiments, we add additional randomly sampled noise to $x'$, and the training is performed to reconstruct the noisy image $x$ (Noise2Noise wasn't used due to its pronounced over-smoothing denoising effect in the DDM2 experiments~\citep{xiang2023ddm}. We want to evaluate our method further by using an advanced version, Noisier2Noise). \textit{(iii)} Patch2Self~\citep{fadnavis2020patch2self} generalizes Noise2Noise~\citep{lehtinen2018noise2noise} and Noise2Self~\citep{batson2019noise2self} for voxel-by-voxel dMRI denoising. In our experiments, we followed their official implementation without adjusting their hyperparameters. \textit{(iv)} DDM2~\citep{xiang2023ddm} proposes a three-stage framework that integrates statistic-based denoising theory into diffusion models and performs denoising through conditional generation. In our experiments, we follow their official repository\footnote{https://github.com/StanfordMIMI/DDM2} without adjusting their hyperparameters.

{Additionally, more comparisons with other denoising methods, including MPPCA \citep{veraart2016denoising}, Noise2Score \citep{kim2021noise2score}, Recorrupted2Recorrupted \citep{pang2021recorrupted}, and Patch2Self2 \citep{fadnavis2024patch2self2}, are provided in Appendix \ref{Supplementary experimental results}. We implemented MPPCA using the code from DIPY \citep{garyfallidis2014dipy}. For Noise2Score (N2S), we utilized their official repository\footnote{{https://github.com/cubeyoung/Noise2Score}}. Recorrupted2Recorrupted (R2R) was implemented using its repository\footnote{{https://github.com/PangTongyao/Recorrupted-to-Recorrupted-Unsupervised-Deep-Learning-for-Image-Denoising}}. For Patch2Self2 (P2S2), we directly used the denoised data provided in their supplementary material\footnote{The denoised data is shared at {https://figshare.com/s/87f6ffee972510bfda76}} \citep{fadnavis2024patch2self2}.}

\subsection{Downstream clinical tasks implementation details}
\label{Modeling implementation details}

\paragraph{On tractography}

To reconstruct white-matter pathways in the brain, one integrates orientation information of the underlying axonal bundles (streamlines) obtained by decomposing the signal in each voxel using a microstructure model~\citep{behrens2014mr,fadnavis2020patch2self,garyfallidis2014dipy}. Noise that corrupts the acquired DWI may impact the tractography results, leading to spurious streamlines generated by the tracking algorithm. We explore the effect of denoising on probabilistic tracking~\citep{girard2014towards} by employing the Fiber Bundle Coherency (FBC) metric~\citep{portegies2015improving}. We first fit the data to the Constrained Spherical Deconvolution (CSD) model~\citep{tournier2007robust}. The fiber orientation distribution information required to perform the tracking is obtained from the Constrained Spherical Deconvolution (CSD) model fitted to the same data. The Optic Radiation (OR) is reconstructed by tracking fibers (3x3x3 voxels ROI cube, and the seed density is 6) from the calcarine sulcus (visual cortex V1) to the lateral geniculate nucleus (LGN). After the streamlines are generated, their coherency is measured with the local FBC algorithm~\citep{portegies2015improving,duits2011left}, with yellow-orange representing - spurious/incoherent fibers and red-blue representing valid/coherent fibers. Since low FBCs indicate which fibers are isolated and poorly aligned with their neighbors, we further clean the results of tractography algorithms by using the stopping criterion outlined in ~\citep{meesters2016cleaning} (the stopping criterion was only performed on noisy data's density map of FBC; thus, its results are captioned by ``Noisy\_filtering'' and can be considered as the reference for high FBCs).

\paragraph{On microstructure model fitting}

Microstructure modeling poses a complicated inverse problem and often leads to degraded parameter estimates due to the low SNR of dMRI~\citep{novikov2018modeling}. Different denoising methods can be compared based on their accuracy in fitting the diffusion signal. We apply two commonly used diffusion microstructure models, the diffusion tensor model (DTI)~\citep{basser1994mr} and Constrained Spherical Deconvolution (CSD)~\citep{tournier2007robust} (DIPY~\citep{garyfallidis2014dipy} has made available of these two models), on raw and denoised data. DTI is a simple model that captures the local diffusion information within each voxel by modeling it in the form of a 6-parameter tensor. CSD is a more complex model using a spherical harmonic representation of the underlying fiber orientation distributions. In order to quantify the results, we perform a 3-fold cross-validation~\citep{hastie2009elements} at two exemplary voxel locations, corpus callosum (CC), a single-fiber structure, and centrum semiovale (CSO), a crossing-fiber structure. The data is divided into three different subsets for the selected voxels, and data from two folds are used to fit the model, which predicts the data on the held-out fold. We quantify the goodness of fit of the models by calculating the $R^2$ score ($R^2$ metric is computed from each model fit on the corresponding data)~\citep{fadnavis2020patch2self}. 

\paragraph{On diffusion signal estimates}

We examine how the denoising quality translates to downstream clinical tasks such as creating DTI~\citep{basser1994mr} diffusion signal estimates using the various denoising methods. To do the comparisons, we use the volumes acquired by the first ten diffusion directions and the ten b-value=0 volumes. Before fitting the data, we perform data pre-processing. We first use the method in ~\citep{ostu1979threshold} to compute a brain mask to avoid unnecessary calculations on the background of the image. Now that we have loaded and pre-processed the data we can go forward with DTI~\citep{basser1994mr} fitting. We can extract the fractional anisotropy (FA), the mean diffusivity (MD), the axial diffusivity (AD) and the radial diffusivity (RD) from the DTI model.

\subsection{SNR and CNR implementation details}
\label{SNR and CNR implementation details}

To quantify the denoising performance, we employ Signal-to-Noise Ratio (SNR) and Contrast-to-Noise Ratio (CNR) metrics, which are also used in DDM2~\citep{xiang2023ddm}. We differentiate foreground and background signals following Patch2Self~\citep{fadnavis2020patch2self}: \textbf{1.} Perform uniform normalization on all the data; \textbf{2.} Use the method in ~\citep{ostu1979threshold} to compute a brain mask; \textbf{3.} fit DTI~\citep{basser1994mr} model to perform corpus callosum segementation; \textbf{4.} $\rm{signal}$ is corpus callosum signal, $\rm{background}$ is the signal out of the brain mask. \textbf{5.} Compute $\rm{SNR}$ and $\rm{CNR}$ using:

\begin{equation} 
    {\rm{SNR}} = \frac{{{\rm{Mean}}\left( {{\rm{signal}}} \right)}}{{{\rm{Var}}\left( {{\rm{background}}} \right)}},\quad {\rm{CNR}} = \frac{{{\rm{Mean}}\left( {{\rm{signal}}} \right) - {\rm{Mean}}\left( {{\rm{background}}} \right)}}{{{\rm{Var}}\left( {{\rm{background}}} \right)}},
\end{equation}

where ${\rm{Mean}}\left(  \cdot  \right)$ is the mean, ${\rm{Var}}\left(  \cdot  \right)$ is the variance; \textbf{6.} Statistics are performed on all computed $\rm{SNR}$ and $\rm{CNR}$ values, and a box plot like Fig. \ref{quantitative results} is created.

\subsection{Simulated data implementation details}
\label{Simulated data implementation details}

\paragraph{Details on making simulated data}
Apart from the experiments on \textit{in-vivo} datasets, we further simulate noisy k-space data to demonstrate that Di-Fusion can still be used for denoising tasks on simulated noisy MRI data, which is done on fastMRI datasets (fastMRI provides raw, complex, multi-echo, and multi-coil k-space MRI data)~\citep{tibrewala2023fastmri,zbontar2018fastMRI}. To simulate the effects of adding additional complex noise to the k-space data, we employ k-space noise addition strategies that have been previously validated in prior work ~\citep{desai2021vortex,desai2021noise2recon,xiang2023ddm}. Specifically, we start by sampling Gaussian noise with different standard deviations (to simulate different noise intensities) and add it to the real and imaginary parts of each coil's k-space data. Next, we utilize the inverse transformation function implemented in fastMRI~\citep{tibrewala2023fastmri,zbontar2018fastMRI} to convert the k-space data into simulated noisy datasets with varying degrees of noise. We simulate five datasets with different noise intensities. 

\paragraph{Declaration}
DIP is not considered as a comparison method due to its long computational time (the need for retraining on each image) and the mild blurring shown in Fig. \ref{qualitative results}, Fig. \ref{morequalitative_stanfordhardi}, \ref{morequalitative_s3sh} and Fig. \ref{morequalitative_ppmi}. The original Patch2Self is not included as a comparison method because it typically requires at least ten 3D volumes to achieve good results~\citep{fadnavispatch2self,fadnavis2020patch2self,garyfallidis2014dipy}. In contrast, the input 3D volumes in the simulated experiments are limited to a maximum of two (two for DDM2, one for Di-Fusion and Noisier2Noise). Directly comparing it with other methods on simulated data would be unfair. However, we still develop a reimplementation of Patch2Self, with modifications to the volume extraction part to limit the input volumes (two in our modified Patch2Self). \textit{It should be noted that our modified Patch2Self is solely utilized in the simulated experiments, where the limited input of two volumes does not yield optimal results. The original Patch2Self is still used in the remaining experiments carried out in this paper.} Nevertheless, Patch2Self remains a useful approach when applied to a larger number of volumes (e.g., ten).

\section{Supplementary experimental results}
\label{Supplementary experimental results}

{In this section, we include comparisons with additional denoising methods, including MPPCA \citep{veraart2016denoising}, Noise2Score (N2S) \citep{kim2021noise2score}, Recorrupted2Recorrupted (R2R) \citep{pang2021recorrupted}, and Patch2Self2 (P2S2) \citep{fadnavis2024patch2self2} (reproduction details are provided in Appendix \ref{Evaluation details}).}

\begin{table}
\caption{↑$R^2$ of microstructure model fitting on CSD \& DTI. \textbf{Bold} and \underline{Underline} fonts denote the best and the second-best performance, respectively. As measured by $R^2$, Di-Fusion achieves the best results across all four different settings.}\label{microstruture table}
\begin{center}
\renewcommand{\arraystretch}{1.0}
\setlength\tabcolsep{2.6pt} 
\begin{tabular}{lcccc}
\toprule
& \multicolumn{2}{c}{CSD} & \multicolumn{2}{c}{DTI}\\
\cmidrule{2-3}\cmidrule(l{2pt}r{2pt}){4-5}
Method & CC & CSO & CC & CSO \\

\midrule
Noisy & 0.797 & 0.614 & 0.789 & 0.484 \\
ASCM & 0.934 & 0.844 & 0.942 & 0.789 \\
DIP & 0.868 & 0.477 & 0.875 & 0.381 \\
Nr2N & \underline{0.959} & \underline{0.908} & \underline{0.961} & \underline{0.872} \\
P2S & 0.927 & 0.754 & 0.725 & 0.675 \\
DDM2 & 0.863 & 0.810 & 0.845 & 0.790 \\
OURS & \textbf{0.967} & \textbf{0.939} & \textbf{0.976} & \textbf{0.876} \\
\bottomrule
\end{tabular}
\end{center}
\end{table}

\begin{figure}
  \centering
  \includegraphics[scale = 0.60]{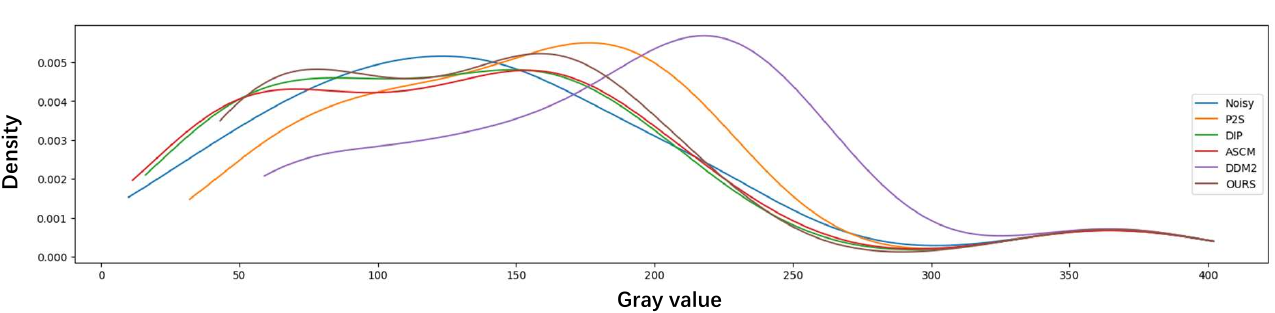}
  \caption{Data distribution plots on raw and denoised data. Note that DDM2 denoised data distribution has shifted from the noisy data.}
  \label{Quantitative comparison of microstructure model fitting}
\end{figure}

\begin{table}
{\caption{↑$R^2$ of microstructure model fitting on CSD \& DTI. \textbf{Bold} and \underline{Underline} fonts denote the best and the second-best performance, respectively. As measured by $R^2$, Di-Fusion achieves the best results across all four different settings.}\label{microstruture sup table}}
\begin{center}
\renewcommand{\arraystretch}{1.0}
\setlength\tabcolsep{2.6pt}
{
\begin{tabular}{lcccccccccccc}
\toprule
Method & Noisy & ASCM & MPPCA & DIP & Nr2N & N2S & R2R & P2S & DDM2 & P2S2 & OURS \\
\midrule
CSD-CC & 0.797 & 0.934 & 0.884 & 0.868 & \underline{0.959} & 0.823 & 0.879 & 0.927 & 0.863 & 0.957 & \textbf{0.967} \\
CSD-CSO & 0.614 & 0.844 & 0.750 & 0.477 & 0.908 & 0.468 & 0.731 & 0.754 & 0.810 & \underline{0.934} & \textbf{0.939} \\
DTI-CC & 0.789 & 0.942 & 0.881 & 0.875 & 0.961 & 0.831 & 0.872 & 0.725 & 0.845 & \underline{0.973} & \textbf{0.976} \\
DTI-CSO & 0.484 & 0.789 & 0.614 & 0.381 & \underline{0.872} & 0.348 & 0.677 & 0.675 & 0.790 & 0.867 & \textbf{0.876} \\
\bottomrule
\end{tabular}
}

\end{center}
\end{table}

\subsection{Microstructure model fitting and data distribution plots}
\label{A-Scatter plots of microstructure model fitting and data distribution plots}

In Table \ref{microstruture table}, as measured by $R^2$, our Di-Fusion achieves the best results across all four different settings. An intriguing observation is that the denoised data from DDM2 exhibits a relatively higher distribution (higher mean value) than other methods in Fig. \ref{Quantitative comparison of microstructure model fitting}. An observation is that the data from DDM2 exhibits a higher distribution than other methods. This may explain the improvement of DDM2 in CNR/SNR metrics in Fig. \ref{quantitative results}, as the foreground signals have higher values. Our experiments on downstream clinical tasks in Section \ref{Impacts on modeling} have shown no correlation between high or low scores on CNR/SNR metrics and the performance of the downstream clinical tasks.

{In Table \ref{microstruture sup table}, we summarize the quantitative \( R^2 \) metrics of all comparison methods. As measured by $R^2$, our Di-Fusion still achieves the best results across all four different settings.}

\subsection{Additional comparisons on tractography}

\begin{figure}
  \centering
  \includegraphics[scale = 0.310]{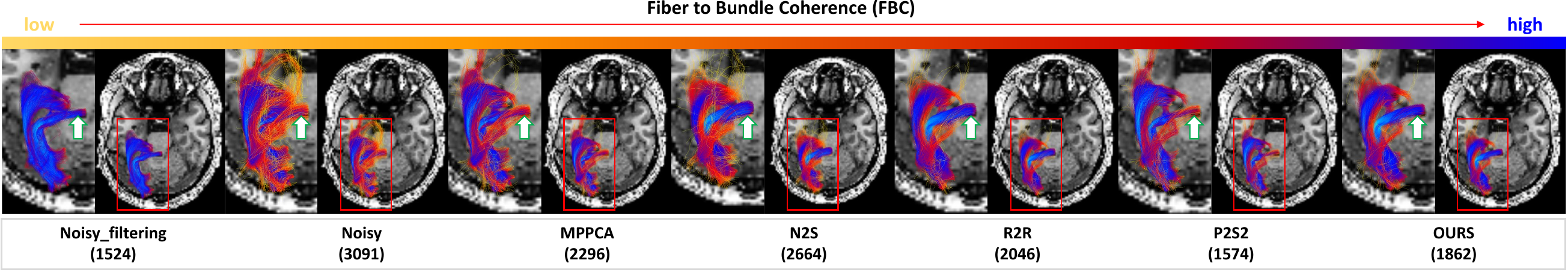}
  {\caption{Density map of FBC projected on the streamlines of the OR bundles. The numbers in parentheses represent the number of streamlines. Di-Fusion maintains high FBCs (consider ``Noisy\_filtering'' as references for high FBCs).}}
  \label{tractography sup}
\end{figure}

{In Fig. \ref{tractography sup}, we illustrate the effect on the tractography of OR using additional denoising methods. Di-Fusion effectively performs denoising while maintaining high FBCs, with "Noisy\_filtering" serving as the reference for high FBCs.}

\subsection{DTI diffusion signal estimates comparisons}
\label{A-DTI diffusion estimate comparisons}

\begin{figure}
  \centering
  \includegraphics[scale = 0.131]{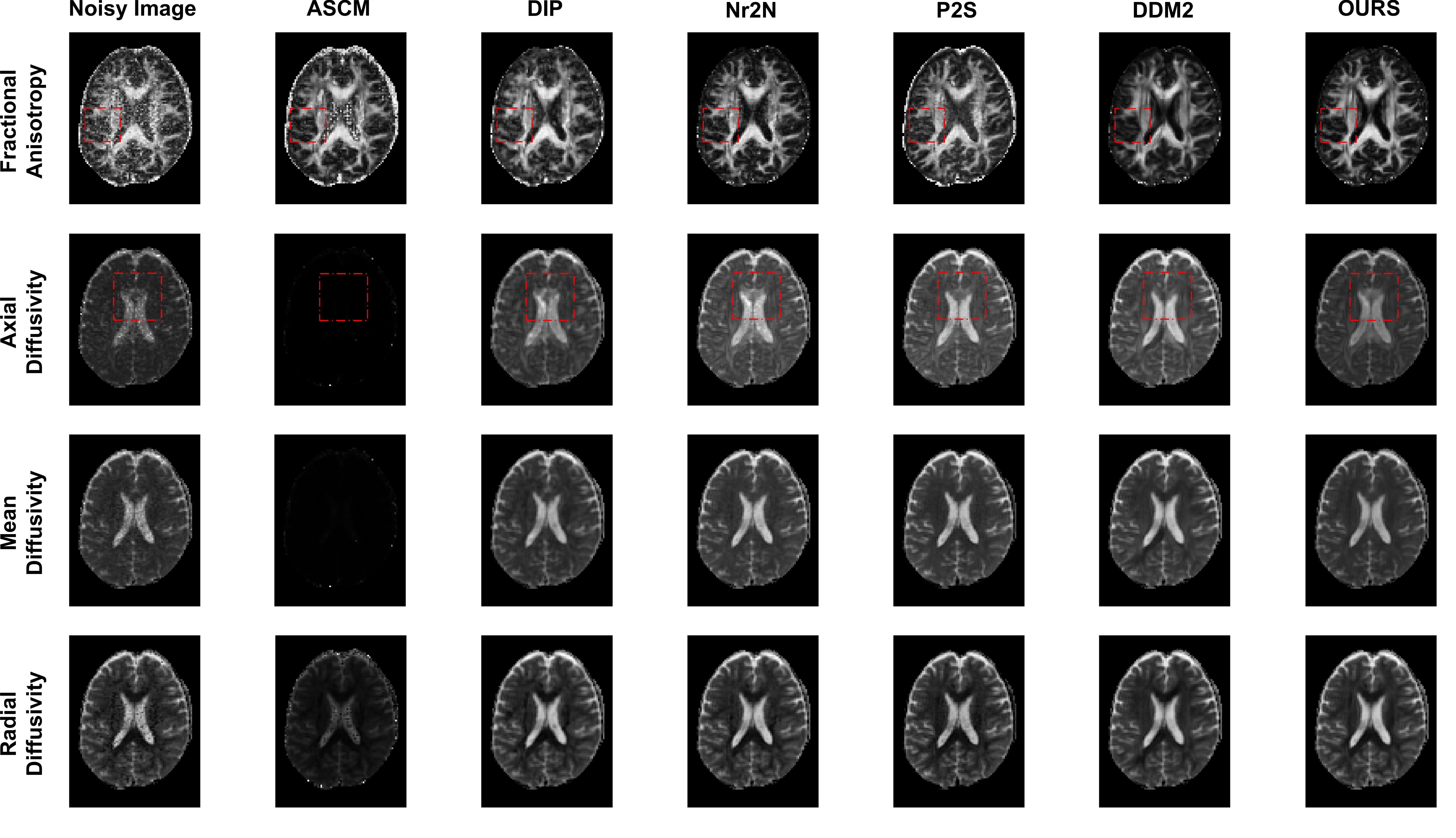}
  \caption{Fractional anisotropy, axial diffusivity, mean diffusivity, and radial diffusivity comparisons. The main differences are highlighted within the red dashed box. Our method effectively suppresses noise and reconstructs fiber tracts while maintaining a grayscale consistency with the original data (No overall brightening of diffusion signal estimates, especially on axial diffusivity)}
  \label{DTI}
\end{figure}

\begin{figure}
  \centering
  \includegraphics[scale = 0.131]{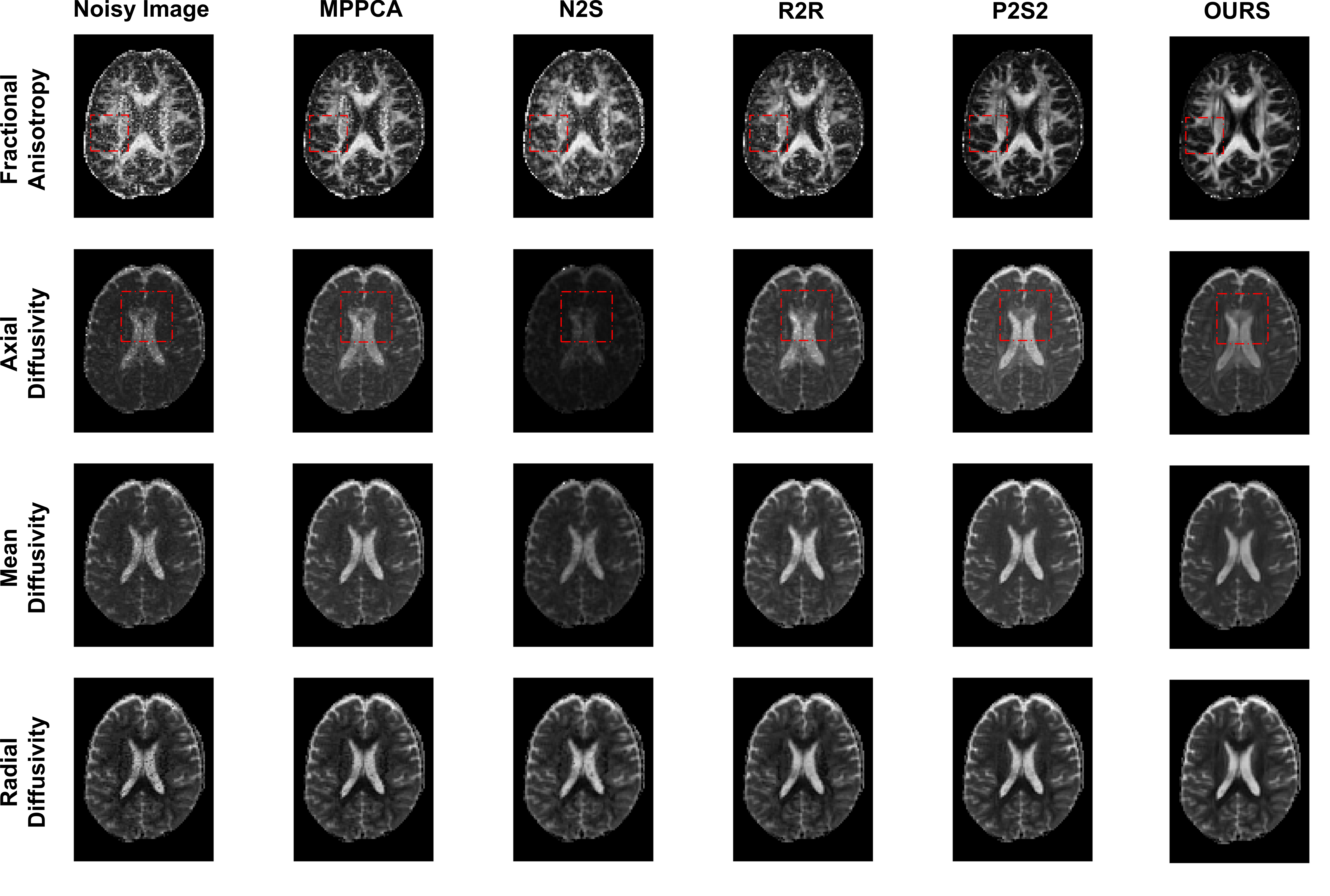}
  {\caption{Fractional anisotropy, axial diffusivity, mean diffusivity, and radial diffusivity comparisons. The main differences are highlighted within the red dashed box. Our method effectively suppresses noise and reconstructs fiber tracts while maintaining a grayscale consistency with the original data (No overall brightening of diffusion signal estimates, especially on axial diffusivity)}}
  \label{DTI sup}
\end{figure}

\begin{figure}
  \centering
  \includegraphics[scale = 0.131]{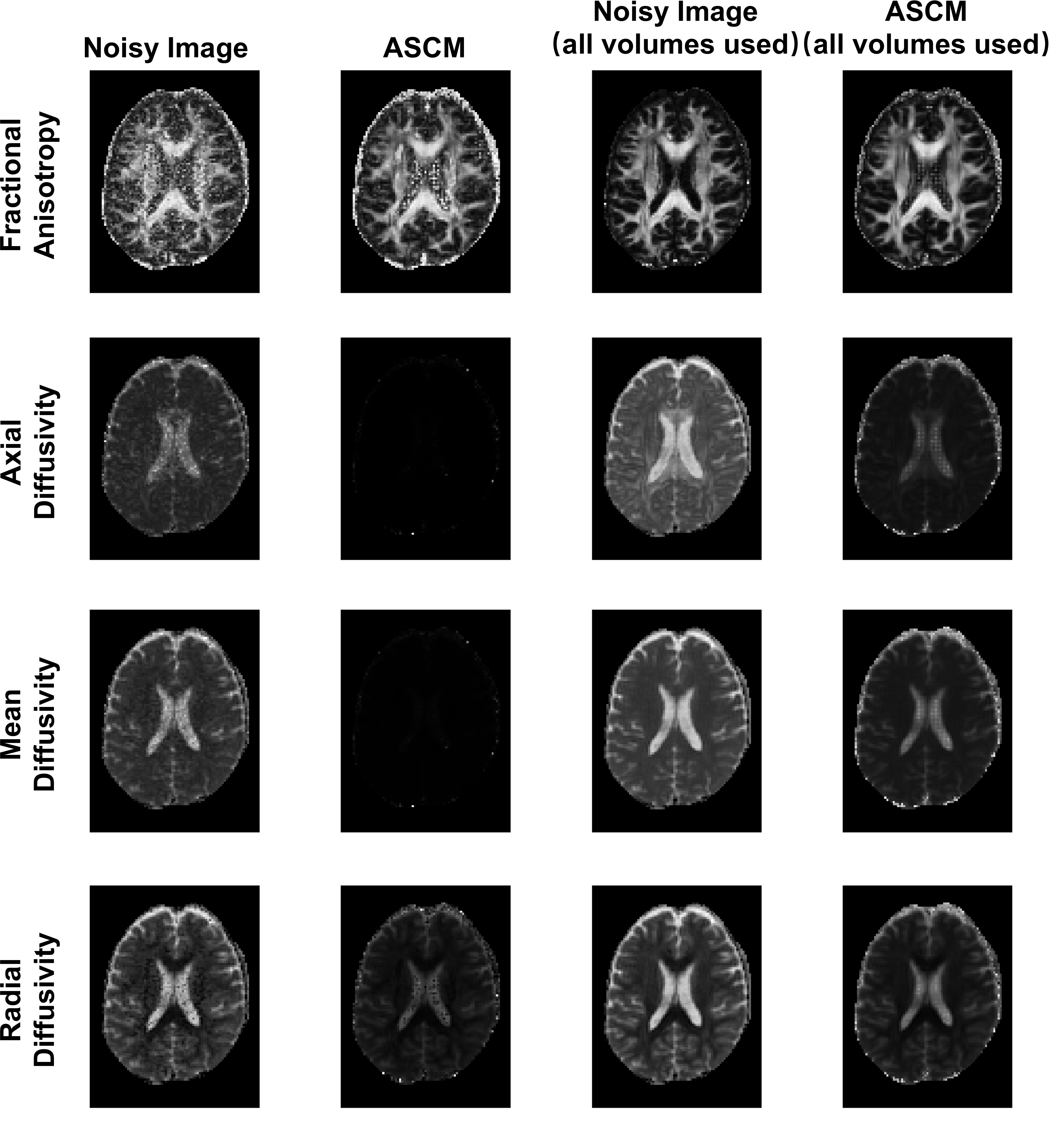}
  \caption{Fractional anisotropy, axial diffusivity, mean diffusivity and radial diffusivity comparisons of previous version results and all volumes used results.}
  \label{ASCMdti}
\end{figure}

We further examine how the denoising quality translates to downstream clinical tasks such as creating DTI~\citep{basser1994mr} diffusion signal estimates using the various denoising methods. Details are in Appendix \ref{Modeling implementation details}. In Fig. \ref{DTI}, we show fractional anisotropy, axial diffusivity, mean diffusivity, and radial diffusivity comparisons. Apart from the poor performance of ASCM, we observe that other methods performed well on diffusion signal estimates. 

On radial diffusivity, all methods exhibited an improved and less noisy representation of the diffusion directions of fiber tracts. However, on fractional anisotropy, and axial diffusivity, DDM2 shows inconsistencies with the noisy image at specific locations (highlighted by the red dashed box), indicating excessive denoising. Our method effectively suppresses noise and reconstructs fiber tracts while maintaining a grayscale consistency with the original data (no overall brightening of diffusion signal estimates, especially on axial diffusivity).

{In Fig. \ref{DTI sup}, we further compare DTI diffusion signal estimates with those obtained using additional denoising methods. We follow the same steps in Appendix \ref{Modeling implementation details} to estimate fractional anisotropy (FA), axial diffusivity (AD), mean diffusivity (MD), and radial diffusivity (RD). Additionally, we computed the reference for FA, AD, MD, and RD using all available dMRI data. Based on this, we performed the calculation of PSNR and SSIM for all slices. The quantitative results are summarized in Table \ref{tab:psnr_ssim}}

Questions on ASCM diffusion signal estimates. In Fig. \ref{DTI}, minimal signals are observed in the ASCM axial and mean diffusivity. We further utilize all volumes to perform diffusion signal estimates and show results in Fig. \ref{ASCMdti}. A noticeable signal should be revealed if all volumes are used in the diffusion signal estimates. This finding demonstrates that the denoising results of ASCM could significantly hinder the DTI diffusion signal estimates.

\begin{table}[ht]
\centering
{
\begin{tabular}{l l l l l l}
\hline
\textbf{Method} & \textbf{Metric} & \textbf{FA} & \textbf{MD} & \textbf{RD} & \textbf{AD} \\ \hline
\multirow{2}{*}{Noisy} & PSNR & 23.47 & 31.79 & 33.56 & 26.39 \\ \cline{2-6} 
                       & SSIM & 0.8760 & 0.9635 & 0.9637 & 0.9247 \\ \hline
\multirow{2}{*}{ASCM} & PSNR & 22.63 & 21.81 & 24.70 & 20.89 \\ \cline{2-6} 
                      & SSIM & 0.8897 & 0.8188 & 0.8921 & 0.8102 \\ \hline
\multirow{2}{*}{DIP}  & PSNR & 24.76 & 33.01 & 32.42 & 30.21 \\ \cline{2-6} 
                      & SSIM & 0.8894 & 0.9675 & 0.9630 & 0.9453 \\ \hline
\multirow{2}{*}{MPPCA} & PSNR & 26.37 & 35.28 & 36.35 & 29.63 \\ \cline{2-6} 
                       & SSIM & 0.9048 & 0.9751 & 0.9780 & 0.9534 \\ \hline
\multirow{2}{*}{Nr2N} & PSNR & 29.47 & 38.42 & 38.48 & 34.87 \\ \cline{2-6} 
                      & SSIM & 0.9354 & 0.9851 & 0.9865 & 0.9712 \\ \hline
\multirow{2}{*}{N2S} & PSNR  & 23.13 & 29.71 & 31.08 & 25.84 \\ \cline{2-6} 
                     & SSIM  & 0.8691 & 0.9469 & 0.9491 & 0.8973 \\ \hline
\multirow{2}{*}{R2R} & PSNR  & 25.93 & 34.54 & 33.82 & 30.21 \\ \cline{2-6} 
                     & SSIM  & 0.8809 & 0.9708 & 0.9678 & 0.9394 \\ \hline
\multirow{2}{*}{P2S}   & PSNR & 24.18 & 32.16 & 30.62 & 33.27 \\ \cline{2-6} 
                       & SSIM & 0.9090 & 0.9708 & 0.9632 & 0.9677 \\ \hline
\multirow{2}{*}{DDM2}  & PSNR & 26.77 & 37.53 & 37.39 & 33.21 \\ \cline{2-6} 
                       & SSIM & 0.9041 & 0.9872 & 0.9848 & 0.9610 \\ \hline
\multirow{2}{*}{P2S2}  & PSNR & 30.08 & 39.28 & 40.23 & 34.06 \\ \cline{2-6} 
                       & SSIM & 0.9432 & 0.9894 & 0.9921 & 0.9701 \\ \hline
\multirow{2}{*}{OURS}  & PSNR & \textbf{30.79} & \textbf{40.26} & \textbf{40.35} & \textbf{35.30} \\ \cline{2-6} 
                       & SSIM & \textbf{0.9450} & \textbf{0.9931} & \textbf{0.9923} & \textbf{0.9763} \\ \hline
\end{tabular}
\caption{PSNR and SSIM comparisons for DTI diffusion signal estimates. Here ``AD'' represents axial diffusivity, ``RD'' represents radial diffusivity, ``MD'' represents mean diffusivity and ``FA'' represents fractional anisotropy.}
}
\label{tab:psnr_ssim}
\end{table}

\subsection{Quantitative results on \textit{in-vivo} data}
\label{Quantitative results on in-vivo data}

\begin{figure}
  \centering
  \includegraphics[scale = 0.415]{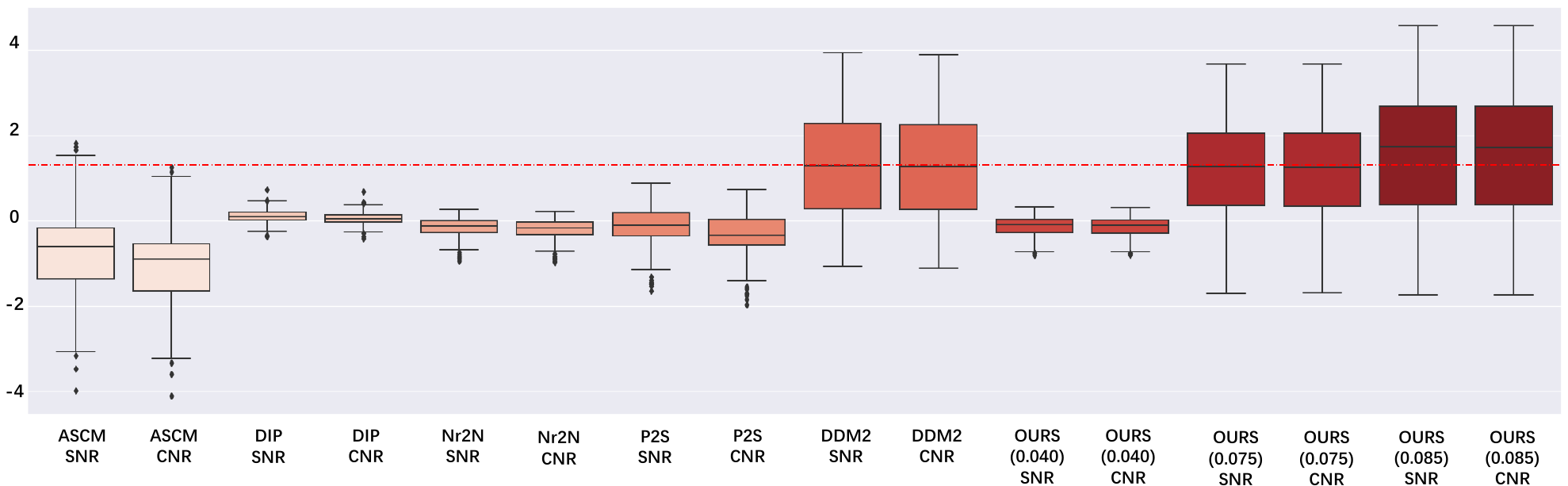}
  \caption{Box plots of Quantitative SNR/CNR metrics scores. The numbers within parentheses under OURS represent the value of ${\mathcal{CSNR}}$ (Section \ref{Towards iterative and adjustable refinement}). Di-Fusion indicates better performance in terms of SNR/CNR metrics.}
  \label{quantitative results}

\end{figure}

\begin{table}
{\caption{Comparison of SNR and CNR. \textbf{Bold} and \underline{Underline} fonts denote the best and the second-best performance, respectively.}\label{snr_cnr_table}}
\begin{center}
\renewcommand{\arraystretch}{1.0}
\setlength\tabcolsep{2.6pt}
{
\begin{tabular}{lcccccccccc}
\toprule
Method & ASCM & MPPCA & DIP & R2R & N2S & Nr2N & P2S & P2S2 & DDM2 & OURS \\
\midrule
SNR & -0.7251 & 0.2372 & 0.1035 & -0.0099 & 0.2266 & -0.1598 & -0.1616 & 0.1526 & \underline{1.3040} & \textbf{1.5735} \\
CNR & -1.0513 & 0.2191 & 0.0567 & -0.1161 & -0.0304 & -0.2004 & -0.3694 & 0.1177 & \underline{1.2725} & \textbf{1.5687} \\
\bottomrule
\end{tabular}
}
\end{center}
\end{table}

Given the absence of a consensus on image quality metrics, particularly in unsupervised reference-free settings~\citep{chaudhari2020utility,woodard2006no}, the task of assessing perceptual MRI quality becomes a challenging research problem~\citep{mittal2011blind}. Considering the infeasibility of using PSNR and SSIM metrics (no ground truth reference images) and their limited correlation with clinical utility~\citep{mason2019comparison}, computing metrics in downstream clinical regions of interest is more reasonable~\citep{adamson2021ssfd}. We follow the procedure outlined in \citep{xiang2023ddm} to calculate SNR/CNR metrics (Details are in \ref{SNR and CNR implementation details}). The quantitative denoising results were reported as mean and standard deviation scores for the complete 4D volumes in Fig. \ref{quantitative results}. Di-Fusion indicates better performance against competing methods. Our experiments in Section \ref{Impacts on modeling} have shown no correlation between high or low scores on CNR/SNR metrics and the performance of the downstream clinical tasks. 

{We summarize the CNR and SNR metrics of all comparison methods in Table \ref{snr_cnr_table}, where our method achieves better results in both CNR and SNR metrics.}

\begin{figure}
  \centering
  \includegraphics[scale = 0.71]{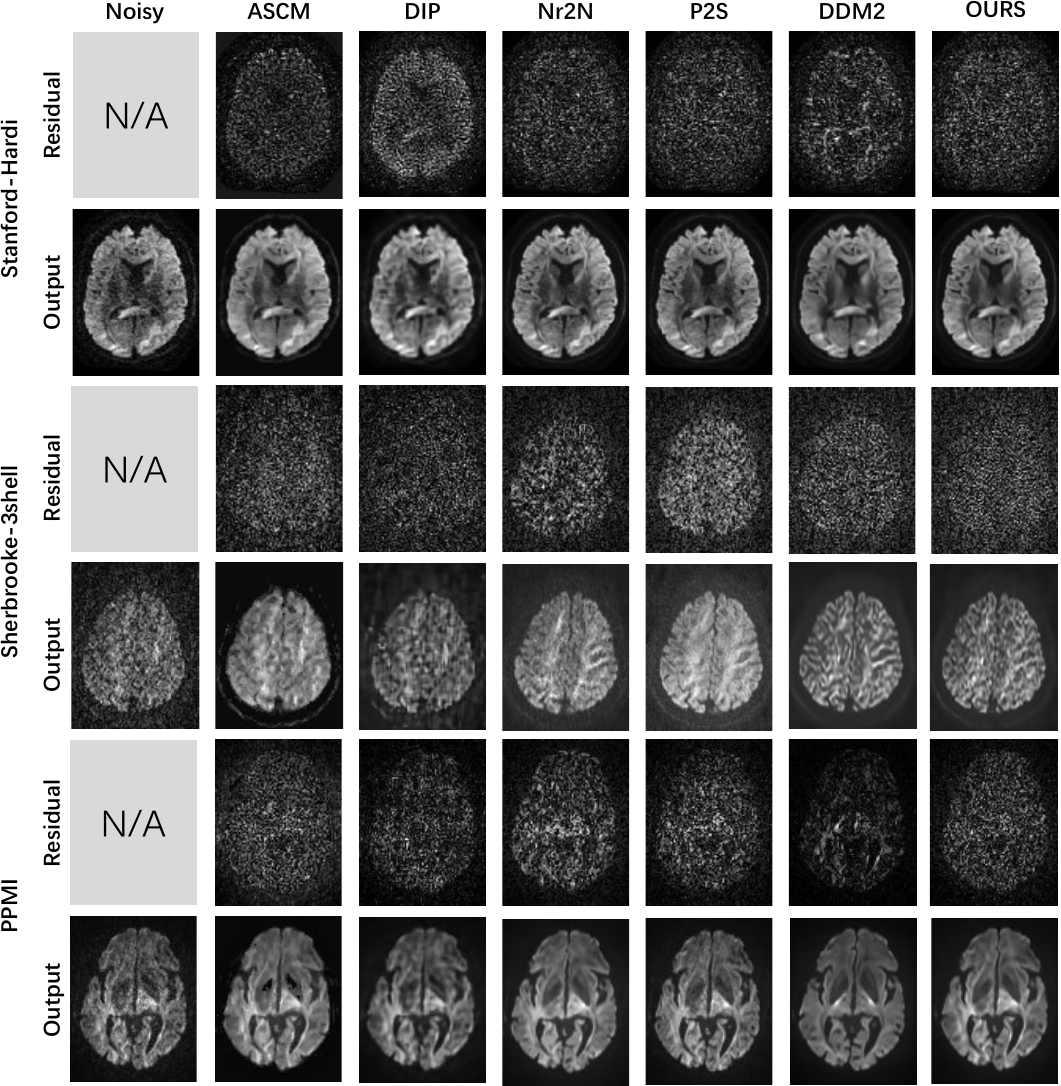}
  \caption{More qualitative results on Stanford-Hardi. ``OURS'' results are obtained when ${\mathcal{CSNR}}=0.040$  (Section \ref{Towards iterative and adjustable refinement}). Notice that Di-Fusion suppresses noise and does not show any anatomical structure in the residual plots.}
  \label{morequalitative_stanfordhardi}
\end{figure}

\begin{figure}
  \centering
  \includegraphics[scale = 0.47]{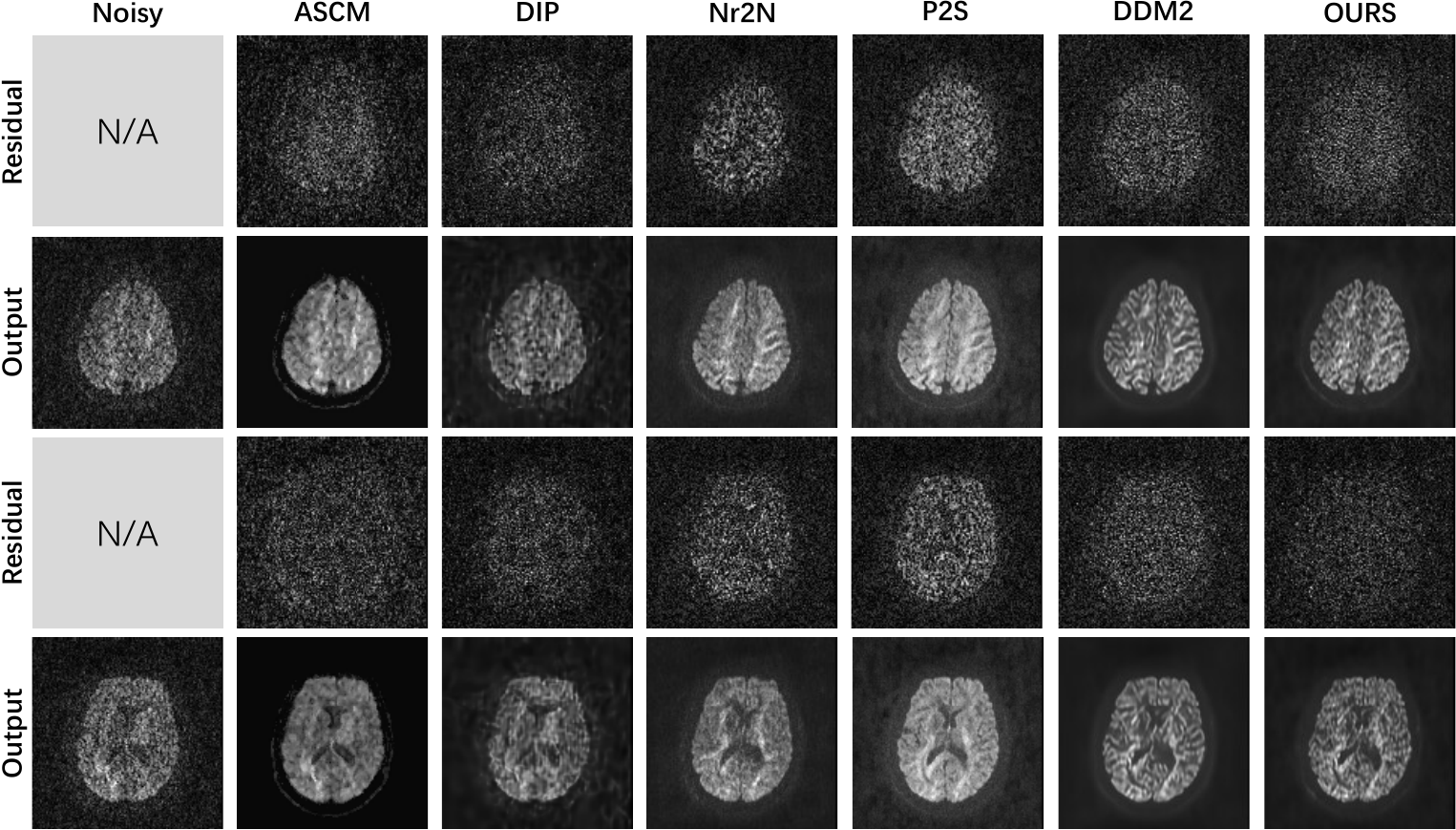}
  \caption{More qualitative results on Sherbrooke 3-Shell. ``OURS'' results are obtained when ${\mathcal{CSNR}}=0.040$ (Section \ref{Towards iterative and adjustable refinement}). Notice that Di-Fusion suppresses noise and does not show any anatomical structure in the residual plots.}
  \label{morequalitative_s3sh}
\end{figure}

\begin{figure}
  \centering
  \includegraphics[scale = 0.48]{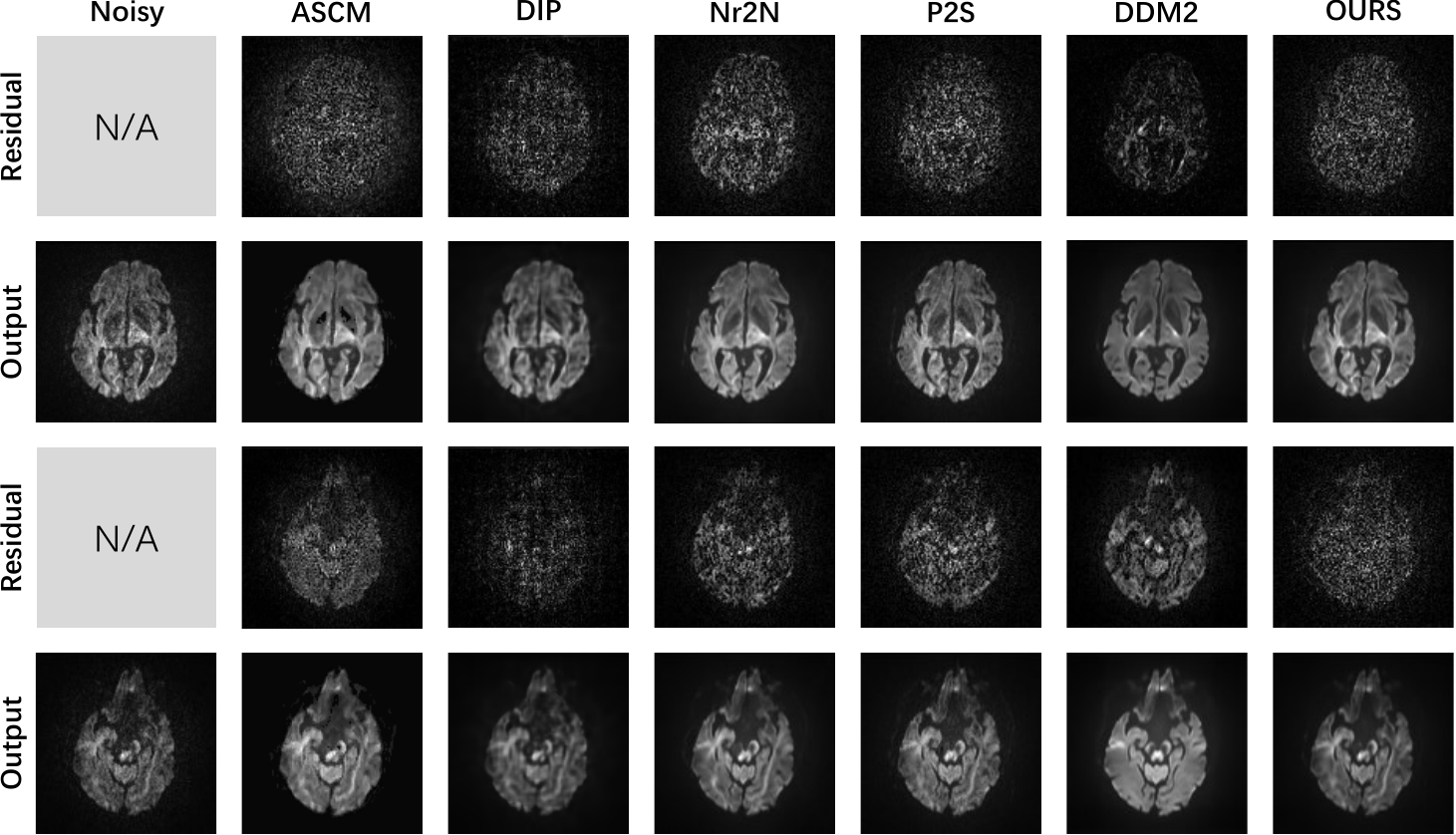}
  \caption{More qualitative results on PPMI. ``OURS'' results are obtained when ${\mathcal{CSNR}}=0.040$ (Section \ref{Towards iterative and adjustable refinement}). Notice that Di-Fusion suppresses noise and does not show any anatomical structure in the residual plots.}
  \label{morequalitative_ppmi}
\end{figure}

\subsection{More qualitative results}
\label{A-More qualitative results}

In Fig. \ref{morequalitative_stanfordhardi}, Fig. \ref{morequalitative_s3sh} and Fig. \ref{morequalitative_ppmi}, we show more qualitative results. For each of the datasets, we show the axial slice of a randomly chosen 3D volume and the corresponding residuals (squared differences between the noisy data and the denoised output). We can observe that the results are generally consistent with those presented in Section \ref{Qualitative results}. From the residuals of ``DDM2'', it can be observed that particular regions are suppressed (especially in Fig. \ref{morequalitative_stanfordhardi}, the DDM2 results for the second slice show that the residuals contain a significant amount of anatomical information). Notice that Di-Fusion suppresses noise and does not show any anatomical structure in the residual plots.

\begin{figure}
  \centering
  \includegraphics[scale = 0.288]{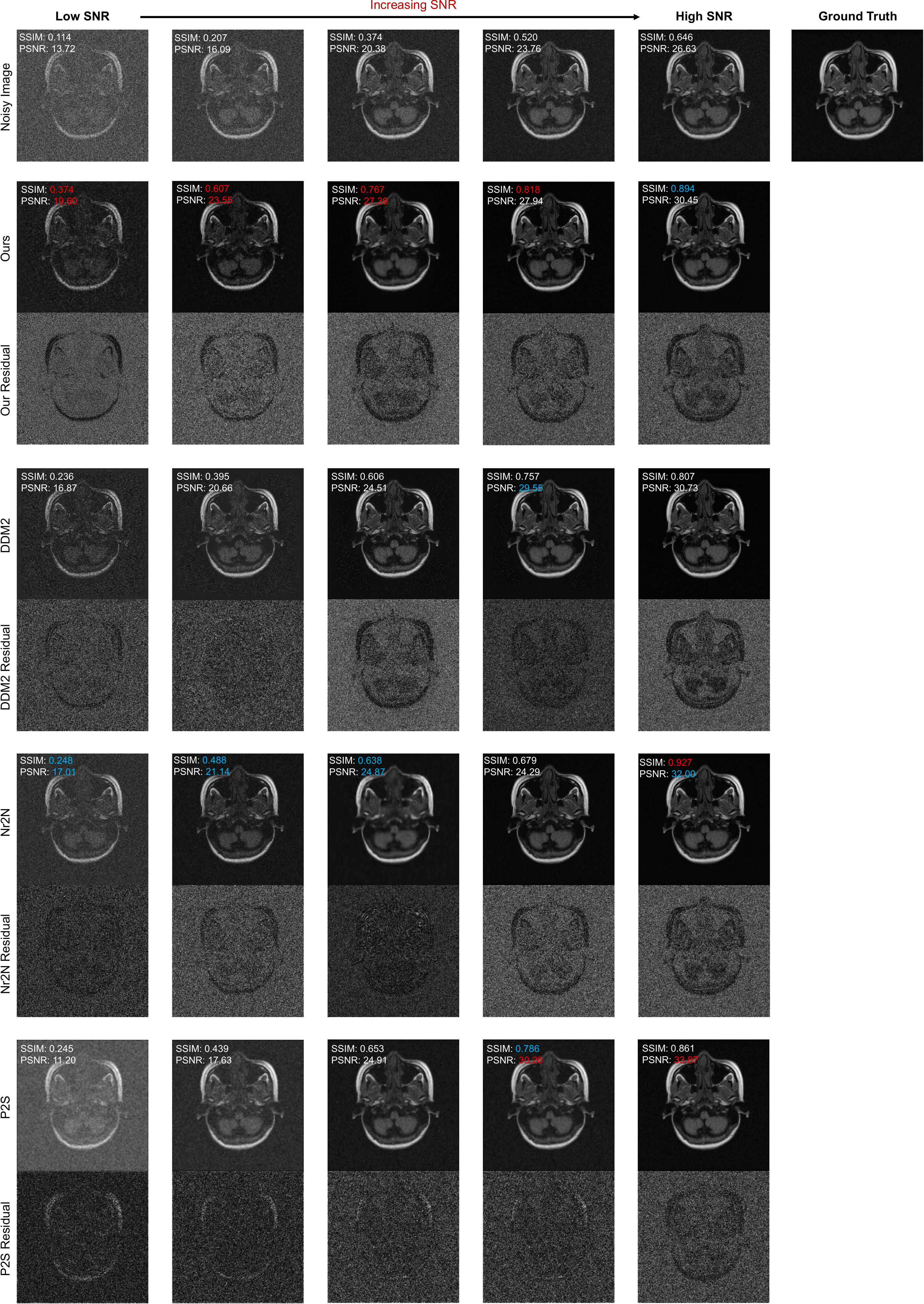}
  \caption{Quantitative and qualitative results on simulated data. In our experiments, $\mathcal{CSNR}=0.040$. The red color represents the highest value for the metric, while the blue color represents the second-highest value. Please note that these are the results of a single round of simulated experiments and their corresponding PSNR and SSIM metrics scores.}
  \label{simulated_experiment}
\end{figure}

\subsection{Quantitative and qualitative results on simulated data}
\label{A-Quantitative and qualitative results on simulated data}

We show the quantitative and qualitative results in Fig. \ref{simulated_experiment}. When the noise intensity is high (left three columns), our method performs the best. When the noise intensity is low (right two columns), denoising results are comparable to other methods. Considering the high PSNR and SSIM in the right two columns, it suggests that in real-world scenarios, such data may not require denoising and can still enable effective clinical use. Di-Fusion has more potential for generalization and applicability as it performs better under high noise intensity.

\subsection{Compare under mixed b-value images}
\label{A-Compare under mixed b-value images}

\begin{figure}
  \centering
  \includegraphics[scale = 0.455]{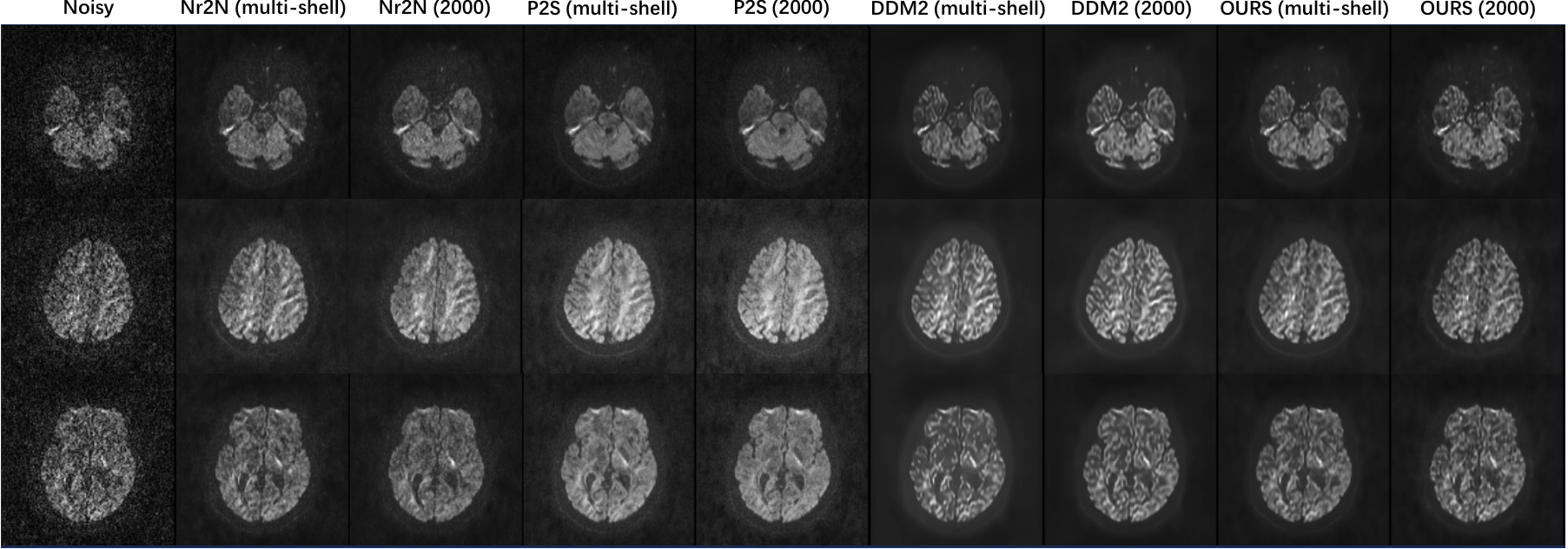}
  \caption{Additional results when training on mixed b-value images (All our results are obtained when $\mathcal{CSNR}=0.040$). ``(2000)'' indicates using data with only a b-value of 2000. ``(multi-shell)'' represents using data with mixed b-values, including 1000, 2000, and 3500. The performance of Di-Fusion is relatively robust and not reliant on specific b-value configurations}
  \label{multi_b}
\end{figure}

In Fig. \ref{multi_b}, we show additional qualitative results when training on mixed b-value images (Sherbrooke 3-Shell has 1000, 2000, and 3500 b-values). Nr2N, P2S, DDM2, and our method both show minimal sensitivity to mixed b-values training data. Minor brightness variations in DDM2 and P2S for multiple b-values have a negligible impact on the overall results. Our method primarily learns the mapping from one volume to another, making it less affected by varying b-values in different volumes (P2S uses all the other different volumes, DDM2 uses two different input volumes at Stage 1, and Di-Fusion only uses one different volume). This suggests that the performance of Di-Fusion is relatively robust and not reliant on specific b-value configurations.

\subsection{Qualitative results when using fewer dMRI volumes}
\label{A-Qualitative results when using fewer dMRI volumes}

\begin{figure}
  \centering
  \includegraphics[scale = 0.6]{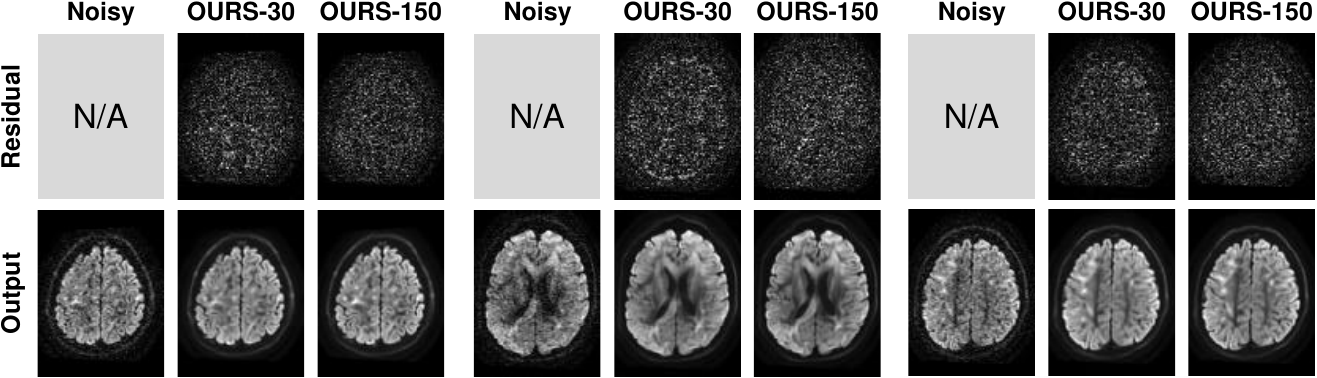}
  {\caption{Qualitative results when using fewer dMRI volumes. OURS-30 indicates using 30 dMRI volumes, while OURS-150 represents using 150 dMRI volumes.}}
  \label{30&150}
\end{figure}

{As shown in Fig. \ref{30&150}, when using fewer dMRI volumes (20\% of original dMRI volumes), Di-Fusion still demonstrates effective denoising capabilities.} Please pay special attention that the 30 dMRI volumes here refer to the total training data. Additionally, using a portion of dMRI data from different individuals for model training is a more clinically feasible approach.

\section{Visualization of Di-Fusion}

\subsection{Fusion process: linear interpolation between the two endpoints}
\label{A-Fusion process: linear interpolation}

The noise schedule can be found in Appendix \ref{Noise schedule}. In Fig. \ref{linear interpolation}, we provide a visual demonstration of ${x_t^*}$ (Eq. (\ref{eq:fusion})). Without the Fusion process, the model output would deviate from ${x_{out}}$, resulting in drifted results. By incorporating the Fusion process, where each linear interpolation ${x_t^*}$ from ${x'}$ to ${x}$ has ${x}$ as the target, the inference process avoids drifted results (Fig. \ref{key feature} (a)). We further conducted ablation studies to demonstrate the significance of the Fusion process in Appendix \ref{On Fusion process}.

\begin{figure}
  \centering
  \includegraphics[scale = 0.5]{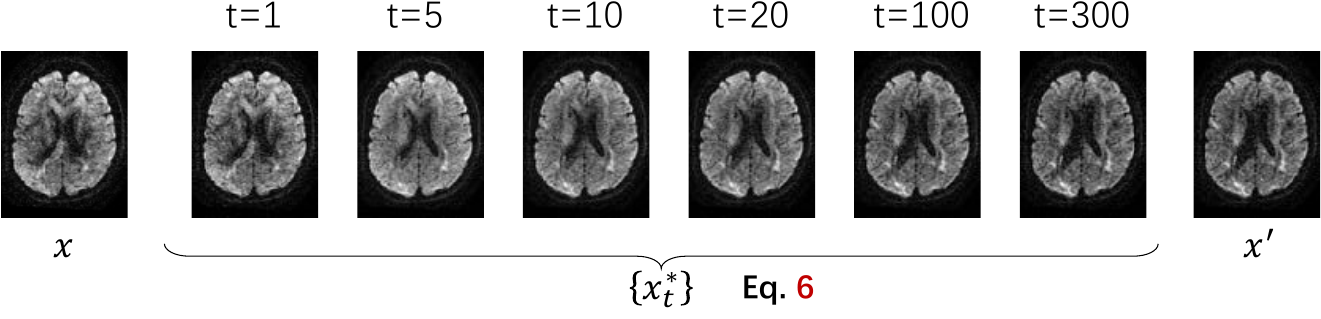}
  \caption{Visual demonstration of ${x_t^*}$ obtained by different $t$.}
  \label{linear interpolation}
\end{figure}

\subsection{``Di-'' process: different noise distribution}
\label{A-Di- process: Different noise distribution}

Experiment details: We computed all the ${\xi _{x - x'}}$ in Stanford HARDI dataset (meaning a total of $76*150=11400$ noisy images), calculated the grayscale histogram, mean and variance of these noisy images, and presented the calculated mean and variance in the form of a histogram. At the same time, we randomly sampled $11400$ Gaussian noisy images and performed the same statistical operation.

\begin{figure}
  \centering
  \includegraphics[scale = 0.42]{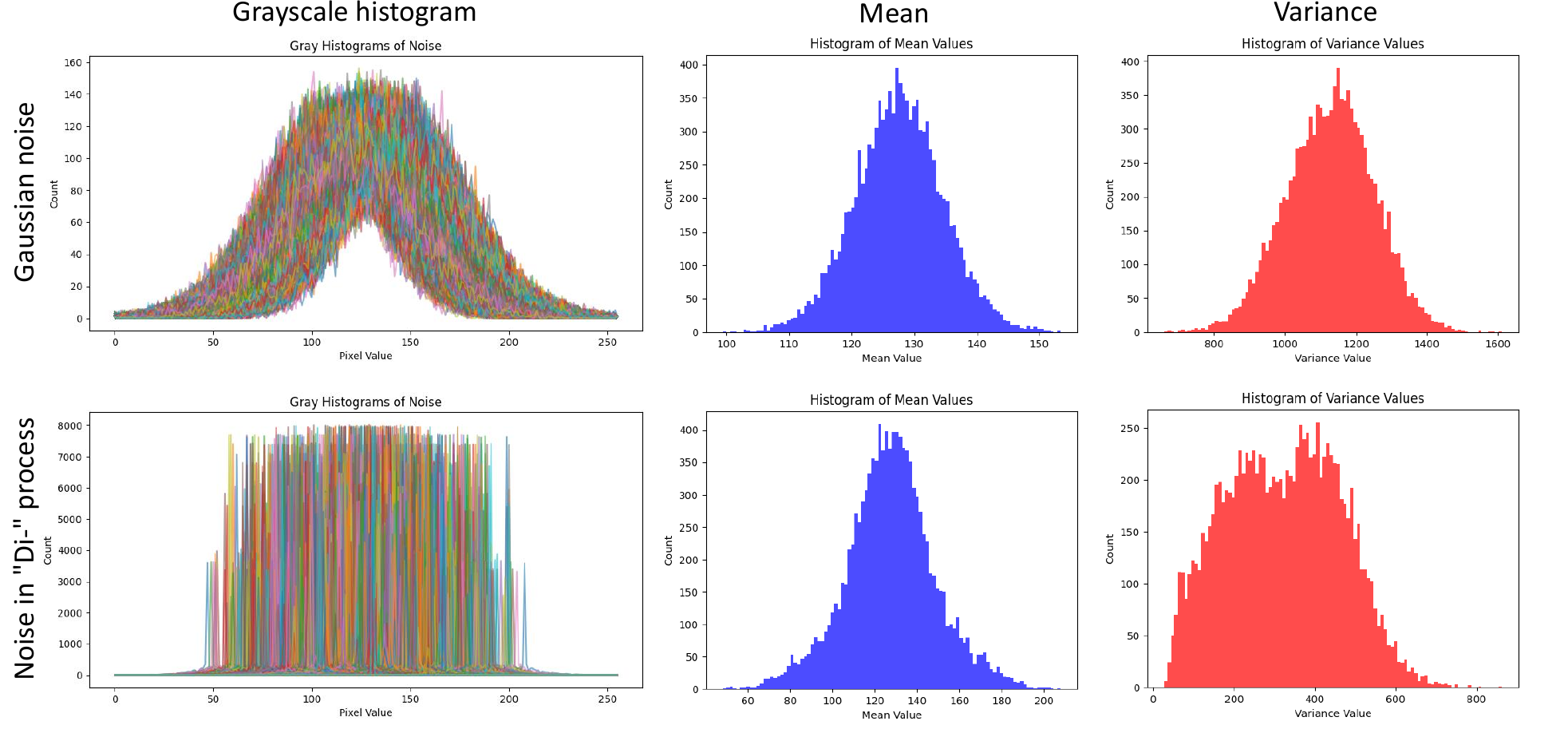}
  \vspace{-2mm}
  \caption{Grayscale histogram, mean and variance of these noisy images. We computed all the ${\xi _{x - x'}}$ in Stanford HARDI dataset (meaning a total of $76*150=11400$ noisy images), calculated the grayscale histogram, mean and variance of these noisy images, and presented the calculated mean and variance in the form of a histogram. At the same time, we randomly sampled $11400$ Gaussian noisy images and performed the same statistical operation. The noise calculated by the ``Di-'' process has significantly different statistical properties from Gaussian noise. This is reflected in the fact that: \textbf{1.} the variance of the calculated noise is relatively small and does not follow a normal distribution \textbf{2.} the counts of each pixel value on the grayscale histogram of ${\xi _{x - x'}}$ are similar, rather than a normal distribution in Gaussian noise.}
  \label{noise}
  \vspace{-2mm}
\end{figure}

Statistical properties of ${\xi_{x - x'}}$: From Fig. \ref{noise}, the noise calculated by the ``Di-'' process has significantly different statistical properties from Gaussian noise. This is reflected in the fact that: \textbf{1.} the variance of the calculated noise is relatively small and does not follow a normal distribution \textbf{2.} the counts of each pixel value on the grayscale histogram of ${\xi _{x - x'}}$ are similar, rather than a normal distribution in Gaussian noise. Different noise distribution makes ${\mathcal{F}_\theta }$ more capable of modeling real-world noise.

\subsection{Sampling process: iterative and stable refinement}
\label{Sampling process: iterative and stable refinement}

\begin{figure}
  \centering
  \includegraphics[scale = 0.52]{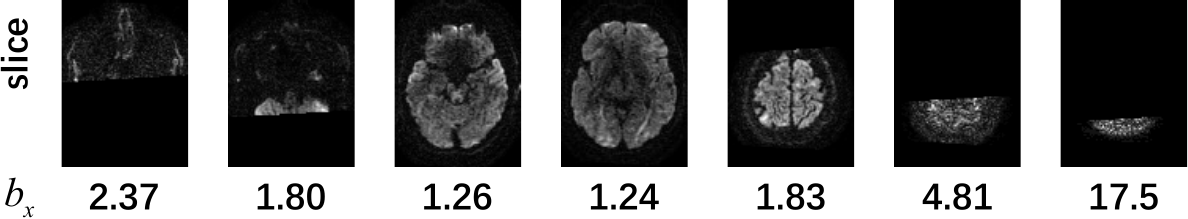}
  \caption{Slices and their corresponding $b_x$. The \( b_x \) values of the edge slices are relatively larger.}
  \label{brainvalue}
\end{figure}

\paragraph{Value of ${b_x}$}
\label{Value of b_x}
In Section \ref{Towards iterative and adjustable refinement}, we adopt a simple definition (Eq. (\ref{eq:brainvalue})) to calculate a coefficient $b_x$ that accounts for the ratio of brain tissue to the entire image. Fig. \ref{brainvalue} displays the slices accompanied by their corresponding $b_x$. It can be observed that Eq. (\ref{eq:brainvalue}) is a simple method for evaluating the proportion of the brain tissue and $b_x$ can be used to correct $d_x$ in Eq. (\ref{eq:reverseloss}).

\begin{figure}
  \centering
  \includegraphics[scale = 0.43]{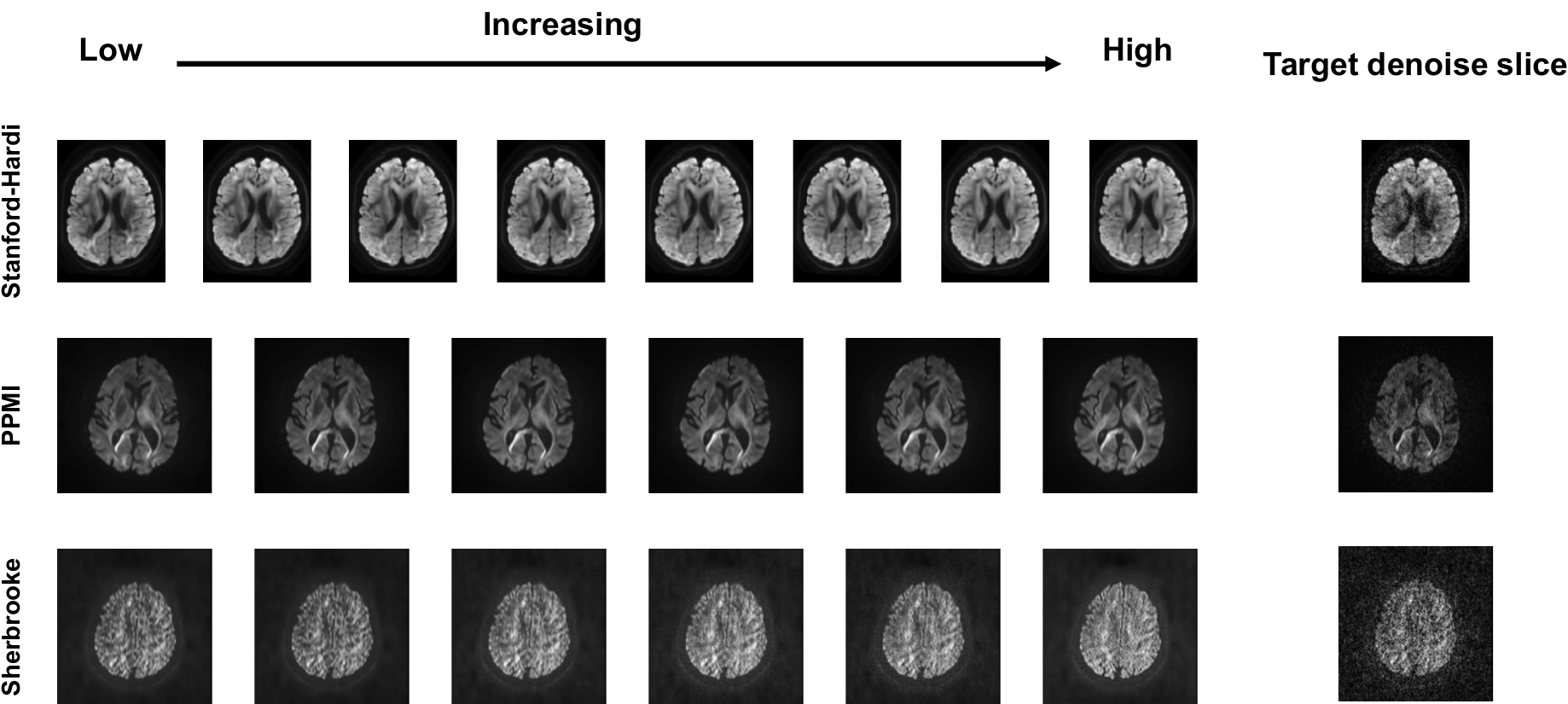}
  \caption{The results of sampling process obtained by different $\mathcal{CSNR}$.}
  \label{sampling}
\end{figure}

\paragraph{Iterative and controllable refinement}
\label{Iterative and stable refinement}
In Section \ref{Towards iterative and adjustable refinement}, we propose an adaptive termination during the sampling process. This allows us to control the sampling process by setting the value of $\mathcal{CSNR}$. In general, setting a lower $\mathcal{CSNR}$ will preserve more anatomical details. On the other hand, setting a higher $\mathcal{CSNR}$ will remove more noise at the cost of losing some anatomical details (see Fig. \ref{sampling} for visual demonstrations). 

\paragraph{${d_x}$ plots}
\label{d_x plots}

\begin{figure}
  \centering
  \includegraphics[scale = 0.61]{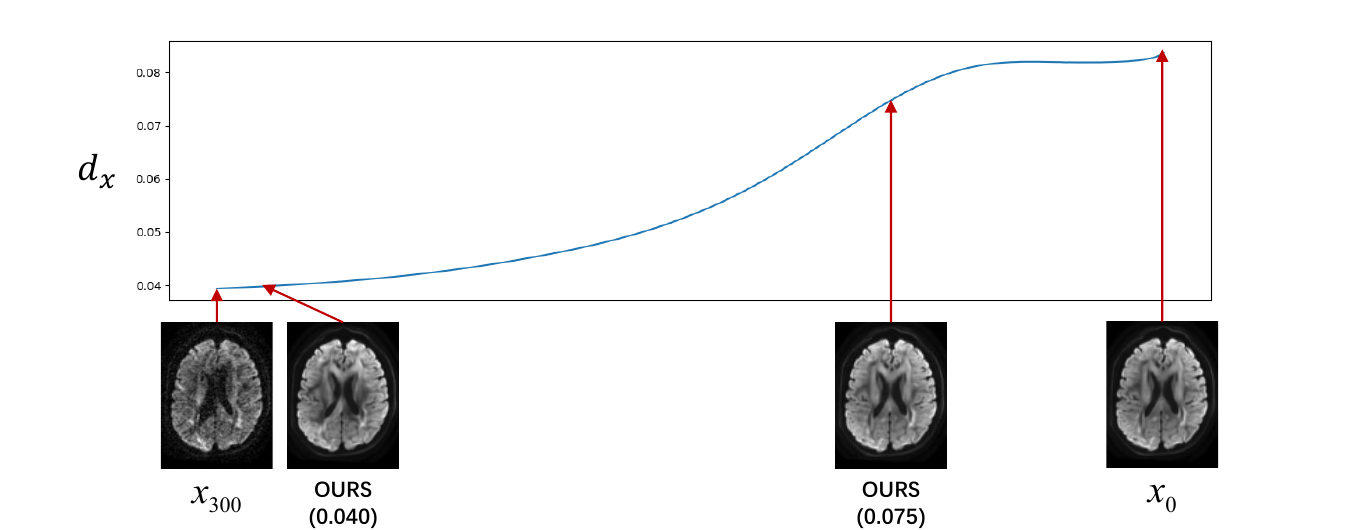}
  \caption{Variation of $d_x$ during the sampling process. The numbers within parentheses below OUR represent the value of ${\mathcal{CSNR}}$ (Section \ref{Towards iterative and adjustable refinement}).}
  \label{dx}
\end{figure}

In Section \ref{Towards iterative and adjustable refinement}, we calculate $d_x$ (Eq. (\ref{eq:reverseloss})) to represent the degree of denoising in ${x_{out}}$. In Fig. \ref{dx}, we illustrate the variation of $d_x$ during the reverse sampling process and present the results when implementing an adaptive termination. It can be observed that with such an adaptive termination, it is possible to quickly obtain denoised results (low $\cal CSNR$ results) or further remove noise effectively (high $\cal CSNR$ results).

\section{Ablation studies}
\label{A-Ablation studies}

\subsection{On training in Di-Fusion}
\label{On training in Di-Fusion}

\begin{figure}
  \centering
  \includegraphics[scale = 0.63]{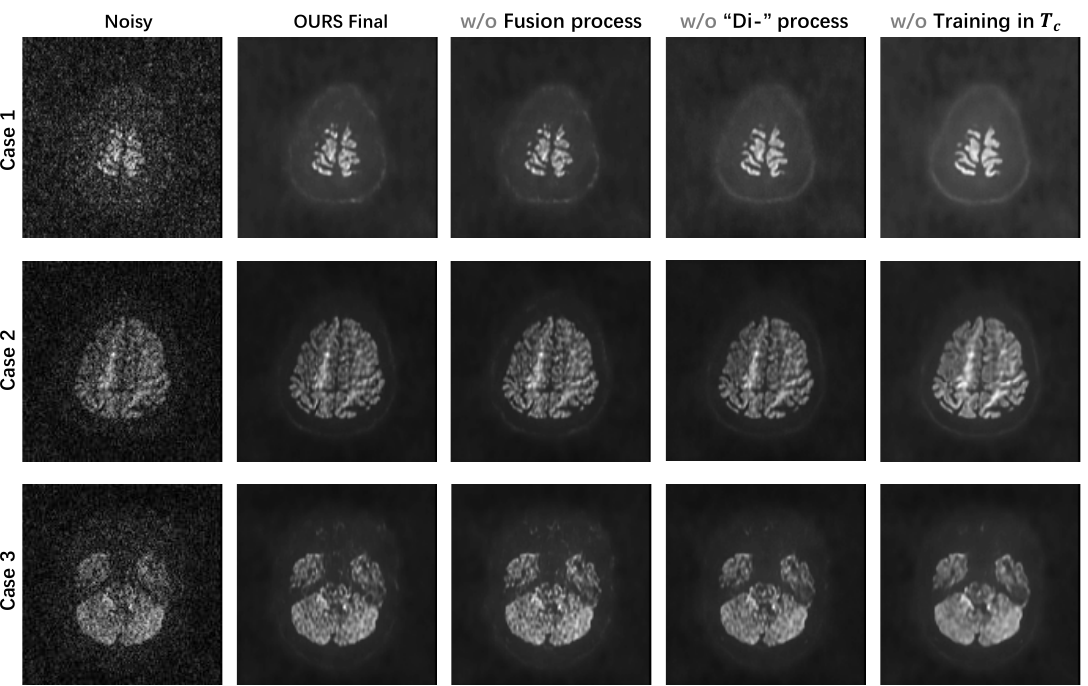}
  \caption{Qualitative results of ablation studies (Implement an adaptive termination mentioned in Section \ref{Towards iterative and adjustable refinement} during the sampling process and all the experiments $\mathcal{CSNR}=0.040$). Headings distinguish results obtained using different ablation settings.}
  \label{ablation_nosample}
\end{figure}

\begin{figure}
  \centering
  \includegraphics[scale = 0.634]{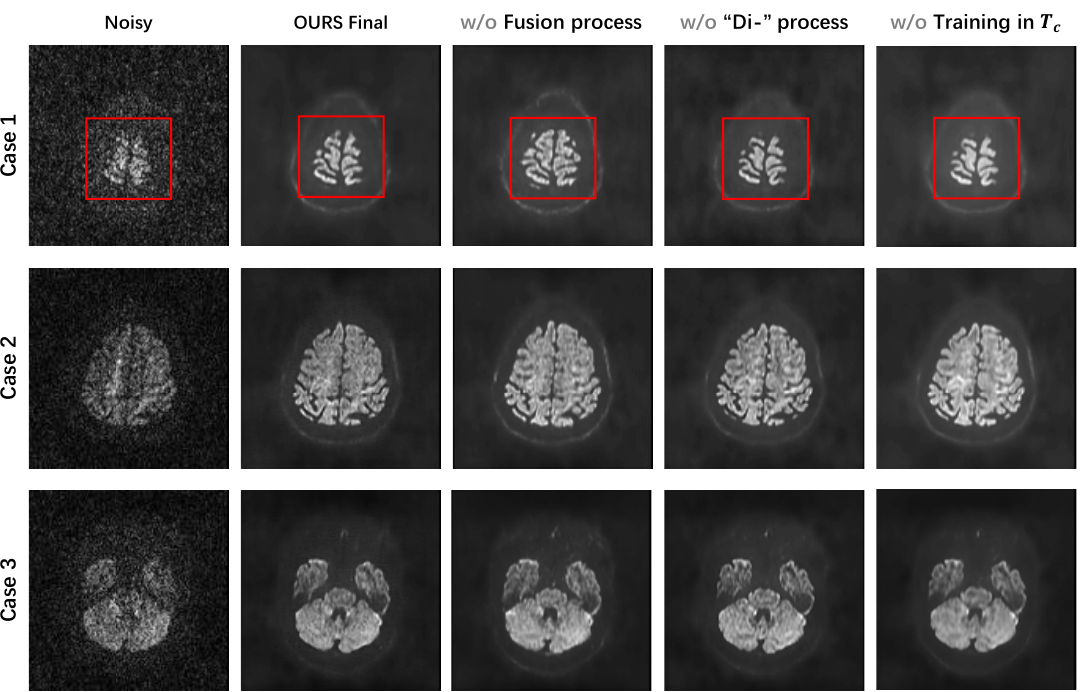}
  \caption{Qualitative results of ablation studies (Didn't implement an adaptive termination mentioned in Section \ref{Towards iterative and adjustable refinement} during the sampling process). Headings distinguish results obtained using different ablation settings. The red box highlights the main differences.}
  \label{ablation}
\end{figure}

\paragraph{On Fusion process}
\label{On Fusion process}
In Section \ref{Fusion process}, we utilize Eq. (\ref{eq:fusion}) to compute linear interpolation from $x'$ to $x$, aiming to reduce drift in final results. We disable the Fusion process by substituting $x'$ for ${x_t^ *}$. In Fig. \ref{ablation_nosample}, when going through several reverse steps (low $\cal SCNR$), the results without the Fusion process do not exhibit significant deviations. However, when the adaptive termination is not implemented (which means completing all the sampling steps), noticeable slice misalignment occurs in Fig. \ref{ablation} (highlighted by the red box).

\paragraph{On ``Di-'' process}
\label{On Di- process}
In Section \ref{"Di-" process}, we utilize Eq. (\ref{eq:Di-}) to compute a noise distribution ${\xi_{x - x'}}$ and use it in $q\left( {{x_{t}}{\rm{|}}x_t^*} \right)$ and ${p_\mathcal{F}}\left( {{x_{t - 1}}|{x_t}} \right)$. We directly replace ${\xi_{x - x'}}$ calculated in the ``Di-'' process with Gaussian noise. Without the ``Di-'' process, results lack some high-frequency information, and the overall gray value of the denoised images has also changed (Case 1 in Fig. \ref{ablation_nosample}). Some may consider using ${\xi_{x - x'}}$ only during the diffusion process $q\left( {{x_{t}}{\rm{|}}x_t^*} \right)$ and Gaussian noise during the sampling process ${p_\mathcal{F}}\left( {{x_{t - 1}}|{x_t}} \right)$. We present the results of this setting in Fig. \ref{Diprocess}, where it can be observed that artifacts occur along the edge slices.

\paragraph{On training the latter diffusion steps}
\label{On conditional training}
In Section \ref{Intuition of conditional training}, we preform training the latter diffusion steps by optimizing ${\mathcal{F}_\theta }$ to condition on ${{\overline \alpha  }_{t}}$, $t \sim {\rm{Uniform}}\left( {\left\{ {1, \cdots ,T_c} \right\}} \right)$, $T_c=300$. We disable training the latter diffusion steps by optimizing ${\mathcal{F}_\theta }$ to condition on ${{\overline \alpha  }_{t}}$, $t \sim {\rm{Uniform}}\left( {\left\{ {1, \cdots ,1000} \right\}} \right)$ and balance the training iterations (training the latter diffusion steps iterations: $1e^5$, training all diffusion steps: $3.5e^5$). Without training the latter diffusion steps, the denoising results are noticeably smoother and have more hallucinations (Fig. \ref{ablation_nosample} and Fig. \ref{ablation}). We recommend training the latter diffusion steps based on its potential advantages, which include \textit{(i)} mitigating hallucinations and \textit{(ii)} reducing training time with improved stability.

\subsection{On sampling in Di-Fusion}
\label{On sampling in Di-Fusion}

\paragraph{On adaptive termination}
\label{On adaptive termination}
In Section \ref{Reverse process}, we introduce an adaptive termination to enable iterative and adjustable refinement. In Fig. \ref{brainvalue}, we show slices and their corresponding $b_x$, the \( b_x \) values of the edge slices are relatively larger. In Fig. \ref{sampling}, we illustrate the impact of $\mathcal{CSNR}$ on the sampling results. In Fig. \ref{dx}, we show variation of $d_x$ during the sampling process.

\paragraph{\textit{Run-Walk} accelerated sampling maintains the sampling quality}
\label{Run-Walk accelerated sampling maintains the sampling quality}

\begin{figure}
  \centering
  \includegraphics[scale = 0.62]{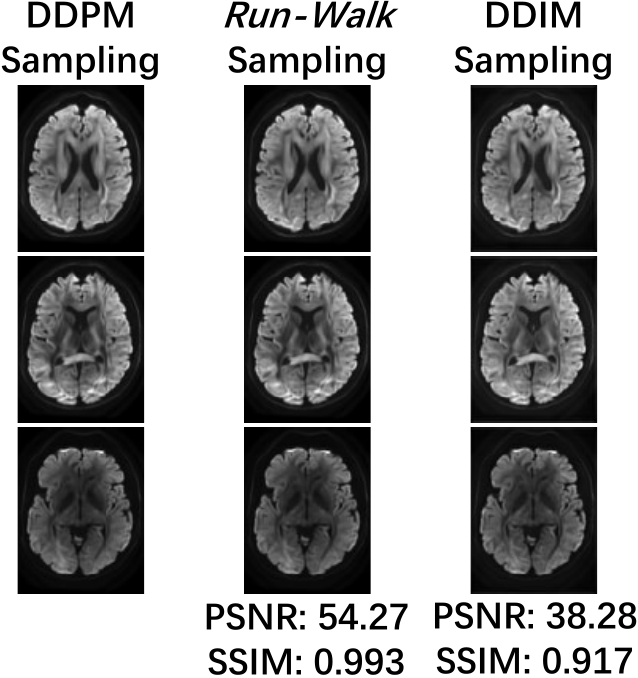}
  \caption{DDPM sampling \textit{v.s.} \textit{Run-Walk} accelerated sampling (didn't implement an adaptive termination in Section \ref{Towards iterative and adjustable refinement}) \textit{v.s.} DDIM sampling. For all results, $\eta = 0$. PSNR, SSIM are calculated using DDPM sampling results as references. This indicates that the sampling results from \textit{Run-Walk} sampling are closer to the sampling results when accelerated sampling is not used.}
  \label{runwalk_sampling}
\end{figure}

In Fig. \ref{runwalk_sampling}, we show results obtained by different sampling strategies and metrics (averaged PSNR and SSIM on all volumes) calculated using DDPM sampling results as references. Directly performing DDIM sampling on a pre-trained model may lead to biased results (use DDPM sampling results as references). \textit{Run-Walk} accelerated sampling significantly improves sampling speed and reduces inference time while maintaining the sampling quality relatively unchanged.

\paragraph{About sampling time}
\label{About sampling time}
We do experiments to demonstrate that the additional computations in Section \ref{Towards iterative and adjustable refinement} do not impact the sampling speed. Firstly, we set $\mathcal{CSNR}$ to 1, which means that all slices undergo the extra computations and the whole sampling process since $\mathcal{CSNR}$ is sufficiently large. Subsequently, we remove the extra computational operations and perform sampling again. The sampling time in the first experiment was 1.19 seconds per slice. In contrast, in the second experiment, it was 1.18 seconds per slice, which indicates that the additional operations have minimal impact on the sampling speed. In Table \ref{Sampling Time per Slice}, the sampling time per individual slice is presented for different $\mathcal{CSNR}$. We find that When $\mathcal{CSNR}$ is low ($\mathcal{CSNR}=0.040$), \textit{the sampling time is 0.0395 seconds per slice}. This indicates that our adaptive termination and \textit{Run-Walk} accelerated sampling greatly reduce the sampling time.

\begin{figure}
  \centering
  \includegraphics[scale = 0.6]{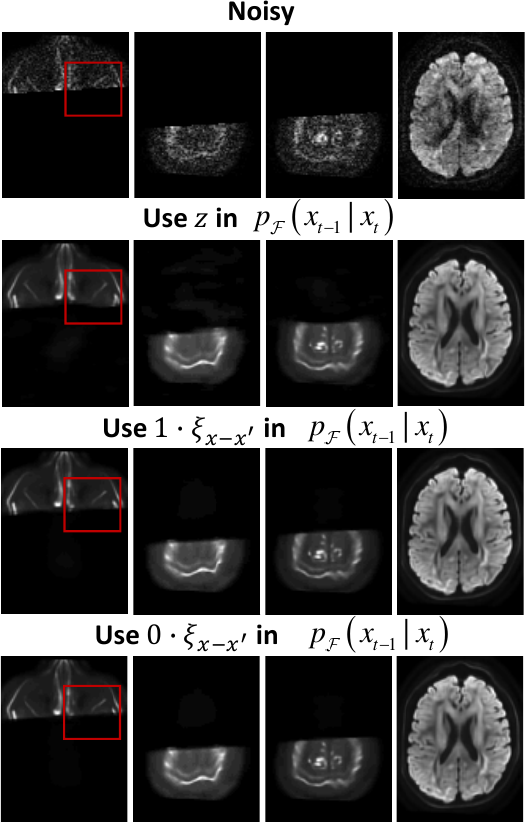}
  \caption{Using $z$ during ${p_\mathcal{F}}\left( {{x_{t - 1}}|{x_t}} \right)$ \textit{v.s.} Using $1 \cdot {\xi_{x - x'}}$ during ${p_\mathcal{F}}\left( {{x_{t - 1}}|{x_t}} \right)$ \textit{v.s.} Using $0 \cdot {\xi_{x - x'}}$ during ${p_\mathcal{F}}\left( {{x_{t - 1}}|{x_t}} \right)$ (all results didn't implement an adaptive termination mentioned in Section \ref{Towards iterative and adjustable refinement} during the sampling process).}
  \label{Diprocess}
\end{figure}

\paragraph{Using ${\xi_{x - x'}}$ in reverse process}
In Fig. \ref{Diprocess}, we demonstrate the importance of using ${\xi_{x - x'}}$ and setting $\eta = 0$ in the sampling process. It can be observed from the final results that in the central slices (with more brain tissue), using $z \sim {\cal N}\left( {\mathbf{0},\mathbf{I}} \right)$ during the sampling process ${p_\mathcal{F}}\left( {{x_{t - 1}}|{x_t}} \right)$ only leads to subtle differences in the denoised results. However, in the edge slices (with less brain tissue), using $z$ significantly impacts the sampling results, resulting in additional regions that appear inexplicably (highlighted by the red box, and these additional regions don't appear in noisy data). During the sampling process, DDM2 uses $z$. Because our sampling process is deterministic, according to the experiments in DDIM~\citep{song2020score}, we set $\eta= 0$. We further demonstrated the sampling results in the figure with $\eta=1$ and $\eta = 0$. When $\eta=1$, although the presence of unexpected regions is reduced, some still remain. However, when $\eta = 0$, such issues don't arise.

\begin{table}[h]
  \centering
  \caption{Sampling time per slice for different $\mathcal{CSNR}$ (Stanford HARDI). We set different \(\mathcal{CSNR}\) parameters for Run-Walk accelerated sampling and DDPM sampling to perform adaptive termination.}
  \label{Sampling Time per Slice}
  \begin{tabular}{ccc}
    \toprule
    $\mathcal{CSNR}$ & Time (s) for Run-walk & Time (s) for DDPM \\
    \midrule
    0.04 & 0.0395 & 0.327 \\
    0.045 & 0.115 & 0.739 \\
    0.05 & 0.626 & 2.01 \\
    0.055 & 1.08 & 5.48 \\
    0.06 & 1.11 & 6.97 \\
    1 & 1.18 & 11.5 \\
    \bottomrule
  \end{tabular}
\end{table}

\section{More comparisons with competing methods}
\label{More comparisons with competing methods}

\subsection{Compare with different Patch2Self settings}
\label{A-Compare with different Patch2Self settings}

\begin{table}
\caption{↑$R^2$ of microstructure model fitting on CSD \& DTI. \textbf{Bold} and \underline{Underline} fonts denote the best and the second-best performance, respectively.}\label{microstruture diffP2S table}
\begin{center}
\renewcommand{\arraystretch}{1.0}
\begin{tabular}{lcccc}
\toprule
& \multicolumn{2}{c}{CSD} & \multicolumn{2}{c}{DTI}\\
\cmidrule{2-3}\cmidrule(l{2pt}r{2pt}){4-5}
Method & CC & CSO & CC & CSO \\

\midrule
Noisy & 0.797 & 0.614 & 0.789 & 0.484 \\
ASCM & 0.934 & 0.844 & 0.942 & 0.789 \\
DIP & 0.868 & 0.477 & 0.875 & 0.381 \\
Nr2N & \underline{0.959} & \underline{0.908} & \underline{0.961} & \underline{0.872} \\
P2S (OLS) & 0.927 & 0.754 & 0.725 & 0.675 \\
P2S (Ridge) & 0.927 & 0.757 & 0.927 & 0.673 \\
P2S (Lasso) & 0.824 & 0.471 & 0.816 & 0.429 \\
P2S (OLS, r=1) & 0.934 & 0.832 & 0.950 & 0.735 \\
DDM2 & 0.863 & 0.810 & 0.845 & 0.790 \\
OURS & \textbf{0.967} & \textbf{0.939} & \textbf{0.976} & \textbf{0.876} \\
\bottomrule
\end{tabular}
\end{center}
\end{table}

\begin{figure}
  \centering
  \includegraphics[scale = 0.65]{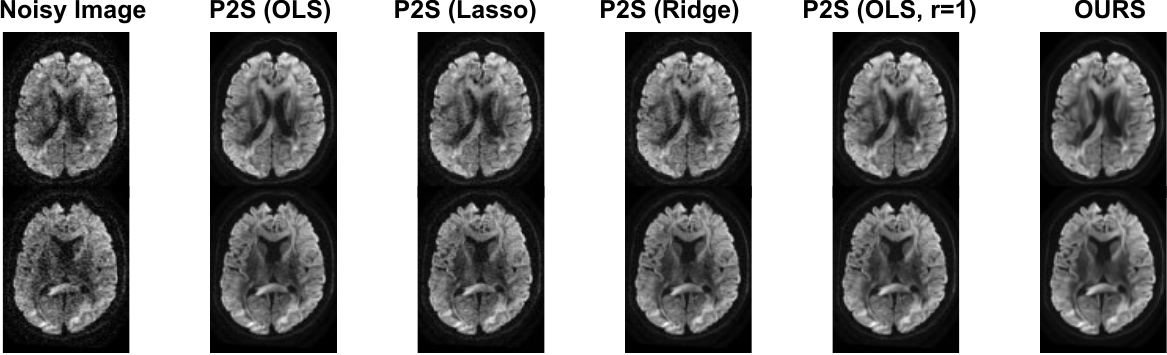}
  \caption{Comparisons with different Patch2Self experimental settings. ``OURS'' results are obtained when ${\mathcal{CSNR}}=0.040$.}
  \label{differentP2S}
\end{figure}

In Fig. \ref{differentP2S}, we show additional results on comparisons with different Patch2Self experimental settings. Our modifications are limited to the denoiser type (OLS, Lasso, Ridge) and patch radius, following the official repository of Patch2Self~\citep{fadnavispatch2self,fadnavis2020patch2self,garyfallidis2014dipy}. The term ``(r=1)'' indicates changing the patch radius to 1, while the patch radius is assumed to be 0 if not specified. Modifying the denoiser type and patch radius in Patch2Self does not yield substantial improvements in the results. Altering the denoiser type does not impact the overall denoising time, whereas changing the patch radius significantly increases the overall denoising time. In our experiments, employing the OLS denoiser required 4 hours, while utilizing OLS with a patch radius of 1 took 26 hours.

\begin{figure}
  \centering
  \includegraphics[scale = 0.39]{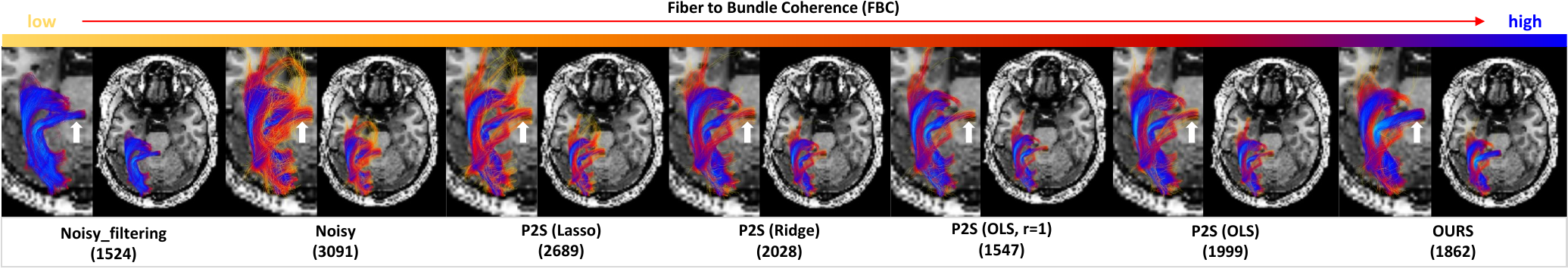}
  \caption{Density map of FBC projected on the streamlines of the OR bundles. The numbers in parentheses represent the number of streamlines.}
  \label{tractography_diffP2S}
\end{figure}

In Fig. \ref{tractography_diffP2S}, we show additional tractography results on comparisons with different Patch2Self experimental settings. Although the number of streamlines is the lowest in the ``(OLS, r=1)'' experimental setting, \textit{it still misses the high FBCs indicated by the white arrows}. There are no significant differences in the results in the remaining experimental settings. Considering the computational burden when setting the patch radius to 1, we suggest setting the patch radius of Patch2Self to 0 to improve efficiency.

In Table \ref{microstruture diffP2S table}, we show quantitative results (on microstructure model fitting) on comparisons with different Patch2Self experimental settings. Varied experimental settings can influence the performance of microstructure model fitting. Nonetheless, these modifications do not change the rankings of the best and second-best results.

\subsection{Compare with Neighbor2Neighbor}
\label{A-Compare with Neighbor2Neighbor}

\begin{figure}
  \centering
  \includegraphics[scale = 0.65]{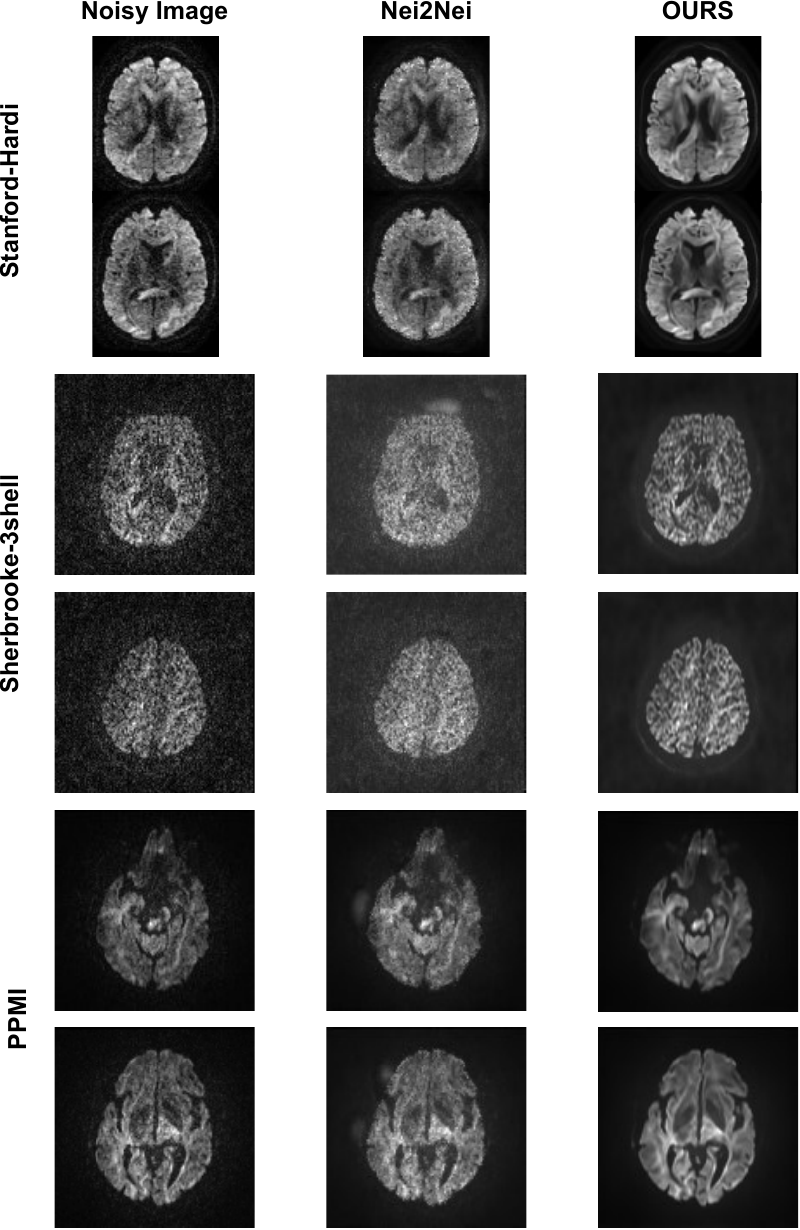}
  \caption{Qualitative comparisons with Neighbor2Neighbor. ``OURS'' results are obtained when ${\mathcal{CSNR}}=0.040$.}
  \label{Nei2Nei}
\end{figure}

In Section \ref{Statistical self-supervised Image Denoising}, the mentioned methods exhibit a significant drop in performance when confronted with real-world noisy images, particularly when the explicit noise model is unknown. To make up for this, Neighbor2Neighbor (Nei2Nei)~\citep{huang2021neighbor2neighbor} and Zero-shot Noise2Noise~\citep{mansour2023zero} sub-sample individual noisy images to create training pairs and are more robust against real-world noise. We compare our method with Nei2Nei using the same model in Di-Fusion. We implement Nei2Nei with parameters set to the default values specified in their official repository\footnote{https://github.com/TaoHuang2018/Neighbor2Neighbor}.

\begin{figure}
  \centering
  \includegraphics[scale = 0.7]{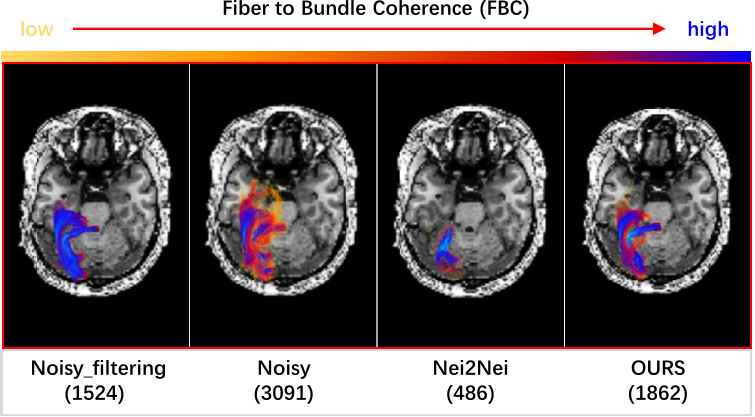}
  \caption{Density map of FBC projected on the streamlines of the OR bundles. The numbers in parentheses represent the number of streamlines.}
  \label{Nei2NeiFBC}
\end{figure}

The qualitative results are in Fig. \ref{Nei2Nei}. It can be observed that Nei2Nei does not perform well in denoising, as there are partial jagged artifacts in the image and significant changes in grayscale. This may be the reason that Nei2Nei denoising relies on the structural similarity of the neighboring regions in the image. This can also be seen in the qualitative results of Nei2Nei, where the images maintain structural similarity in sub-sampled noisy images, leading to better denoising results. In Fig. \ref{Nei2NeiFBC}, it can be observed that the density map of FBC projected on the streamlines of the OR bundles is missing a significant number of FBCs; thus the denoising results of Nei2Nei are unsuitable for modeling tasks.

\section{DDM2: stage 1 has a huge impact on final results}
\label{A-DDM2: Stage 1 has a huge impact on final results}

\begin{figure}
  \centering
  \includegraphics[scale = 0.70]{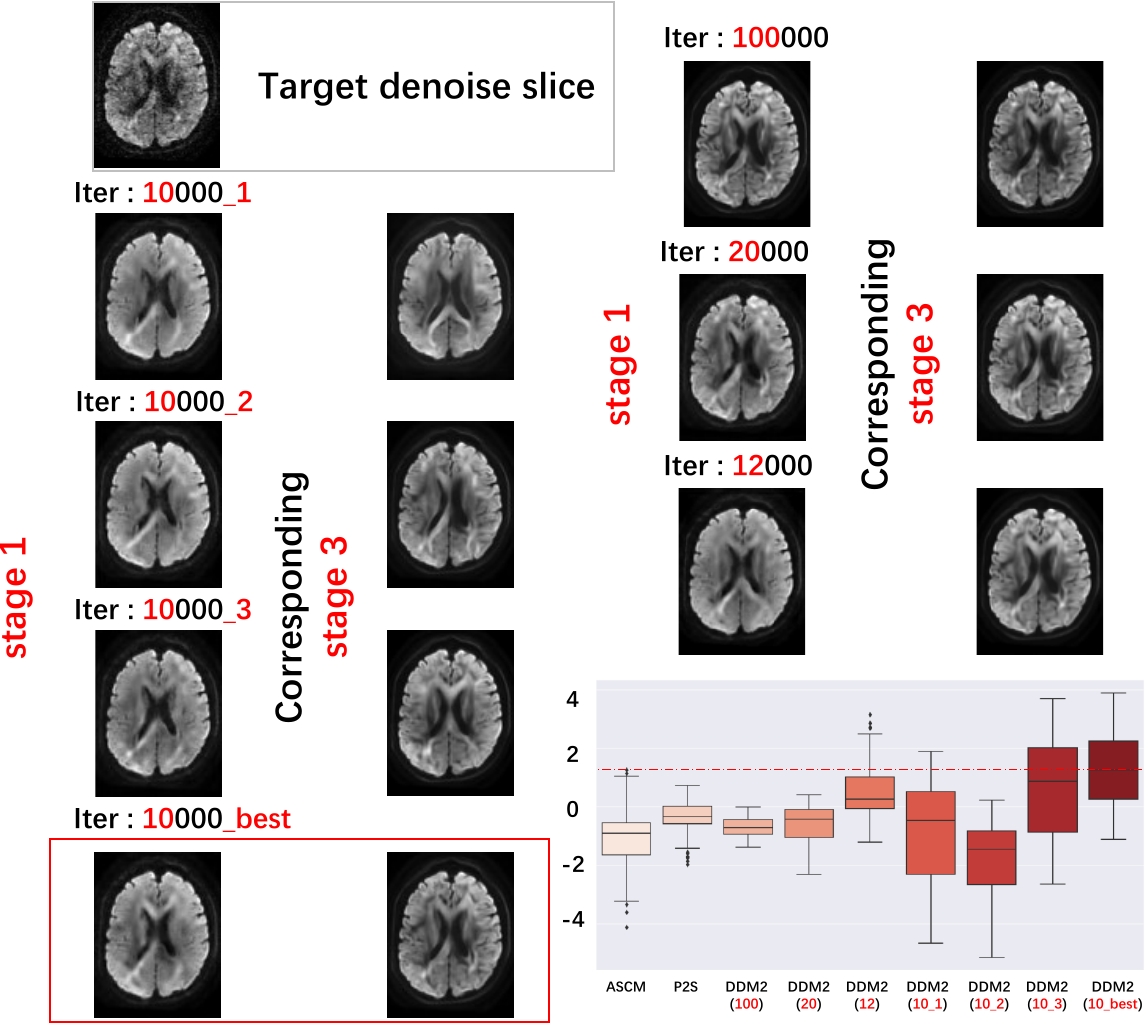}
  \caption{DDM2 unstable model outcomes. (right below) Show the CNR metric of different experiment settings; CNR/SNR metrics show the same trend. The red box highlights the best results obtained using the parameters from the official code repository. Having a stable Stage 1 often leads to poor performance in CNR/SNR metrics, the red color within the parentheses represents the settings corresponding to each experiment.}
  \label{DDM2BAD}
\end{figure}

In Fig. \ref{DDM2BAD}, we present the results of DDM2's first stage and corresponding third stage on the Stanford HARDI dataset. By utilizing the hyperparameters from the DDM2 official repository and conducting experiments (\textit{only the training iteration in stage 1 was modified}, the official training iteration is set to $10e^4$), we have observed that coarser outcomes in the first stage yield more striking yet less stable denoising results in the final stage. Conversely, deterministic outcomes in the first stage result in more stable but uninteresting denoising results in the final stage. Different first stage results lead to drastically distinct outcomes in the third stage. The CNR and SNR scores show significant differences between different first stage results (Fig. \ref{DDM2BAD} (right below)). Please note that the subsequent experiments we conduct on DDM2 are using their best results.

\end{document}